\documentclass[12pt,epsf,qsymbols]{article}
\pdfoutput=1
\usepackage[utf8]{inputenc}
\usepackage[normalem]{ulem}
\usepackage[english]{babel}
\usepackage{tabularx}
\usepackage{array}
\usepackage{graphics}
\usepackage[pdftex]{graphicx}
\usepackage{psfrag}
\usepackage{epsfig}
\usepackage{subfig}
\usepackage{mathtools}
\usepackage{amssymb}
\usepackage{bbold}
\usepackage{setspace}
\usepackage{rotating}
\usepackage{colortbl}
\usepackage{longtable}
\usepackage{slashed}
\usepackage{braket}
\usepackage{lineno}
\makeatletter
\usepackage{textcomp}
\usepackage[usenames,dvipsnames]{xcolor}
\usepackage{relsize}
\usepackage{cite}

\usepackage{float}

\usepackage{bbm}
\usepackage{bm}
\usepackage{enumerate}
\usepackage{amsmath}
\usepackage{verbatim}

\usepackage{hyperref}
\hypersetup{colorlinks,citecolor=nicegreen,linkcolor=niceblue}
\hypersetup{colorlinks=true}

\allowdisplaybreaks

\setlongtables

\newcommand{\msbar}{{\overline{\rm MS}}}

\newcommand{\cb}{{\mathcal B}}
\newcommand{\bea}{\begin{eqnarray}}
\newcommand{\eea}{\end{eqnarray}}
\newcommand{\beq}{\begin{equation}}
{
\newcommand{\eeq}{\end{equation}}
\newcommand{\ec}{\end{center}}
\newcommand{\bc}{\begin{center}}

\newcommand{\gev}{{\rm GeV}}

\newcommand{\pdir}{p\kern -5.2pt\raise 0.2ex\hbox {/}}

\newcommand{\vdir}{v\kern -5.75pt\raise 0.15ex\hbox {/}}
\newcommand{\kdir}{k\kern -5.75pt\raise 0.15ex\hbox {/}}
\newcommand{\epsdir}{\epsilon\kern -5.0pt\raise 0.15ex\hbox {/}}
\newcommand{\bvdir}{\bar{v}\kern -5.75pt\raise 0.15ex\hbox {/}}
\newcommand{\Ddir}{D\kern -7.75pt\raise 0.20ex\hbox {/}}
\newcommand{\Adir}{A\kern -7.75pt\raise 0.20ex\hbox {/}}
\newcommand{\ldir}{l\kern -5.0pt\raise 0.2ex\hbox{/}}
\newcommand{\varepsdir}{\varepsilon\kern -5.5pt\raise 0.15ex\hbox{/}}

\newcommand{\nn}{\nonumber}

\makeatother

\definecolor{niceblue}{rgb}{0.15,0.15,0.6}
\definecolor{nicegreen}{rgb}{0.1,0.5,0.1}
\definecolor{Red}{rgb}{1.,0.,0.}

\definecolor{Green}{rgb}{0.2,.7,0.2}

\begin{document}
\unitlength = 1mm

\thispagestyle{empty} 
\begin{flushright}
\begin{tabular}{l}
{\tt \footnotesize LPT-Orsay-17-05}\\
\end{tabular}
\end{flushright}
\begin{center}
\vskip 3.4cm\par
{\par\centering \textbf{\LARGE  
\Large \bf Two Higgs Doublet Models and $b\to s$ exclusive decays }}
\vskip 1.2cm\par
{\scalebox{.85}{\par\centering \large  
\sc Pere Arnan$^a$, Damir Be\v{c}irevi\'c$^b$, Federico Mescia$^a$ and Olcyr Sumensari$^{b,c}$}
{\par\centering \vskip 0.7 cm\par}
{\sl 
$^a$~Departament de F\'isica Qu\`antica i Astrof\'isica (FQA),\\
Institut de Ci\`encies del Cosmos (ICCUB), Universitat de Barcelona (UB), Spain.}\\
{\par\centering \vskip 0.25 cm\par}
{\sl 
$^b$~Laboratoire de Physique Th\'eorique (B\^at.~210)\\
CNRS, Univ. Paris-Sud, Universit\'e Paris-Saclay, 91405 Orsay, France.}\\
{\par\centering \vskip 0.25 cm\par}
{\sl 
$^c$~Instituto de F\'isica, Universidade de S\~ao Paulo, \\
 C.P. 66.318, 05315-970 S\~ao Paulo, Brazil.}\\

{\vskip 1.65cm\par}}
\end{center}

\vskip 0.85cm
\begin{abstract}
We computed the leading order Wilson coefficients relevant to all the exclusive $b\to s\ell^+\ell^-$ decays in the framework of the Two Higgs Doublet Model (2HDM) with a softly broken $\mathbb{Z}_2$ symmetry by including the $\mathcal{O}(m_b)$ corrections. We elucidate the issue of appropriate matching between the full and the effective theory when dealing with the (pseudo-)scalar operators for which keeping the external momenta different from zero is necessary. We then make a phenomenological analysis by using the measured $\cb(B_s\to \mu^+\mu^-)$ and $\cb(B\to K \mu^+\mu^-)_{\mathrm{high}-q^2}$, for which the hadronic uncertainties are well controlled, and discuss their impact on various types of 2HDM. 
A brief discussion of the decays with $\tau$-leptons in the final state is provided too.
\end{abstract}
\newpage
\setcounter{page}{1}
\setcounter{footnote}{0}
\setcounter{equation}{0}
\noindent

\renewcommand{\thefootnote}{\arabic{footnote}}

\setcounter{footnote}{0}

\tableofcontents

\newpage


\section{Introduction}

Physical processes driven by the flavor changing neutral currents (FCNC) are forbidden in the Standard Model (SM) at tree level. Since they occur through loops, 
their measurements offer a low energy window to the particle content in the loops. In other words, they do not only represent a fine test of validity of the SM, but they also offer an opportunity 
to look for the effects of physics (particles) beyond the SM (BSM) at low energies. The main obstacle to the accurate comparison between the SM theory and the experimental data lies in the fact that the non-perturbative QCD effects are not under full theoretical control. While the solution to non-perturbative QCD is lacking, in some situations the hadronization effects can be solved by means of numerical simulations of QCD on the lattice (LQCD). Over the past couple of decades we witnessed a huge progress in reducing the uncertainties in the LQCD results. Nowadays, an excellent theoretical control of the neutral meson mixing processes promoted those FCNC processes to viability tests of the New Physics (NP) model candidates. 
Besides the oscillation frequencies of the neutral meson systems, the processes based on $b\to s$ transitions received a great deal of attention in the particle physics community. While the inclusive and exclusive processes based on the penguin-induced $b\to s\gamma$ decay have been, and still are, a very significant constraint when building a NP model, the processes based on $b\to s\ell^+\ell^-$ received a huge attention because they allow to access another types of penguin and box diagrams. With the advent of the Large Hadron Collider (LHC) the measurement of $\cb(B_s\to \mu^+\mu^-)$ became possible~\cite{CMS:2014xfa} and the result appeared to be somewhat lower than predicted~\cite{Bobeth:2013uxa}. The spectrum of $d\cb(B\to K\mu^+\mu^-)/dq^2$ has been measured~\cite{Aaij:2014pli} too and in the range of large $q^2$'s it appears to be larger than predicted~\cite{Bobeth:2011nj,Becirevic:2015asa}. A full angular analyses of $\cb(B\to K^\ast \mu^+\mu^-)$~\cite{Aaij:2014pli,Aaij:2016flj} and $\cb(B_s\to \phi \mu^+\mu^-)$~\cite{Aaij:2015esa} revealed discrepancies in several observables with respect to their SM predictions~\cite{Altmannshofer:2014rta}. Moreover, the measurement of $R_K= \cb^\prime (B\to K \mu^+\mu^-)/\cb^\prime (B\to K e^+e^-)$~\cite{Aaij:2014ora} was shown to be significantly lower than predicted (by $2.4\sigma$)~\cite{Bordone:2016gaq}.~\footnote{We use $\cb^\prime (B\to K\ell^+\ell^-)$ to indicate that the decay rate has not been fully integrated but only within the window $q^2\in [1,6]\ \gev^2$.} Those new experimental data helped discarding several NP models and are currently used as constraints in building a NP model. 

Simultaneously with the research of FCNC processes, the LHC experiments allowed observing the missing ingredient of the SM, the Higgs boson, the mass of which has been found to be $m_h = 125.09(24)$~GeV~\cite{Aad:2015zhl}. While this was a milestone of the LHC, the pending question of hierarchy of scales remains open and a quest for physics BSM continues. One of the minimalistic approaches to building a model of physics BSM is to extend the Higgs sector by introducing an extra Higgs doublet. Phenomenology in the scenarios with two Higgs doublets appears to be very rich and the associated models are generically called the Two Higgs Doublet Models (2HDM), cf. e.g.~\cite{Gunion:1989we,Gunion:2002zf,Branco:2011iw}. Nowadays the experimental search of the additional Higgs bosons is one of the main goals at LHC, in particular that of the charged Higgs boson~\cite{Akeroyd:2016ymd}. Like in the SM, introducing fermions to the 2HDM context results in a plethora of new parameters. To restrain the number of those parameters and to prevent from appearance the FCNC at tree level it is common to assume a peculiar pattern of Yukawa couplings. To test those assumptions one needs to compare many measured observables with theoretical expressions derived in SM with the extended Higgs sector. In this paper we elaborate a few lessons one can learn from the measured $b\to s\mu^+\mu^-$ processes about 2HDM with a softly broken $\mathbb{Z}_2$ symmetry. In doing so we will use two observables, namely $\cb(B_s\to \mu^+\mu^-)$ and $\cb(B\to K \mu^+\mu^-)_{\mathrm{high}-q^2}$, which are very well measured experimentally and for which the theoretical control of the corresponding hadronic uncertainties is established by the LQCD computations~\cite{Aoki:2016frl}. For other observables the theoretical uncertainties are not as well assessed and one might run a risk of interpreting the unknown hadronic  uncertainties as signals of physics BSM. We should also emphasize that 2HDM alone cannot explain $R_K^\mathrm{SM}>R_K^\mathrm{exp}$, and in this paper we will ignore the channels with electrons in the final state. 
A study along the line we are pursuing here has been initiated in Ref.~\cite{Li:2014fea} in which the authors computed the Wilson coefficients in the Aligned 2HDM (A2HDM),  for the operators relevant to the $B_s\to \mu^+\mu^-$ decay. In this paper we revisit their computation and extend it to include the operators that are needed for the phenomenological analysis of $B\to K^{(\ast)}\ell^+\ell^-$ and other similar decays. While we broadly agree with the results of Ref.~\cite{Li:2014fea}, there are a couple of points in which we disagree. We will examine those points, compute the remaining Wilson coefficients and use our results to discuss the phenomenological consequences on the 2HDM scenarios by comparing $\cb(B_s\to \mu^+\mu^-)^\mathrm{2HDM}$ and $\cb(B\to K \mu^+\mu^-)_{\mathrm{high}-q^2}^\mathrm{2HDM}$ with their experimental values. We will then discuss the consequences on the similar decays with $\tau$-leptons in the final state.

The remainder of this paper is organized as follows: In Sec.~\ref{sec:2hdm} we remind the reader of the main general constraints on the spectrum of scalars in 2HDM and perform a scan of parameters by assuming the lowest CP-even Higgs state to be the one measured at LHC. In Sec.~\ref{sec:eff} we write the low energy effective theory and present our results for all the Wilson coefficients in Sec.~\ref{sec:wc}. We compare our results with the existing ones (in the limits in which the comparison can be made) in Sec.~\ref{sec:compare} and elucidate the subtleties related to the matching procedure in the between the full (2HDM) and effective theories in Sec.~\ref{sec:MATCH}. Phenomenological discussion is made in Sec.~\ref{sec:pheno0} and Sec.~\ref{sec:pheno}. We briefly conclude in Sec.~\ref{sec:concl}.

\section{General constraints on 2HDM}
\label{sec:2hdm}

In this Section we remind the reader of the basic ingredients of 2HDM, enumerate the parameters of the model and list the main general constraints 
on the spectrum of scalars which are then used to perform a scan of allowed parameters to obtain the allowed ranges of the Higgs masses and couplings.

\subsection{2HDM}
We consider a general CP-conserving 2HDM with a softly broken $\mathbb{Z}_2$ symmetry.
The most general potential can then be written as

\begin{align}
\label{eq:V2hdm}
V(\Phi_1,\Phi_2) = m_{11}^2\Phi_1^\dagger\Phi_1 & +m_{22}^2\Phi_2^\dagger\Phi_2 + m_{12}^2(\Phi_1^\dagger\Phi_2+\Phi_2^\dagger\Phi_1)+\dfrac{\lambda_1}{2}(\Phi_1^\dagger \Phi_1)^2+\dfrac{\lambda_2}{2}(\Phi_2^\dagger \Phi_2)^2\nonumber\\
&+\lambda_3 \Phi_1^\dagger\Phi_1 \Phi_2^\dagger\Phi_2+\lambda_4 \Phi_1^\dagger\Phi_2 \Phi_2^\dagger\Phi_1+\dfrac{\lambda_5}{2}\left[ (\Phi_1^\dagger\Phi_2)^2+(\Phi_2^\dagger\Phi_1)^2\right],
\end{align}
where the term proportional to $m_{12}^2$ accounts for the soft breaking of $\mathbb{Z}_2$.~\footnote{
We remind the reader that the $\mathbb{Z}_2$ symmetry ($\Phi_1 \to \pm \Phi_1,\, \Phi_2 \to \mp \Phi_2$) of the Lagrangian forbids  
transitions $\Phi_1 \leftrightarrow \Phi_2$. Soft breaking of $\mathbb{Z}_2$ means that such transitions may occur only due to dimension-$2$ operators 
(terms proportional to $m_{12}^2$ in Eq.~\eqref{eq:V2hdm}) so that $\mathbb{Z}_2$ remains preserved at very short distances, cf. discussion in Ref.~\cite{ginzburg}.} The scalar doublets $\Phi_a$ ($a=1,2$) can be parameterized as

\begin{align}
\Phi_a(x) = \begin{pmatrix}
\phi_a^+(x) \\ 
\frac{1}{\sqrt{2}}\left[v_a+\rho_a(x)+i \eta_a(x)\right]
\end{pmatrix}, 
\end{align}
with $v_{1,2}\geq 0$ being the vacuum expectation values satisfying $v^\mathrm{SM}=\sqrt{v_1^2+v_2^2}$, already known from experiments, $v^\mathrm{SM}=246.22$~GeV~\cite{Olive:2016xmw}. In the following, for notational simplicity, we will drop the argument  of the Higgs fields. Two of the six fields are Goldstone bosons, while the remaining ones are four massive scalars: two CP-even states ($h$, $H$), one CP-odd state ($A$), and one charged Higgs ($H^\pm$). These fields are defined as 
\begin{align}
\begin{pmatrix}
\phi_1^+\\ \phi_2^+
\end{pmatrix}
&=\begin{pmatrix}
\cos \beta & - \sin \beta\\
\sin\beta  & \cos \beta
\end{pmatrix}
\begin{pmatrix}
G^+\\ H^+
\end{pmatrix},\qquad
\begin{pmatrix}
\eta_1\\ \eta_2
\end{pmatrix}
=\begin{pmatrix}
\cos \beta & - \sin \beta\\
\sin\beta  & \cos \beta
\end{pmatrix}
\begin{pmatrix}
G^0\\ A
\end{pmatrix},
\end{align}
and
\begin{align}
\begin{pmatrix}
\rho_1\\ \rho_2
\end{pmatrix}
&=\begin{pmatrix}
\cos \alpha & - \sin \alpha\\
\sin\alpha  & \cos \alpha
\end{pmatrix}
\begin{pmatrix}
H\\ h
\end{pmatrix},
\end{align}
The mixing angles $\alpha$ and $\beta$ satisfy
\begin{align}
\tan \beta = \frac{v_2}{v_1},\qquad \tan 2\alpha = \dfrac{2(-m_{12}^2+\lambda_{345}v_1 v_2)}{m_{12}^2(v_2/v_1-v_1/v_2)+\lambda_1 v_1^2-\lambda_2 v_2^2},
\end{align}
with $\lambda_{345}\equiv \lambda_3+\lambda_4+\lambda_5$.
The masses of the physical scalars can be written in terms of parameters which appear in the potential as

\begin{align}
\label{eq:massesH}
	m_H^2 &= M^2 \sin^2(\alpha-\beta)+\left(\lambda_1 \cos^2\alpha \cos^2\beta+\lambda_2 \sin^2\alpha \sin^2\beta+\frac{\lambda_{345}}{2}\sin 2\alpha \sin 2\beta\right)v^2,\\
\label{eq:massesh}
	m_h^2 &= M^2 \cos^2(\alpha-\beta)+\left(\lambda_1 \sin^2\alpha \cos^2\beta+\lambda_2 \cos^2\alpha \sin^2\beta-\frac{\lambda_{345}}{2}\sin 2\alpha \sin 2\beta\right)v^2,\\
\label{eq:massesA}
	m_{A}^2 &= M^2-\lambda_5 v^2,\\
\label{eq:massesHp}
	m_{H^\pm}^2 &= M^2-\frac{\lambda_{4}+\lambda_5}{2} v^2,
\end{align}
where the $\mathbb{Z}_2$ breaking term is now parameterized via $M^2\equiv \dfrac{m_{12}^2}{\sin \beta \cos \beta}$.

In the Yukawa sector, the $\mathbb{Z}_2$ symmetry becomes particularly important as it prevents the flavor changing processes to appear at tree level~\cite{Glashow:1976nt}. Furthermore it 
enforces that each type of the right-handed fermion couples to a single Higgs doublet. Four choices are then possible and they are called Type I, II, X (Lepton Specific) and Z (Flipped) 2HDM~\cite{Branco:2011iw}.~\footnote{The model that we call Type~Z or Flipped 2HDM is sometimes referred to as Type~Y.} To be more specific, we first write the Yukawa Lagrangian as 
\begin{align}
\label{eq:lyuk}
\mathcal{L}_Y = &- \dfrac{\sqrt{2}}{v} H^+ \Big{\lbrace} \bar{u}~[\zeta_d \, V m_d P_R-\zeta_u \, m_u V P_L]~d +\zeta_\ell \, \bar{\nu} m_\ell P_R \ell \Big{\rbrace}\nonumber \\
&-\dfrac{1}{v}\sum_{f,\varphi_i^0\in\{h,H,A\}} \xi_f^{\varphi_i^0} \varphi_i^0 \, \Big{[}\bar{f} m_f P_R f \Big{]}+\mathrm{h.c.,}
\end{align}
where $u$ and $d$ stand for the up- and down-type quark, $\ell$ is a lepton flavor, $f$ stands for a generic fermion, $V$ for the Cabibbo--Kobayashi-Maskawa (CKM) matrix, and $P_{L,R}= (1\mp\gamma_5)/2$. A specific choice of parameters $\zeta_f$ corresponds to the above mentioned  types of 2HDM, which we also summarize in Table~\ref{tab:y2hdm}. Notice that the couplings $\xi_f^{\varphi_i^0}$ appearing in the neutral Lagrangian part can be mapped onto the charged ones via
\begin{align}
	\xi^h_f &= \sin(\beta-\alpha) +\cos(\beta-\alpha) \zeta_f, \nonumber \\
	\xi^H_f &= \cos(\beta-\alpha) -\sin(\beta-\alpha) \zeta_f, \nonumber \\
	\xi^A_{u}&=-i\zeta_u,\qquad \xi^A_{d,\ell}=i \zeta_{d,\ell}.
\end{align}
\begin{table}[ht!]
\renewcommand{\arraystretch}{1.5}
\centering
\begin{tabular}{|c|c|c|c|}
\hline 
Model & $\zeta_d$ & $\zeta_u$ & $\zeta_\ell$ \\ \hline\hline
Type I & $\cot \beta$  &  $\cot \beta$ & $\cot \beta$ \\  
Type II & $-\tan \beta$  &  $\cot \beta$ & $-\tan \beta$ \\  
Type X (lepton specific) & $\cot \beta$  &  $\cot \beta$ & $-\tan \beta$  \\
Type Z (flipped) & $-\tan \beta$  &  $\cot \beta$ & $\cot \beta$ \\
  \hline
\end{tabular}
\caption{ \sl Couplings $\zeta_f$ in various types of 2HDM.}
\label{tab:y2hdm} 
\end{table}

\subsection{General Constraints and Scan of Parameters}
\label{sec:scan}
To perform a thorough scan of scalars in a general 2HDM we use the general constraints summarized below. 

\begin{itemize}
	\item \textbf{\underline{Stability}}:
	
	 To ensure that the scalar potential is bounded from below, the quartic couplings should satisfy the relations~\cite{Gunion:2002zf}

\begin{equation}
	\lambda_{1,2}>0,\qquad \lambda_3>-(\lambda_1 \lambda_2)^{1/2},\qquad \mathrm{and} \qquad \lambda_3+\lambda_4- \vert \lambda_5 \vert  >-(\lambda_1 \lambda_2)^{1/2}.
\end{equation}		

Furthermore, the stability of the electroweak vacuum implies that

\begin{align}
m_{11}^2+\dfrac{\lambda_1 v_1^2}{2}+\dfrac{\lambda_3 v_2^2}{2} &= \frac{v_2}{v_1} \left[ m_{12}^2 - (\lambda_4+\lambda_5)\dfrac{v_1 v_2}{2}\right],\\
m_{22}^2+\dfrac{\lambda_2 v_2^2}{2}+\dfrac{\lambda_3 v_1^2}{2} &= \frac{v_1}{v_2} \left[ m_{12}^2 - (\lambda_4+\lambda_5)\dfrac{v_1 v_2}{2}\right],
\end{align} 

which then allows us to express $m_{11}^2$ and $m_{22}^2$ in terms of the soft $\mathbb{Z}_2$ breaking term $m_{12}^2$ and the quartic couplings $\lambda_{1-5}$. These constraints should be combined with the necessary and sufficient condition that the minimum developed at $(v_1,v_2)$ is global~\cite{Barroso:2013awa}:

\begin{equation}
m_{12}^2 \left(m_{11}^2-m_{22}^2 \sqrt{\lambda_1/\lambda_2} \right) \left( \tan \beta - \sqrt[4]{\lambda_1/\lambda_2}\right) >0.
\end{equation} 

\item \textbf{\underline{Perturbative Unitarity}}:
	
An important constraint on the spectrum of scalars within 2HDM stems from the unitarity requirement of the $S$-wave component of the scalar scattering amplitudes. That condition implies the following inequalities~\cite{Kanemura:1993hm,Swiezewska:2012ej}
\begin{equation}
|a_\pm|, |b_\pm|, |c_\pm|, |f_\pm|, |e_{1,2}|, |f_1|, |p_1| < 8 \pi,
\end{equation}
where
\begin{align}
\begin{split}
a_\pm &= \dfrac{3}{2}(\lambda_1+\lambda_2)\pm \sqrt{\dfrac{9}{4}(\lambda_1-\lambda_2)^2+(2\lambda_3+\lambda_4)^2},\\
b_\pm &= \dfrac{1}{2}(\lambda_1+\lambda_2)\pm \dfrac{1}{2} \sqrt{(\lambda_1-\lambda_2)^2+4\lambda_4^2},\\
c_\pm &= \dfrac{1}{2}(\lambda_1+\lambda_2)\pm \dfrac{1}{2} \sqrt{(\lambda_1-\lambda_2)^2+4\lambda_5^2},\\
e_1 &= \lambda_3 + 2 \lambda_4 -3\lambda_5,\hspace*{3cm}
e_2 = \lambda_3-\lambda_5,\\
f_+ &= \lambda_3+2 \lambda_4+3\lambda_5, \hspace*{2.9cm} f_- =\lambda_3+\lambda_5,\\
f_1 &= \lambda_3+\lambda_4, \hspace*{4.3cm}p_1 = \lambda_3-\lambda_4.
\end{split}
\end{align}
\item \textbf{\underline{Electroweak Precision Tests}}:
	
Finally, the additional scalars contribute to the gauge boson vacuum polarization. As a result, the electroweak precision data provide important constraint. In particular  
the $T$ parameter bounds the mass splitting between $m_H$ and $m_{H^\pm}$ in the scenario in which $h$ is identified with the SM-like Higgs, 
cf.~Ref.~\cite{Becirevic:2015fmu} for example. The general expressions for the parameters $S$, $T$ and $U$ in 2HDM can be found in Ref.~\cite{Barbieri:2006bg}. 
To derive the bounds on the scalar spectrum we consider the following values and the corresponding correlation matrix~\cite{Baak:2014ora},
\begin{equation}
  \begin{aligned}
    & \Delta S^{\rm SM} = 0.05\pm 0.11, \\
    & \Delta T^{\rm SM} = 0.09\pm 0.13, \\
    & \Delta U^{\rm SM} = 0.01\pm 0.11, \\
  \end{aligned}
\qquad\qquad
\mathrm{corr} = \left(\begin{array}{ccc}
1 & 0.90 & -0.59 \\
0.90 & 1 & -0.83 \\
-0.59 & -0.83 & 1
\end{array}\right).
\end{equation}

\noindent The $\chi^2$ function is then expressed as

\begin{equation}
  \chi^2= \sum_{i,j}(X_i - X_i^{\rm SM})(\sigma^2)_{ij}^{-1}(X_j - X_j^{\rm SM}),
\end{equation}
where the vector of central values and uncertainties are denoted as $X=(\Delta S, \Delta T, \Delta U)$ and $\sigma =(0.11,0.13,0.11)$, while the elements of the 
covariance matrix are obtained via $\sigma_{ij}^2\equiv \sigma_i \mathrm{corr}_{ij} \sigma_j$. 

\end{itemize}

\

As mentioned above, we identify the lightest CP-even state $h$ with the SM-like scalar observed at the LHC with mass $m_h=125.09(24)$~GeV~\cite{Olive:2016xmw}. To forbid the dangerous decays $h\to A A$ which could over-saturate the total width of $h$ ($\simeq \Gamma_h^\mathrm{SM}$), we assume that $m_A> m_h/2$. Moreover, we impose the alignment condition $|\cos(\beta-\alpha)|\leq 0.3$, in order to ensure that the couplings of $h$ to $V=W,Z$ remain consistent with the values measured so far, which appear to be in good agreement with the SM predictions~\cite{Corbett:2015ksa}. The above-mentioned constraints are then imposed onto a set of randomly generated points in the intervals: 

\begin{align}
\begin{split}
	&\tan \beta \in (0.2,50),\qquad \hspace*{1.5cm}\alpha\in\left(-\frac{\pi}{2},\frac{\pi}{2}\right), \qquad \hspace*{1cm} \left|M^2\right| \leq (1.2~\mathrm{TeV})^2,\\[0.6em]
	&m_{H^\pm}\in (m_W, 1.2~\mathrm{TeV}),\qquad m_{H}\in (m_h, 1.2~\mathrm{TeV}),\qquad m_{A}\in \left(m_h/2, 1.2~\mathrm{TeV}\right).
\end{split}
\end{align}

\

\begin{figure}[ht]
  \centering
  \includegraphics[width=0.5\textwidth]{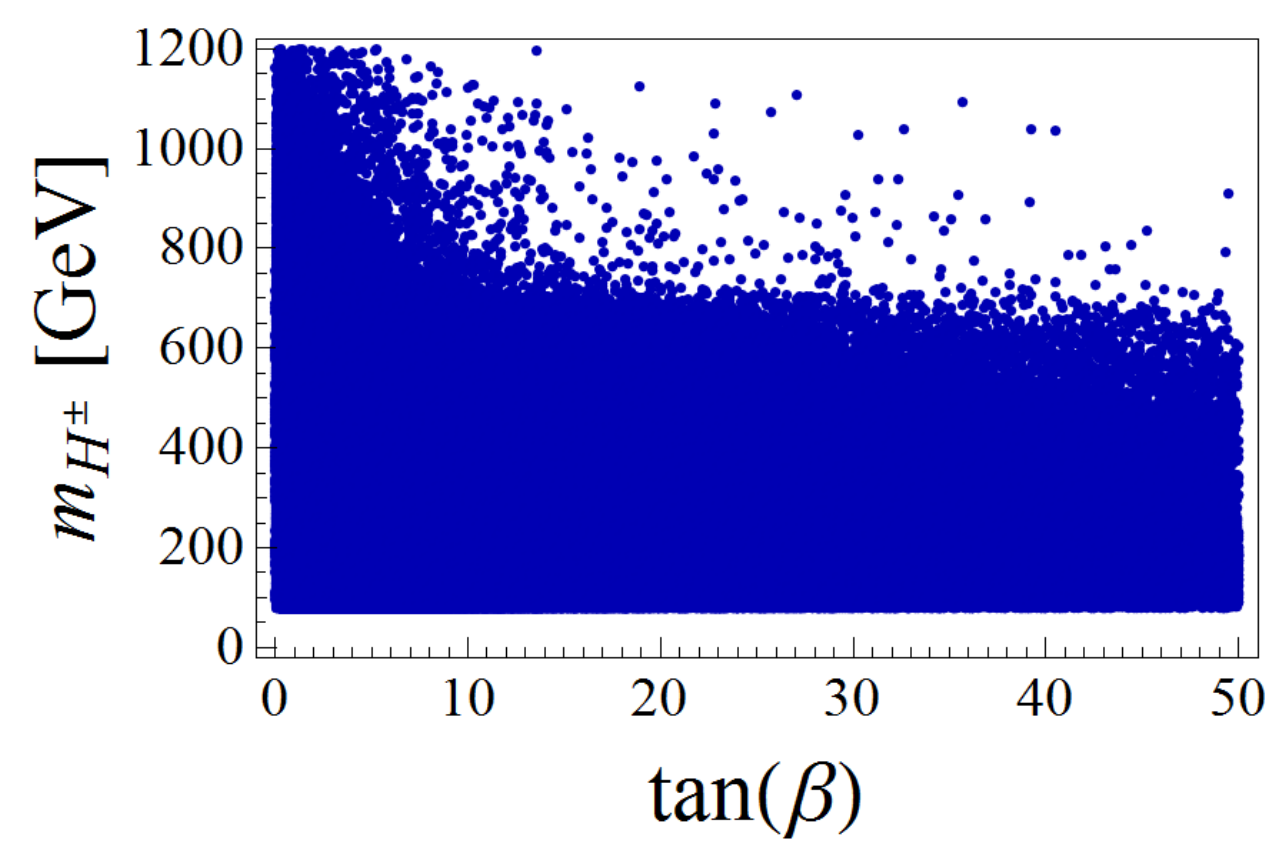}~\includegraphics[width=0.5\textwidth]{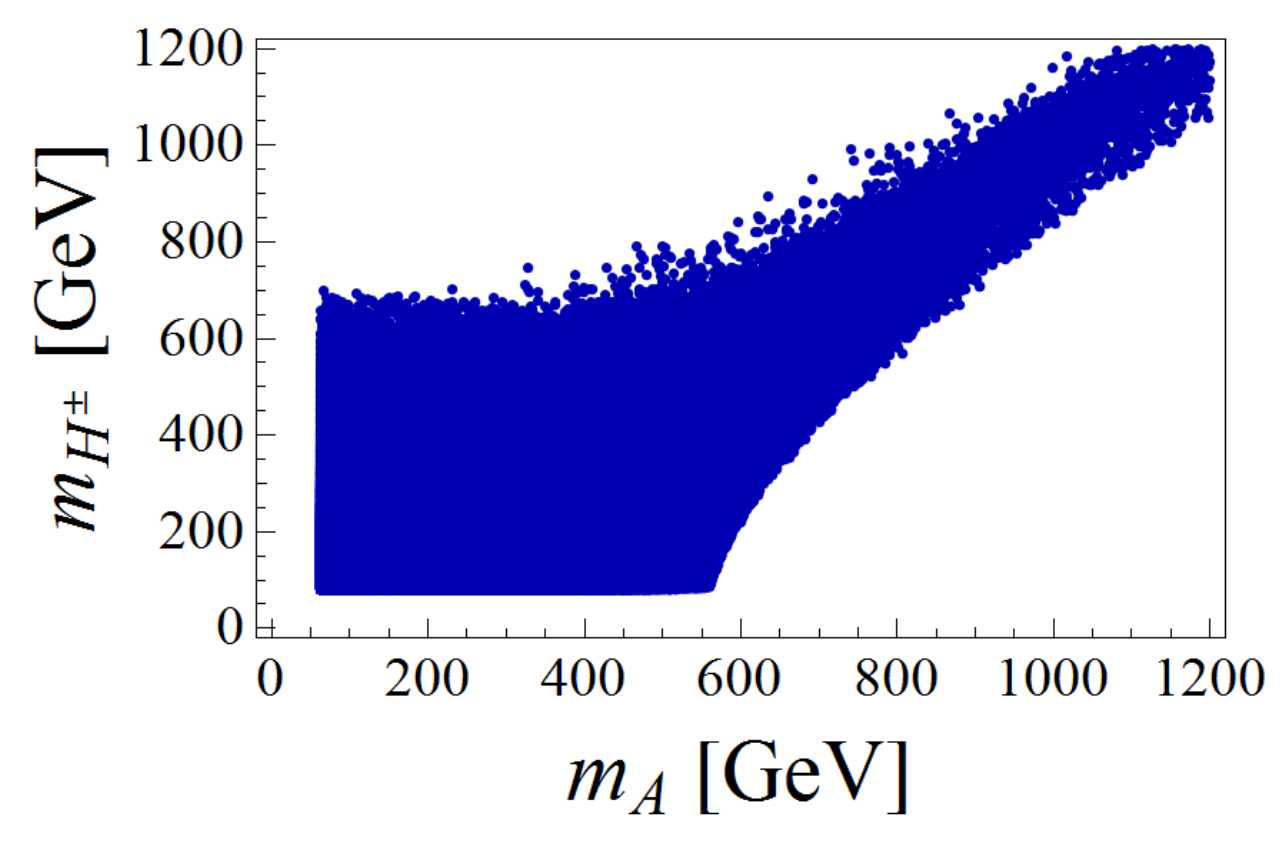}
  \caption{ \sl Results of the scan described in the text.}
  \label{fig:scan}
\end{figure}

\noindent A scan of parameters consistent with the constraints listed above favors the moderate and small values of $\tan\beta \in (0.2,15]$. To see that the larger values of $\tan \beta$ cannot be discarded it is sufficient to examine Eq.~\eqref{eq:massesH} in the alignment limit. For that reason, and in addition to the free scan, we perform a second scan with $m_H\approx|M|$, which helps us probing higher values of $\tan\beta$, and we then combine results of both scans. The combined results are shown in Fig.~\ref{fig:scan} in two planes, $(\tan \beta, m_{H^\pm})$ and $(m_A, m_{H^\pm})$. From the right panel of Fig.~\ref{fig:scan} we observe that the additional scalars become mass degenerate in the decoupling region ($M^2 \gg v^2$), as it can be easily deduced from Eqs.~\eqref{eq:massesH}--\eqref{eq:massesHp}. We should also emphasize that the results of our scans agree with what has been previously reported in the literature, cf.~\cite{scans}.

In Sec.~\ref{sec:pheno} we will confront the points allowed by our scan with the experimental measurements of exclusive $b\to s$ decays. 

\section{Effective Hamiltonian}
\label{sec:eff}

The most general effective Hamiltonian describing the $b\to s \ell \ell$ transitions, made of dimension six operators, is given by~\cite{Altmannshofer:2008dz}

\begin{equation}
\label{eq:heff}
\mathcal{H}_\mathrm{eff} = - \frac{4 G_F}{\sqrt{2}}V_{tb}V_{ts}^\ast \sum_{i}\Bigg{(}C_i(\mu)\mathcal{O}_i(\mu)+C_i^\prime(\mu)\mathcal{O}_i^\prime(\mu)\Bigg{)} + \mathrm{h.c.},
\end{equation}

\noindent where
\begin{alignat}{3}
\label{eq:basisA}
\mathcal{O}_9 &= \frac{e^2}{(4\pi)^2}(\bar{s}\gamma_\mu P_L b)(\bar{\ell}\gamma^\mu \ell), \qquad\qquad  && \mathcal{O}_S &&= \frac{e^2}{(4\pi)^2}(\bar{s} P_R b)(\bar{\ell} \ell),  \\
\mathcal{O}_{10} &= \frac{e^2}{(4\pi)^2}(\bar{s}\gamma_\mu P_L b)(\bar{\ell}\gamma^\mu \gamma_5 \ell) ,\qquad\qquad && \mathcal{O}_P &&= \frac{e^2}{(4\pi)^2}(\bar{s} P_R b)(\bar{\ell}\gamma_5 \ell),  \\
\mathcal{O}_T &= \frac{e^2}{(4\pi)^2}(\bar{s}\sigma_{\mu\nu} b)(\bar{\ell}\sigma^{\mu\nu} \ell), \qquad \qquad && \mathcal{O}_{T5}  &&= \frac{e^2}{(4\pi)^2}(\bar{s}\sigma_{\mu\nu} b)(\bar{\ell}\sigma^{\mu\nu} \gamma_5\ell),
\end{alignat}
and $\mathcal{O}_7=e/(4\pi)^2 m_b (\bar{s} \sigma_{\mu\nu}P_R b)F^{\mu\nu}$ is the electromagnetic penguin operator. The operators with a flipped chirality, $\mathcal{O}^\prime_{7,9,10,S,P}$, are obtained from $\mathcal{O}_{7,9,10,S,P}$ by replacing $P_L \leftrightarrow P_R$ in the quark current. 

The dimension six operators appearing in Eq.~\eqref{eq:heff} are sufficient to match the one-loop amplitude when the external fermion momenta are neglected. 
This, however, is not true if the computation is made with external momenta different from zero which is, in general, {\it necessary} when dealing with {\bf (pseudo-)scalar operators}. 
For example, in order to get a correct expression for the Wilson coefficient $C_P$ one needs to consider the external momenta, which then can give rise to the contributions coming from the dimension-seven operators. One class of such terms can be related to the operators of basis~\eqref{eq:basisA} by equations of motion. For example, 
\begin{align}
\frac{\alpha}{4\pi} \frac{1}{m_W}  \left(\bar{s}\slashed{q} P_L b  \right) \left( \bar{\ell}\gamma_5 \ell \right) &= 
\frac{\alpha}{4\pi} \frac{m_b}{m_W}  \left(\bar{s} P_R b  \right) \left( \bar{\ell}\gamma_5 \ell \right) - 
\frac{\alpha}{4\pi} \frac{m_s}{m_W}  \left(\bar{s} P_L b  \right) \left( \bar{\ell}\gamma_5 \ell \right) \nn\\
& = \frac{m_b}{m_W} \mathcal{O}_P -\frac{m_s}{m_W} \mathcal{O}_P^\prime \simeq  \frac{m_b}{m_W} \mathcal{O}_P
\,.
\end{align}
A complication arises when encountering the operators with insertion of 
$\slashed{p}_{b} +\slashed{p}_s$ in the leptonic current, with the convention $b(p_b)\to s(p_s) \ell^-(p_-)\ell^+(p_+)$, where we also use $q=p_b-p_s=p_++p_-$. 
A way to deal with that, adopted in Ref.~\cite{Li:2014fea}, consists in setting $p_s=0$, so that $\slashed{p}_{b} +\slashed{p}_s= \slashed{q} + 2 \slashed{p}_{s} = \slashed{q}=\slashed{p}_{+} +\slashed{p}_-$, and in this way one can again, like in the previous example, use the equations of motion. That way to deal with the problem in hands, however, leads to a wrong expression for $C_P$, for example. If, instead, one keeps all the momenta non-zero, we get a correct result. At this point we just emphasize that the matching should be performed by keeping all the external momenta different from zero 
and the contributions stemming from dimension-seven operators can be neglected at the very end of computation. We further elucidate this problem in Sec.~\ref{sec:MATCH} where we also propose a general framework for the appropriate matching between the full and effective theories in a case in which the (pseudo-)scalar bosons are explicitly taken into account.

\section{Wilson Coefficients}
\label{sec:wc}

After unambiguously matching the full with the effective theories we obtain the one-loop expressions for the Wilson coefficients generated by the additional scalar particles. We summarize our results in this Section. For clarity we will write them as,

\begin{align}
\label{eq:c7}
C_{7} &= C_7^{\mathrm{NP}\,,\gamma} , \\
\label{eq:c9}
C_{9} &= C_9^{\mathrm{NP}\,,\gamma}+C_9^{\mathrm{NP},\,Z}, \\
\label{eq:c10}
C_{10} &= C_{10}^{\mathrm{NP},\,Z},\\
\label{eq:cp}
C_{P} &= C_{P}^{\mathrm{NP},\,\mathrm{box}}+C_{P}^{\mathrm{NP},\,Z}+C_{P}^{\mathrm{NP},\,A}\\
\label{eq:cs}
C_{S} &= C_{S}^{\mathrm{NP},\,\mathrm{box}}+C_S^{\mathrm{NP},\,h}+C_S^{\mathrm{NP},\,H}
\end{align}

\noindent where the superscripts denote the types of diagrams that contributes to a given Wilson coefficient, namely, the box diagrams, the $\gamma,Z$-penguins and the (pseudo-)scalar penguins. These coefficients should be added to the (effective) ones obtained in the SM: $C_{7}=-0.304$, $C_9=4.211$, $C_{10}=-4.103$, and $C_{S,P}\simeq 0$~\cite{Bobeth:1999mk}.~\footnote{Special attention should be paid to the scalar penguin with the SM-like Higgs to avoid the double counting since it also appears with modifications in A2HDM.}

Henceforth, we neglect the $s$-quark mass and give all our results in the unitary gauge. To check the consistency of our formulas, we also performed the computation in the Feynman gauge. In the 
remainder of this Section we present our resulting expressions for each of the coefficients appearing in Eqs.~\eqref{eq:c10}--\eqref{eq:cs}. We use the standard notation, 
\begin{equation}
x_q = \dfrac{m_q^2}{m_W^2},\qquad\quad x_{H^\pm} = \dfrac{m_{H^\pm}^2}{m_W^2},\qquad\quad x_{\varphi_i^0}=\dfrac{m_{\varphi_i^0}^2}{m_W^2},
\end{equation}
where $q\in\{ b,t\}$,  and $\varphi_i^0 \in \lbrace h,H,A \rbrace$. 

\subsection{$\gamma$-penguins in 2HDM}
\label{sec:wc-gamma-penguins}

\begin{figure}[h]
 \captionsetup[subfigure]{labelformat=empty}
 \centering
\subfloat[2.1]{\includegraphics[scale=0.42]{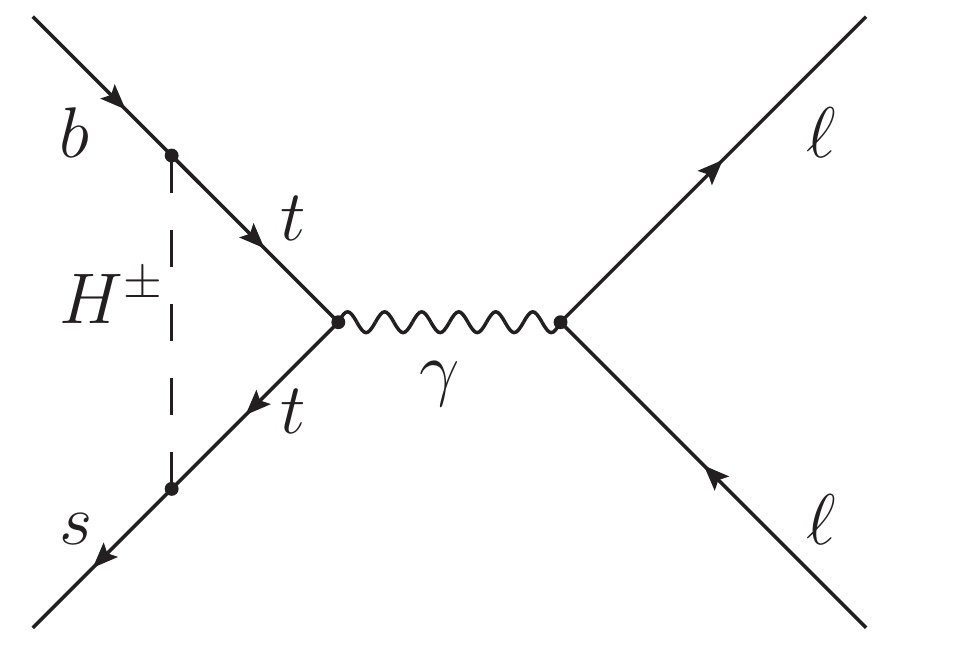}}
\subfloat[2.2]{\includegraphics[scale=0.42]{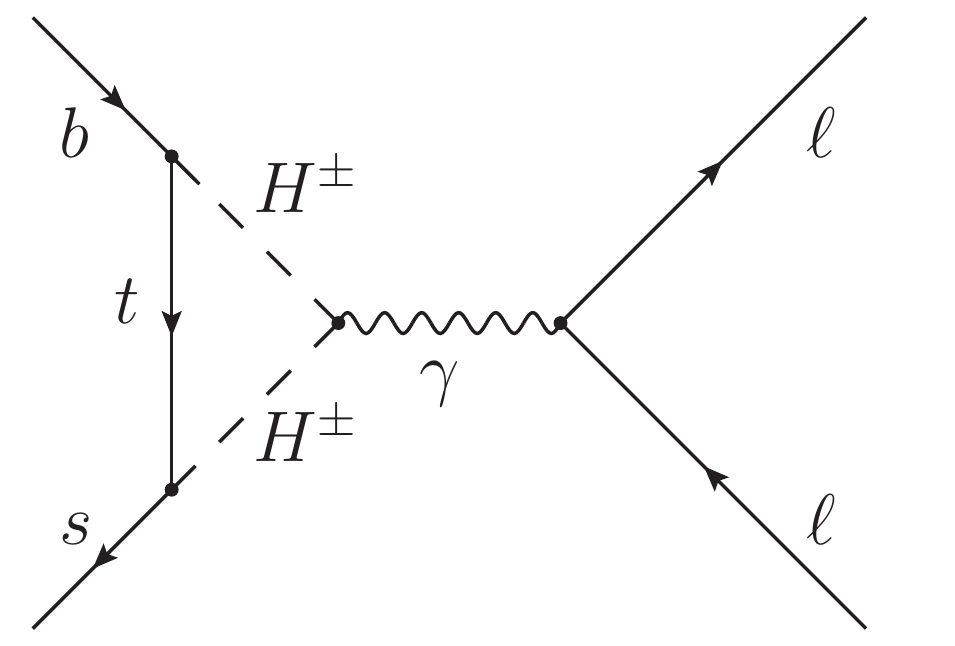}}
\subfloat[2.3]{\includegraphics[scale=0.42]{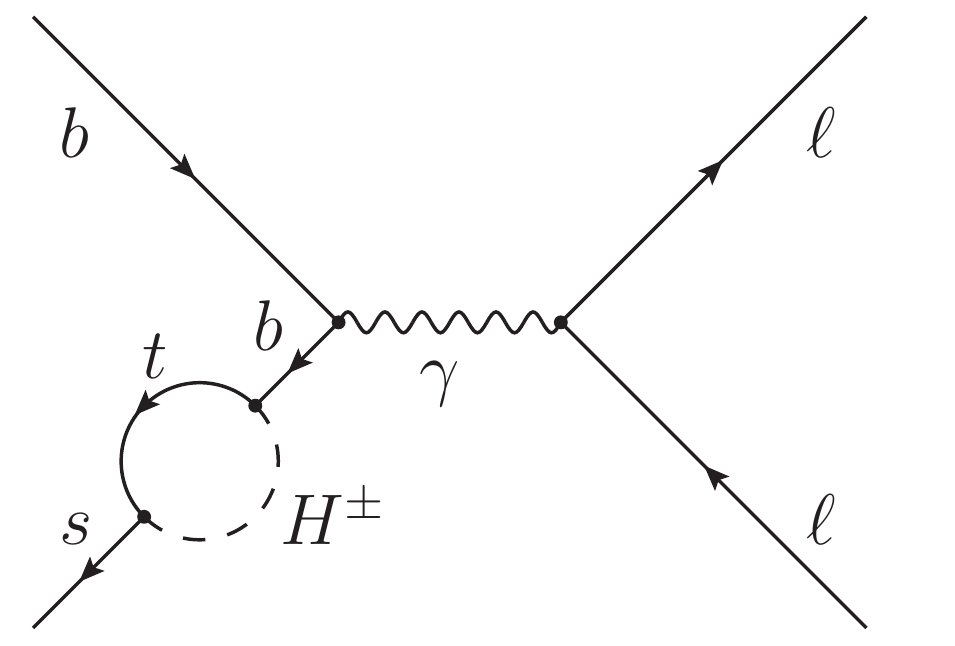}}
\subfloat[2.4]{\includegraphics[scale=0.42]{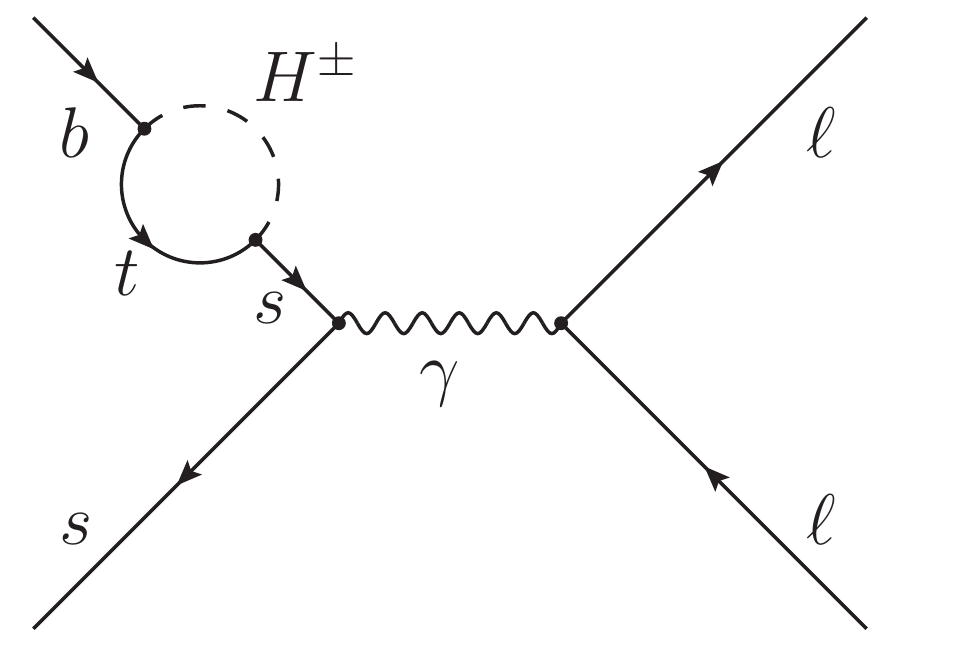}}
 \caption{\sl Photon penguin diagrams generated by the charged Higgs bosons. }
  \label{fig:gamma-penguins}
\end{figure}

The $\gamma$--penguin diagrams induced by the charged Higgs are shown in Fig.~\ref{fig:gamma-penguins}. The off-shell and on-shell contributions can be matched onto the Wilson coefficients $C_7$ and $C_9$, respectively, we obtain,

\begin{equation}
\begin{split}
C_7^{\mathrm{NP}, \gamma} =  &-|\zeta_u|^2 \frac{x_t}{72}\Bigg{[}\frac{7 x_{H^\pm}^2-5 x_{H^\pm}x_t-8 x_t^2}{ (x_{H^\pm}-x_{t})^3}+\dfrac{6 x_{H^\pm}x_t(3x_t-2 x_{H^\pm})}{(x_{H^\pm}-x_{t})^4}\log \left( \frac{x_{H^\pm}}{x_{t}} \right)\Bigg{]}\\
&-\zeta_u^\ast \zeta_d \frac{x_t}{12} \Bigg{[} \frac{3 x_{H^\pm}-5 x_t}{(x_{t}-x_{H^\pm})^2}+\dfrac{2 x_{H^\pm}(3x_{t}-2x_{H^\pm})}{(x_{t}-x_{H^\pm})^3}\log \left( \frac{x_{t}}{x_{H^\pm}} \right)\Bigg{]},
\end{split}
\end{equation}
and
\begin{equation}
\begin{split}
C_9^{\mathrm{NP}, \gamma} &= |\zeta_u|^2 \frac{x_t}{108 }\Bigg{[} \frac{38 x_{H^\pm}^2-79 x_{H^\pm}x_t+47 x_t^2}{(x_{H^\pm}-x_{t})^3}-\dfrac{6(4 x_{H^\pm}^3-6 x_{H^\pm}^2 x_t+3 x_t^3)}{(x_{H^\pm}-x_{t})^4}\log \left(\frac{x_{H^\pm}}{x_{t}}\right)\Bigg{]}\\
&\hspace*{-0.95cm}+\zeta_u^\ast \zeta_d\dfrac{x_t x_b}{108}\Bigg{[}\frac{-37 x_{H^\pm}^2+8x_{H^\pm}x_t+53 x_t^2}{(x_{H^\pm}-x_t)^4}+\frac{6(2 x_{H^\pm}^3+6 x_{H^\pm}^2 x_t-9 x_{H^\pm}x_t^2-3 x_t^3)}{(x_{H^\pm}-x_t)^5}\log \left(\frac{x_{H^\pm}}{x_t}\right)\Bigg{]}.
\end{split}
\end{equation}
The dominant terms in both $C_7^{\mathrm{NP}, \gamma}$ and $C_9^{\mathrm{NP}, \gamma}$ come from the top quark contribution and are proportional to $|\zeta_u|^2$. The terms proportional to $\zeta_u^\ast\zeta_d$ are suppressed by $m_b^2$, thus indeed subdominant.

\subsection{$Z$-penguins in 2HDM}
\label{sec:wc-Z-penguins}

\begin{figure}[h]
  \centering
 \captionsetup[subfigure]{labelformat=empty}
  \subfloat[3.1]{\includegraphics[scale=0.42]{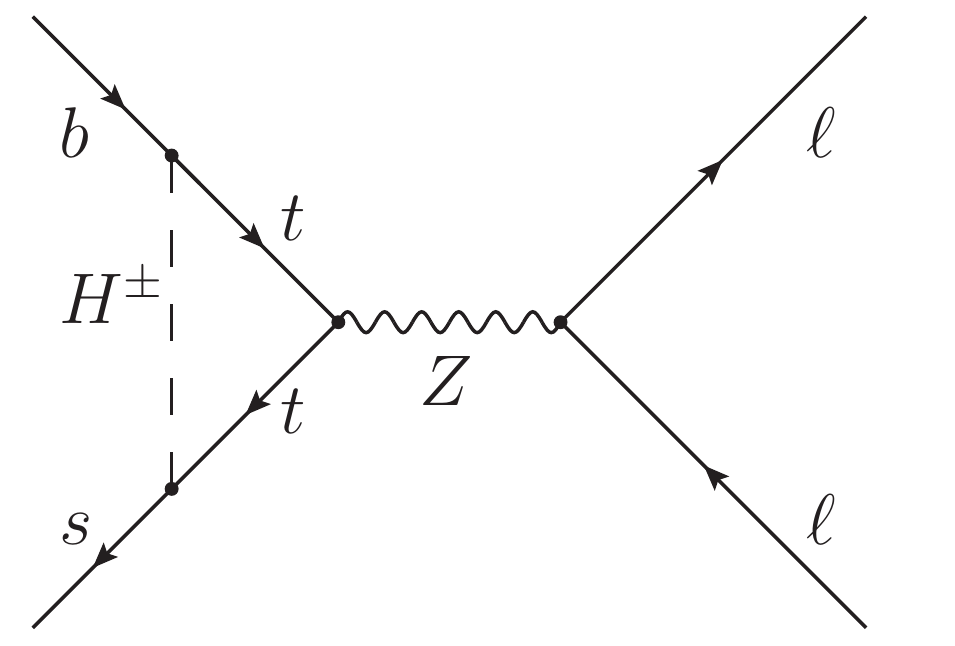}}
\subfloat[3.2]{\includegraphics[scale=0.42]{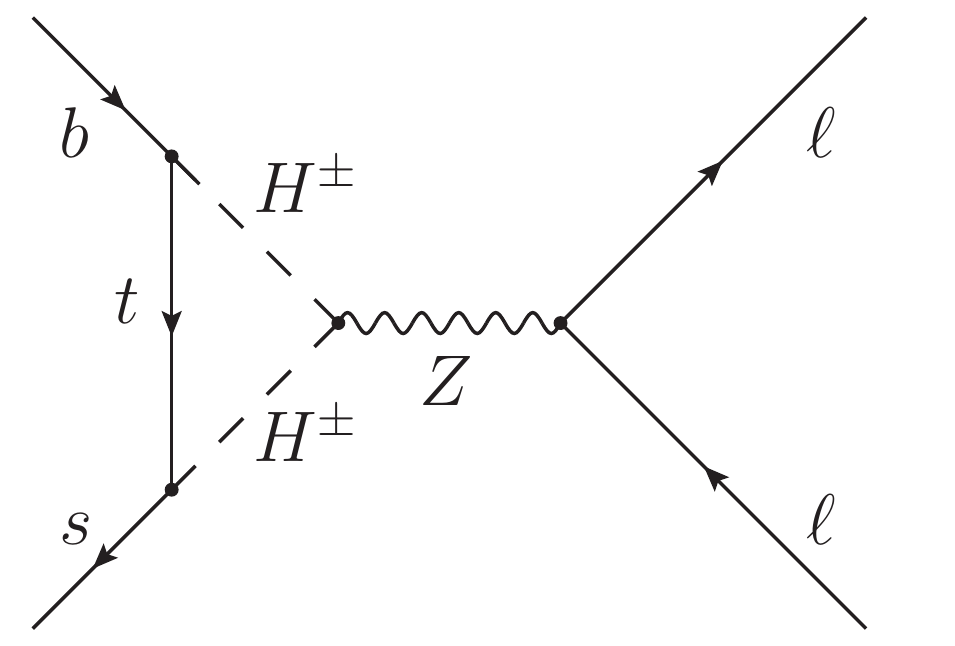}}
\subfloat[3.3]{\includegraphics[scale=0.42]{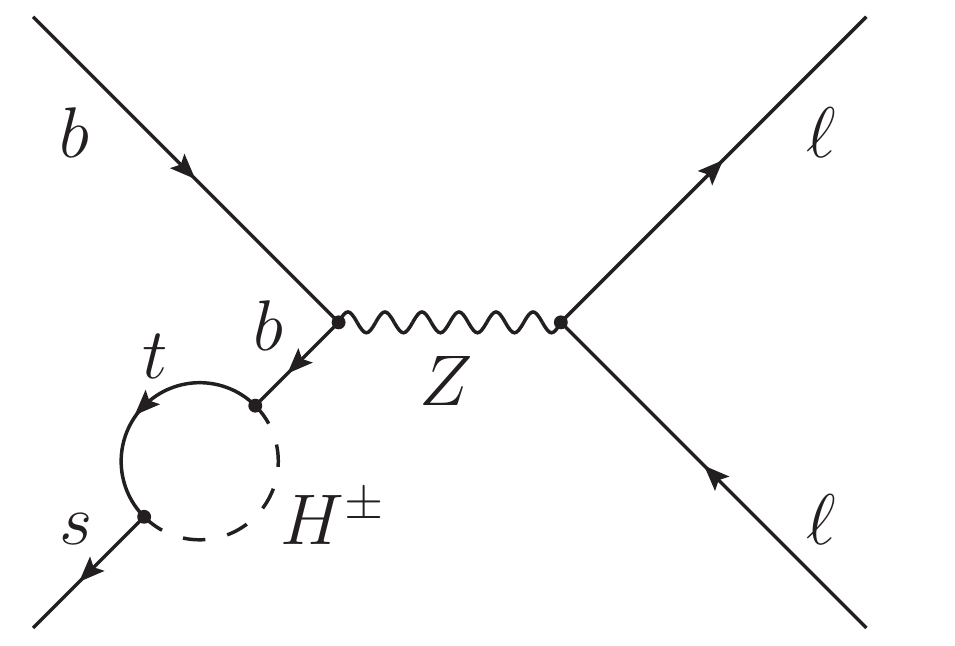}}
\subfloat[3.4]{\includegraphics[scale=0.42]{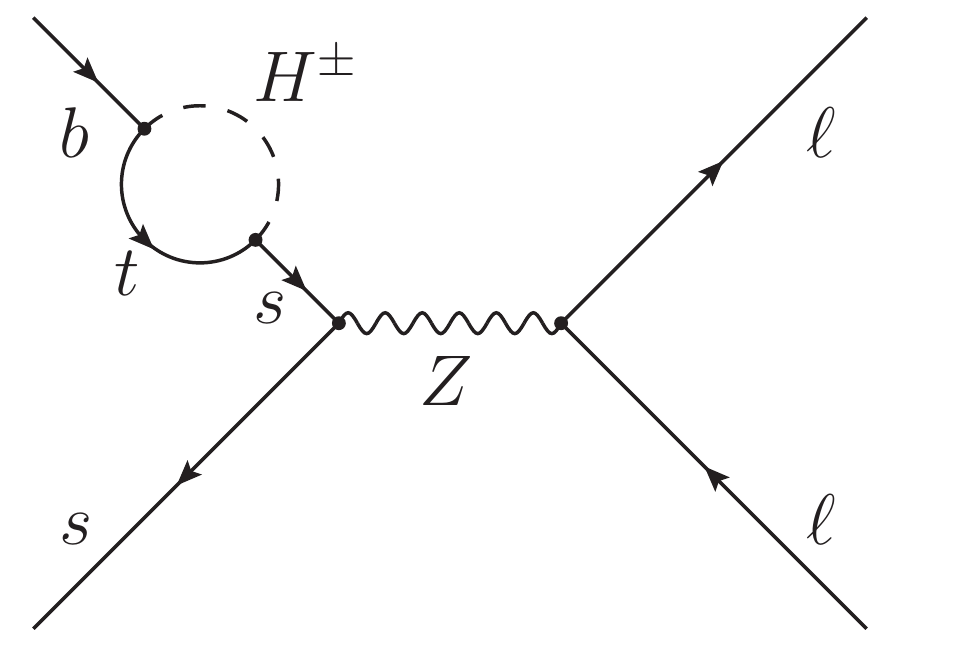}}
 \caption{\sl $Z$ penguin diagrams generated by the additional scalars.}
  \label{fig:Z-penguins}
\end{figure}

The $Z$-penguin diagrams contribute significantly to the Wilson coefficients $C_P$, $C_9$ and $C_{10}$ through the diagrams shown in Fig.~\ref{fig:Z-penguins}. The leading order expressions for $C_9$ and $C_{10}$ read,

\begin{align}
C_{9}^{\mathrm{NP},Z} &= C_{10}^{\mathrm{NP},Z}(-1+4\sin^2 \theta_{W}), \\
C_{10}^{\mathrm{NP},Z} &= |\zeta_u|^2 \frac{x_t^2}{8 \sin^2\theta_W }\Bigg{[}\frac{1}{x_{H^\pm}-x_{t}}-\frac{x_{H^\pm}}{(x_{H^\pm}-x_{t})^2}\log\left(\frac{x_{H^\pm}}{x_{t}}\right)\Bigg{]}\nn \\
&+\zeta_u^\ast\zeta_d \frac{x_t x_b}{16 \sin^2\theta_W}\Bigg{[}\frac{x_{H^\pm}+x_{t}}{(x_{H^\pm}-x_{t})^2}-\frac{2 x_t x_{H^\pm}}{(x_{H^\pm}-x_{t})^3}\log\left(\frac{x_{H^\pm}}{x_{t}}\right)\Bigg{]}.
\end{align}

\noindent Similarly, for $C_P$ we obtain, 

\begin{align}
\begin{split}
C_P^{\mathrm{NP},Z} &= \zeta_u^\ast \zeta_d \dfrac{\sqrt{x_b x_\ell} \,x_t}{16 \sin^2\theta_W}\Bigg{[}\frac{x_t-3 x_{H^\pm}}{(x_{H^\pm}-x_t)^2}+\frac{2 x_{H^\pm}^2}{(x_{H^\pm}-x_t)^3}\log\left(\frac{x_{H^\pm}}{x_t}\right)\Bigg{]} \\
&+|\zeta_u|^2\frac{\sqrt{x_b x_\ell} \,x_t}{216}\Bigg{\lbrace} \frac{38 x_{H^\pm}^2+54 x_{H^\pm}^2 x_t-79 x_{H^\pm}x_t-108 x_{H+} x_t^2+47 x_t^2+54 x_t^3}{(x_{H^\pm}-x_t)^3 }  \\
&-\dfrac{6(4 x_{H^\pm}^3+9 x_{H^\pm}^3 x_t-6 x_{H^\pm}^2 x_t-18 x_{H^\pm}^2x_t^2+9 x_{H^\pm}x_t^3+3 x_t^3)}{(x_{H^\pm}-x_t)^4}\log\left(\frac{x_{H^\pm}}{x_t}\right)\\
&-\dfrac{3}{2\sin^2{\theta_W}}\Bigg{[}\dfrac{2 x_{H^\pm}^2 +36 x_{H^\pm}^2 x_t- 7 x_{H^\pm}x_t-72 x_{H^\pm} x_t^2+11 x_t^2+36 x_t^3}{(x_{H^\pm}-x_t)^3}\\
&-\dfrac{6 x_t(6 x_{H^\pm}^3-12 x_{H^\pm}^2x_t+6 x_{H^\pm}x_t^2+x_t^2}{(x_{H^\pm}-x_t)^4}\log\left(\frac{x_{H^\pm}}{x_t}\right)\Bigg{]}\Bigg{\rbrace}.
\end{split}
\end{align}

\subsection{Charged Higgs Boxes in 2HDM}
\label{sec:wc-box}

\begin{figure}[h]
  \centering
  \captionsetup[subfigure]{labelformat=empty}
  \subfloat[4.1]{\includegraphics[scale=0.47]{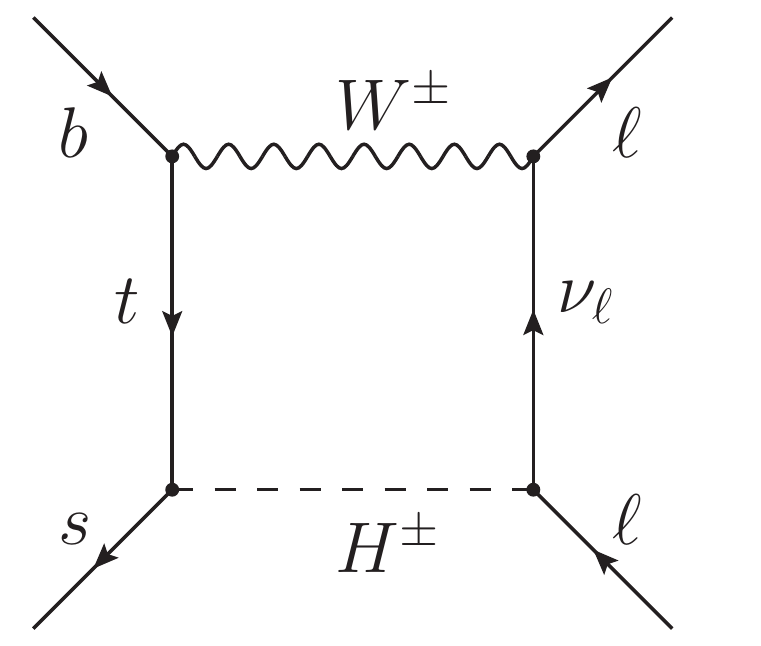}}
\subfloat[4.2]{\includegraphics[scale=0.47]{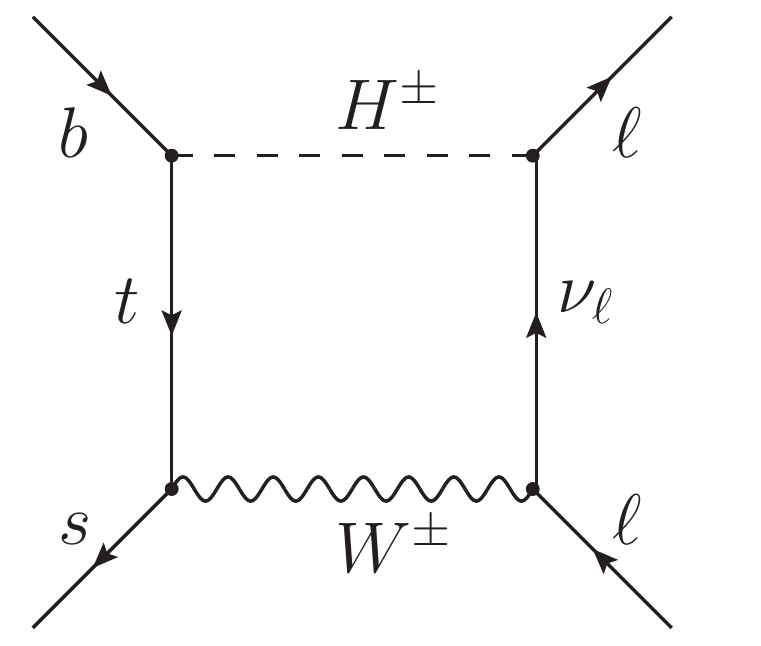}}
\subfloat[4.3]{\includegraphics[scale=0.47]{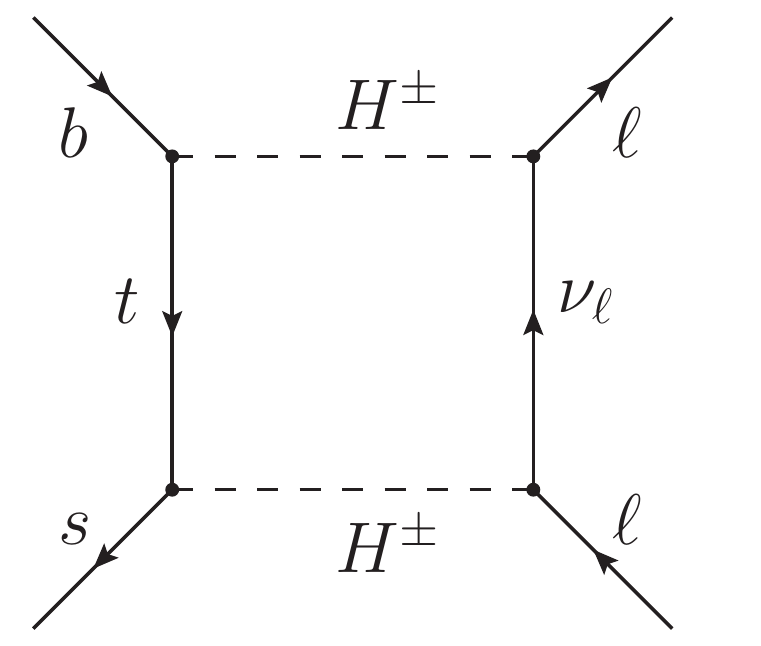}}
\subfloat[4.4]{\includegraphics[scale=0.47]{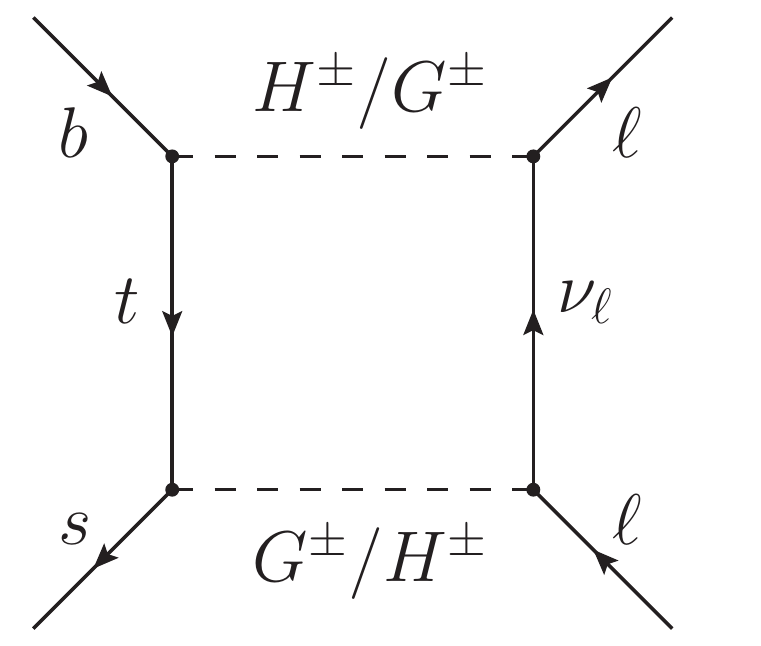}}
 \caption{\sl Box diagrams generated by the additional scalars.}
  \label{fig:boxes}
\end{figure}

The box diagrams, peculiar for 2HDM, are drawn in Fig.~\ref{fig:boxes}. At low-energy they contribute to the Wilson coefficients $C_{S}$ and $C_{P}$ as,
\begin{align}
\begin{split}
C_S^\mathrm{NP,\,box} &= \dfrac{\sqrt{x_\ell x_b}\,x_t}{8(x_{H^\pm}-x_t)\sin^2\theta_W}\Bigg{\lbrace}\zeta_\ell \zeta_u^\ast \left(\frac{x_t}{x_t-1}\log x_t-\frac{x_{H^\pm}}{x_{H^\pm}-1}\log x_{H^\pm}\right)\\
&+\zeta_u \zeta_\ell^\ast\left[1-\frac{x_{H^\pm}-x_t^2}{(x_{H^\pm}-x_t)(x_t-1)}\log x_t - \frac{x_{H^\pm}(x_t-1)}{(x_{H^\pm}-x_t)(x_{H^\pm}-1)}\log x_{H^\pm} \right]\\
&+2\zeta_d\zeta_\ell^\ast \log \left(\dfrac{x_t}{x_{H^\pm}}\right)\Bigg{\rbrace},
\end{split}
\end{align}
and
\begin{align}
\begin{split}
C_P^\mathrm{NP,\,box} &= \dfrac{\sqrt{x_\ell x_b}\,x_t }{8(x_{H^\pm}-x_t)\sin^2\theta_W}\Bigg{\lbrace}\zeta_\ell \zeta_u^\ast \left(\frac{x_t}{x_t-1}\log x_t-\frac{x_{H^\pm}}{x_{H^\pm}-1}\log x_{H^\pm}\right) \\
&-\zeta_u \zeta_\ell^\ast\left[1-\frac{x_{H^\pm}-x_t^2}{(x_{H^\pm}-x_t)(x_t-1)}\log x_t - \frac{x_{H^\pm}(x_t-1)}{(x_{H^\pm}-x_t)(x_{H^\pm}-1)}\log x_{H^\pm} \right]\\
&-2\zeta_d\zeta_\ell^\ast \log\left( \frac{x_t}{x_{H^\pm}}\right)\Bigg{\rbrace}.
\end{split}
\end{align}

\noindent In addition to $C_{S,P}^\mathrm{NP,\,box}$, the tensor and (axial-)vector operators receive contributions but suppressed by the lepton mass, i.e. by $x_\ell=m_\ell^2/m_W^2$. These coefficients are negligible even for decays with $\tau$'s in the final state as it can be verified by using the expressions we provide in Appendix~\ref{app:wc-xl-suppr}.

\subsection{Scalar penguins in 2HDM}
\label{sec:s-penguins}

\begin{figure}[!htbp]
  \captionsetup[subfigure]{labelformat=empty}
  \subfloat[5.1]{\includegraphics[scale=0.42]{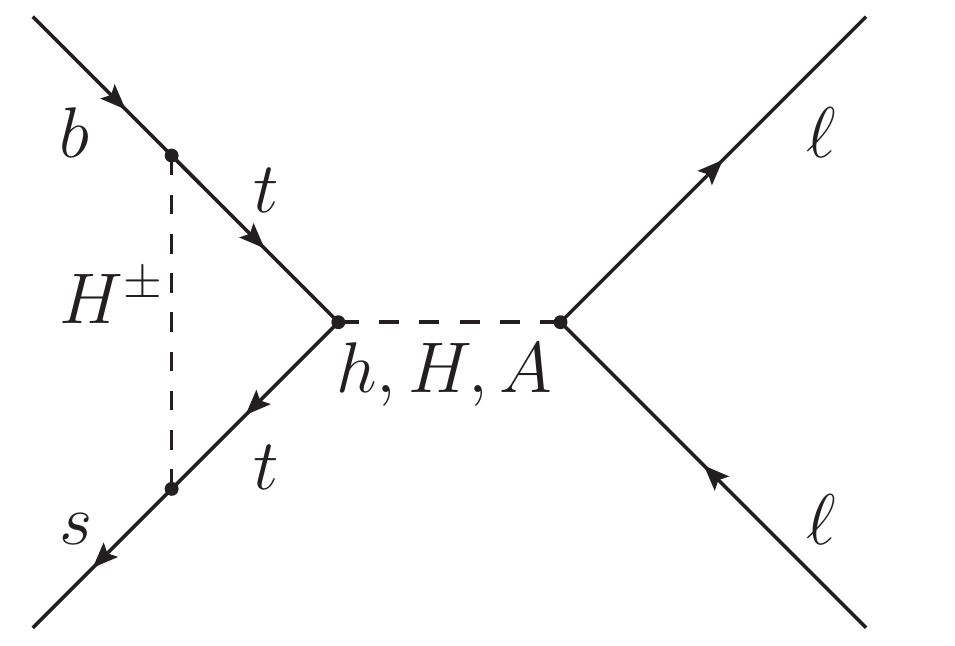}}
\subfloat[5.2]{\includegraphics[scale=0.42]{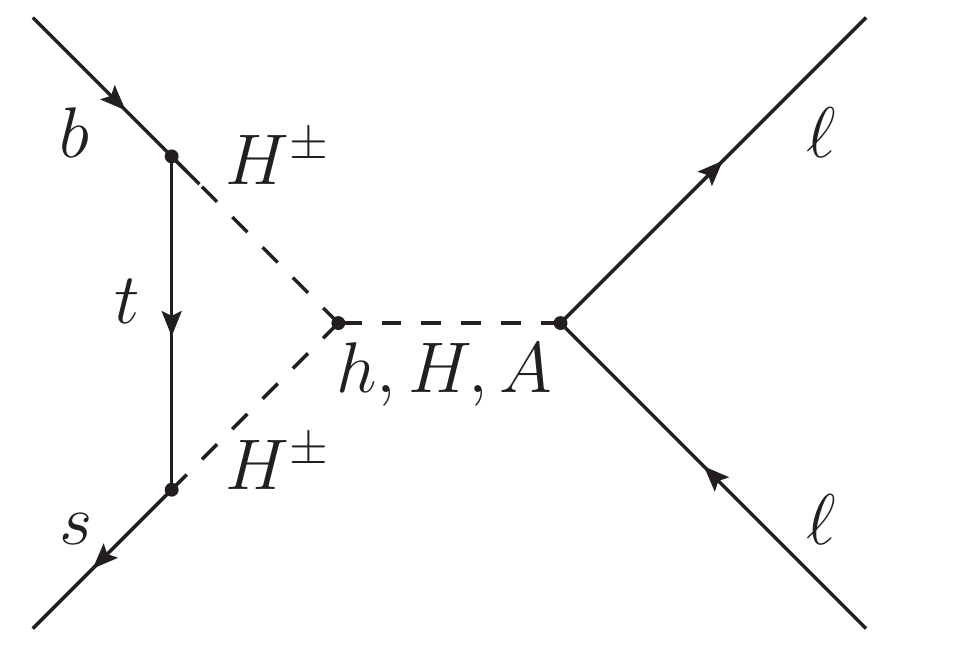}}
\subfloat[5.3]{\includegraphics[scale=0.42]{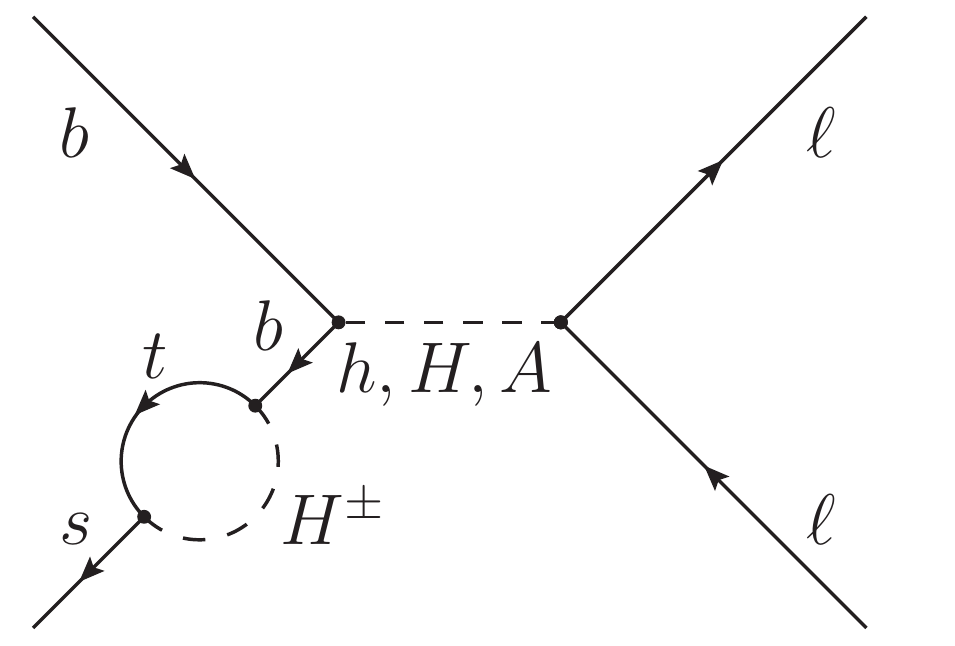}}
\subfloat[5.4]{\includegraphics[scale=0.42]{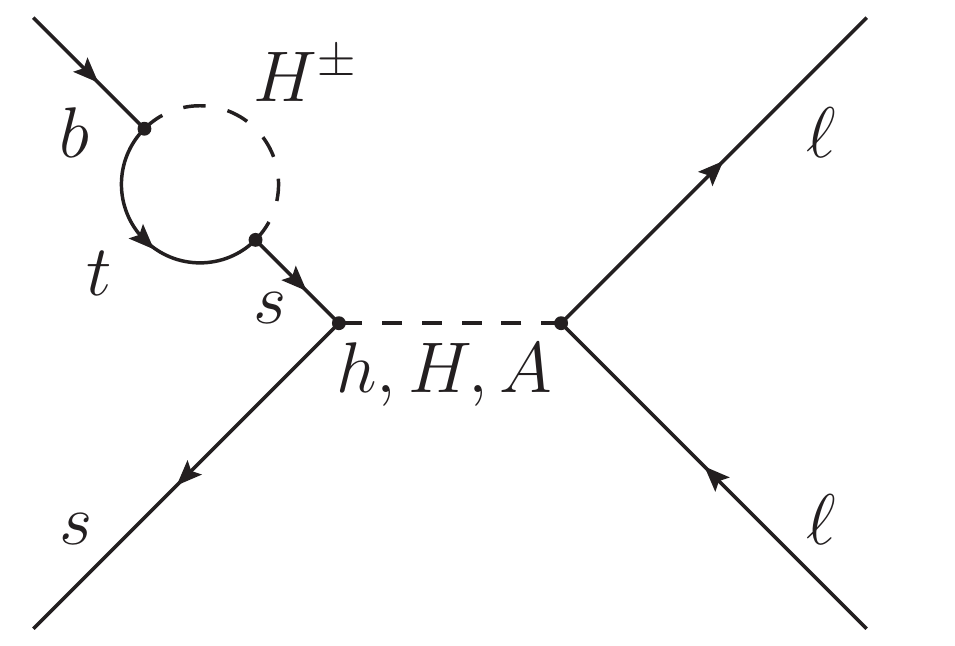}}

\subfloat[5.5]{\includegraphics[scale=0.42]{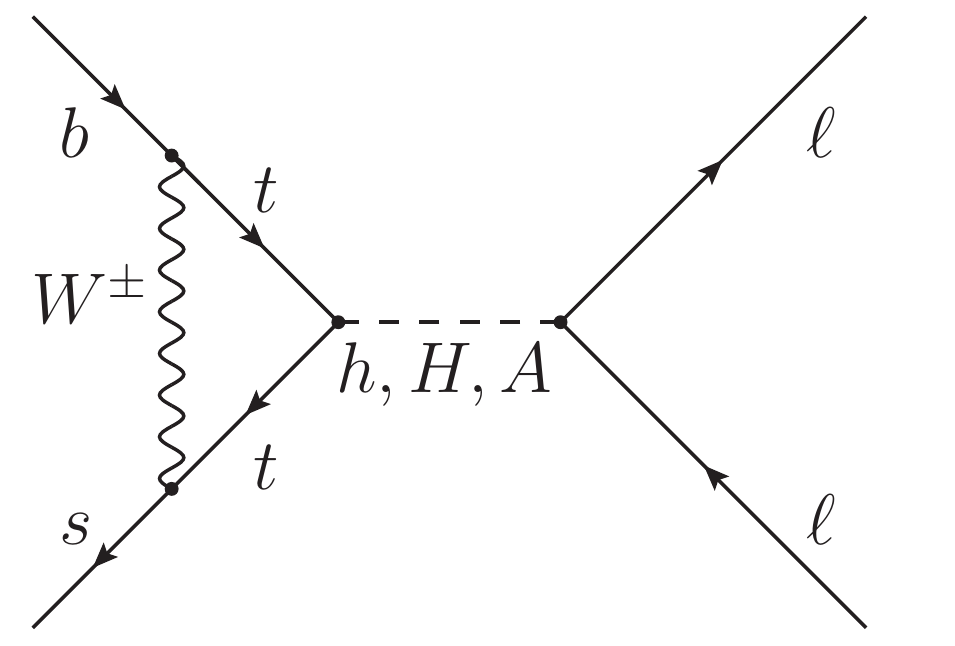}}
\subfloat[5.6]{\includegraphics[scale=0.42]{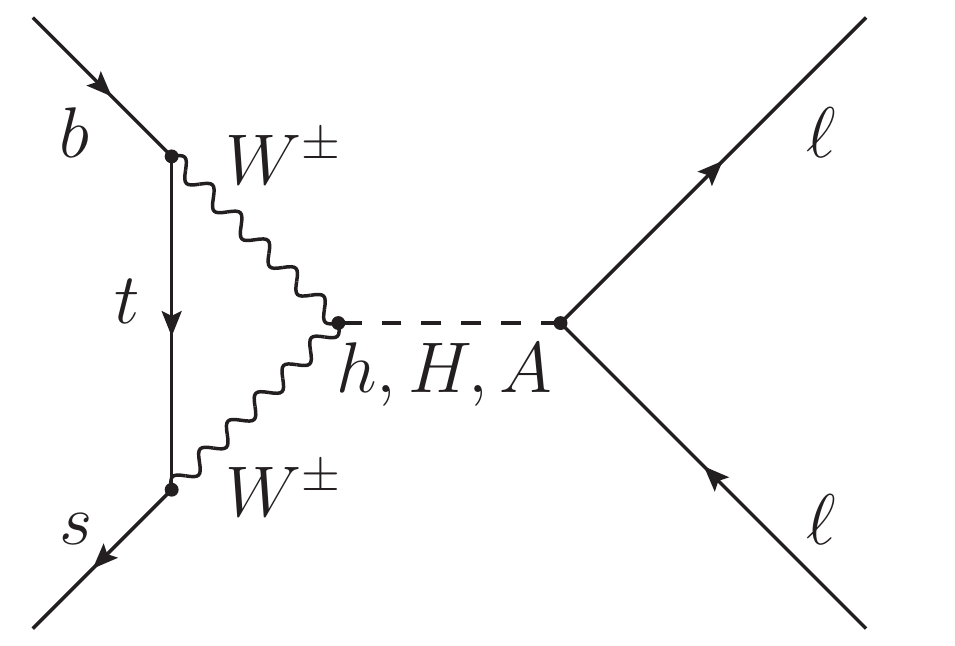}}
\subfloat[5.7]{\includegraphics[scale=0.42]{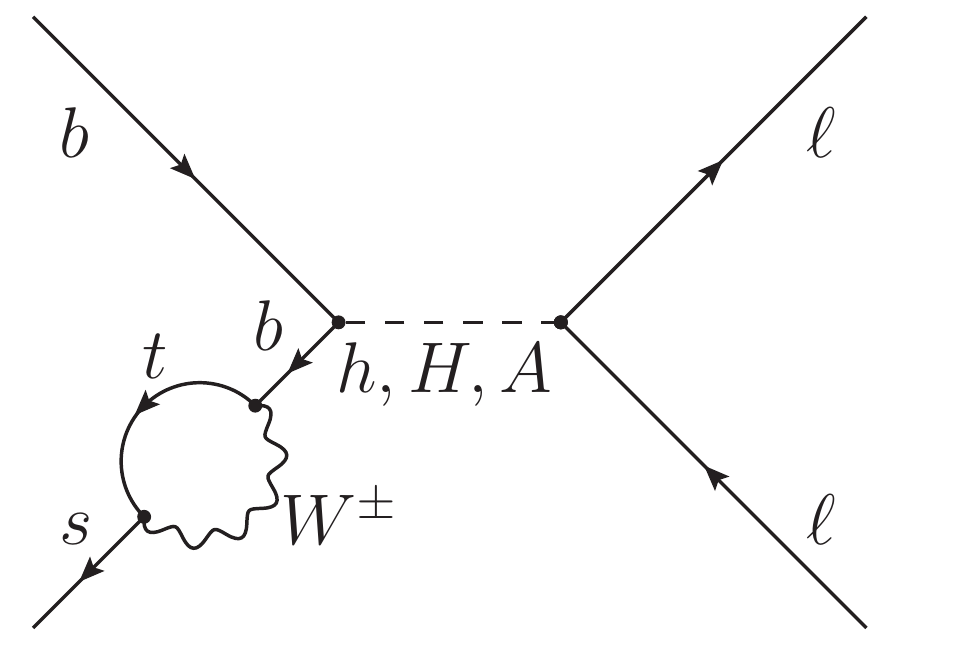}}
\subfloat[5.8]{\includegraphics[scale=0.42]{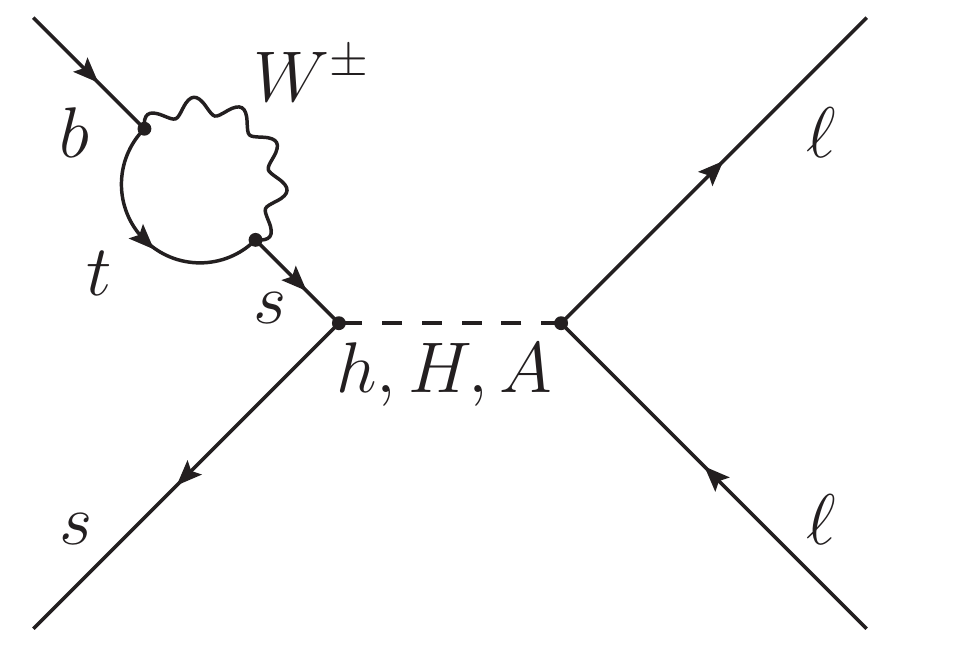}}

\subfloat[5.9]{\includegraphics[scale=0.42]{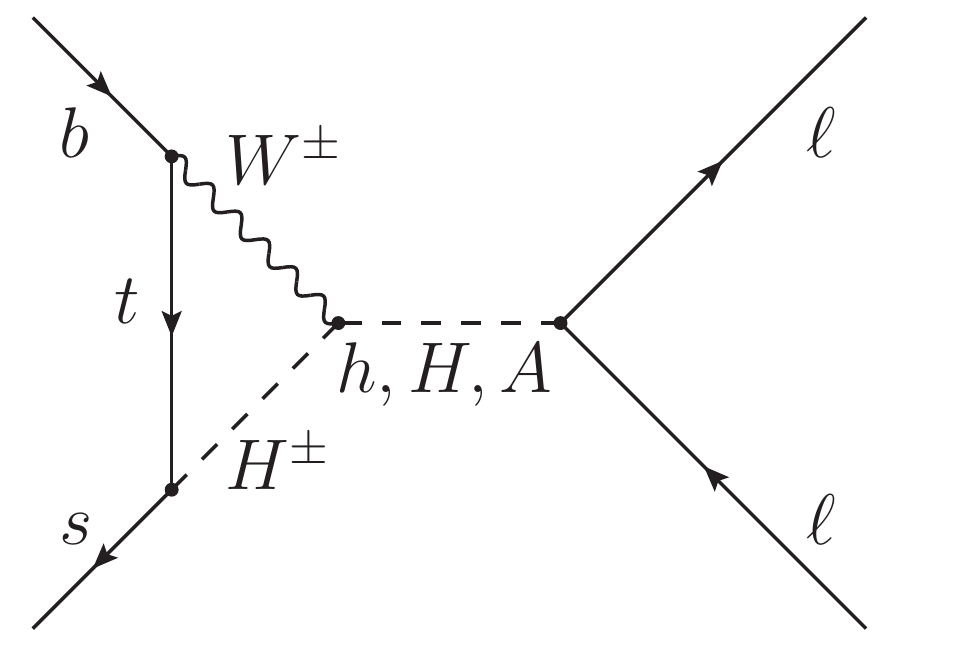}}
\subfloat[5.10]{\includegraphics[scale=0.42]{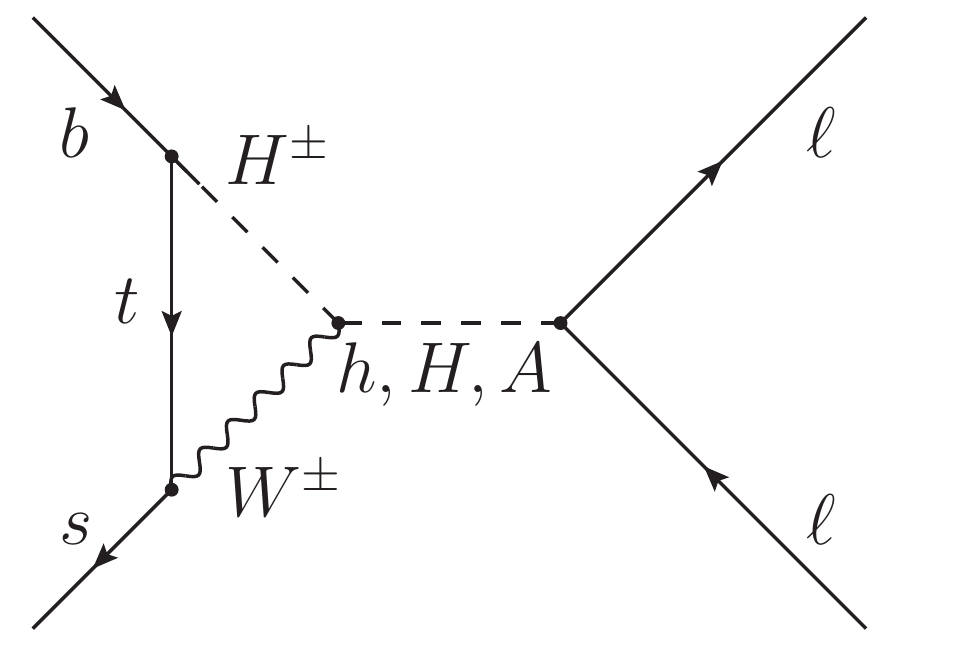}}

\subfloat[5.11]{\includegraphics[scale=0.42]{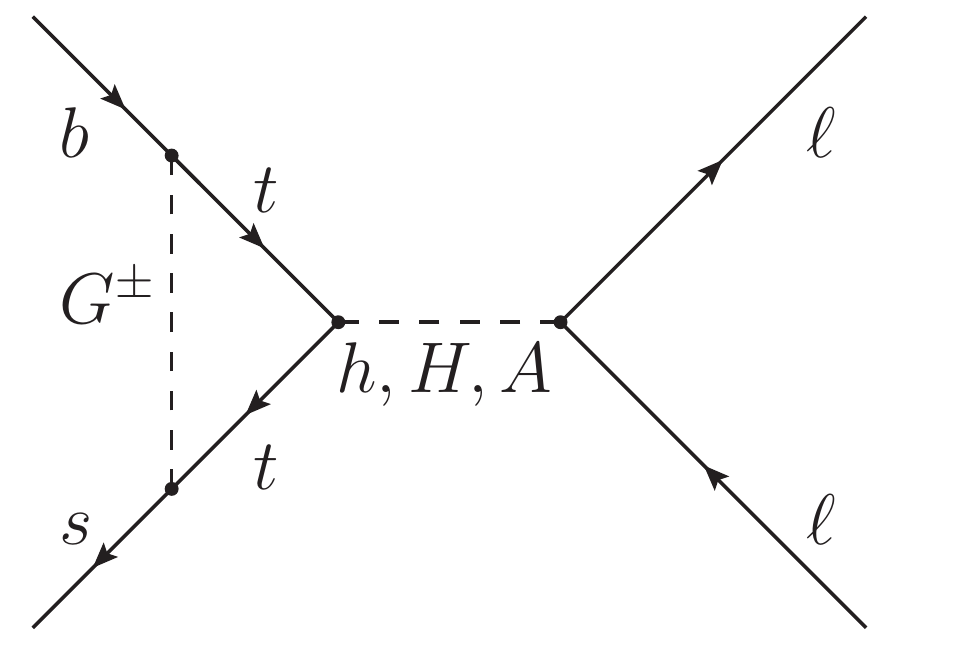}}
\subfloat[5.12]{\includegraphics[scale=0.42]{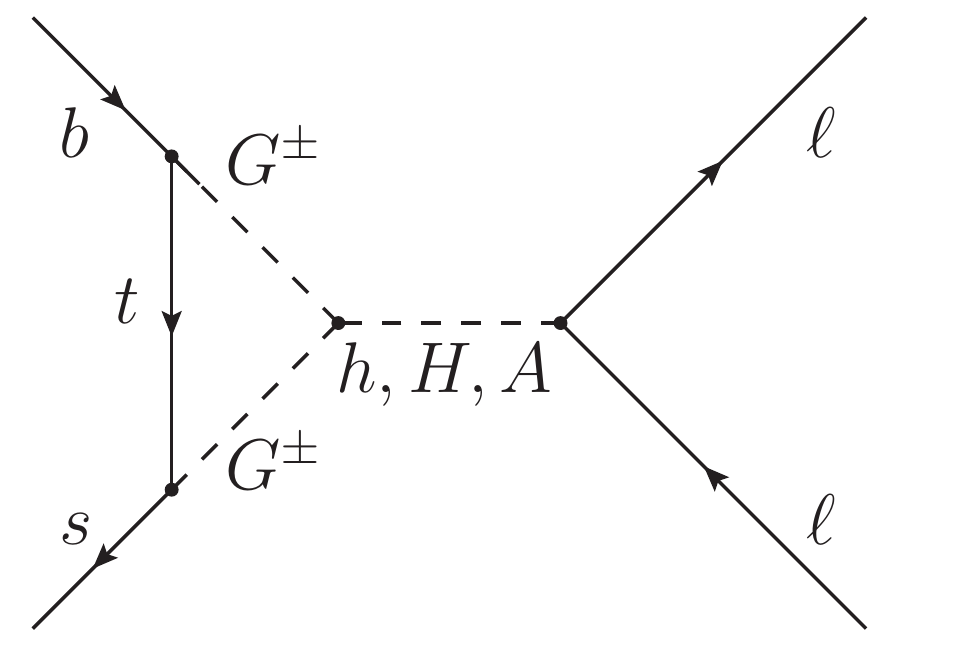}}
\subfloat[5.13]{\includegraphics[scale=0.42]{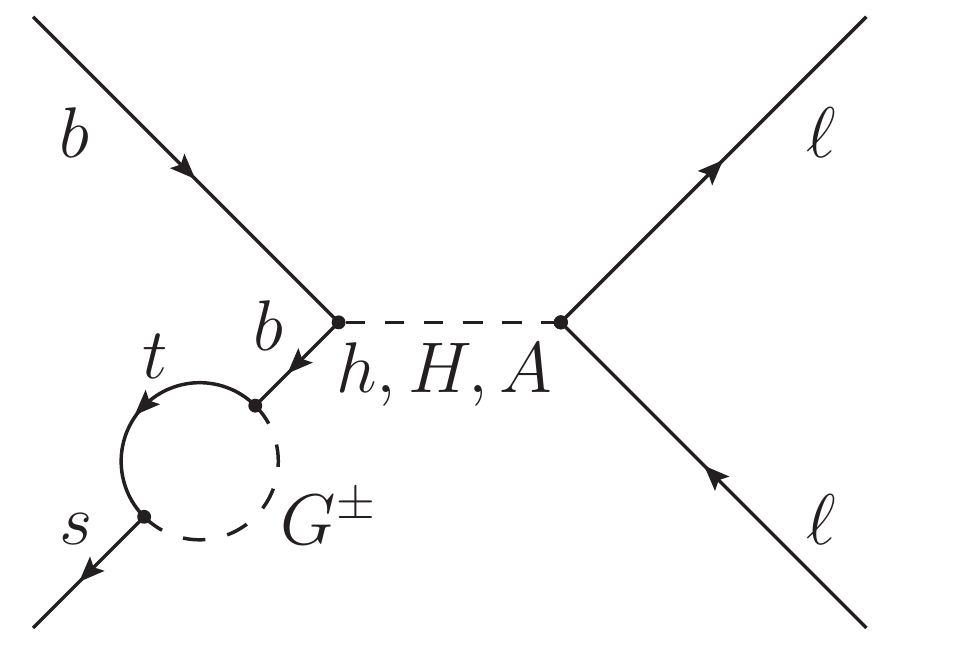}}
\subfloat[5.14]{\includegraphics[scale=0.42]{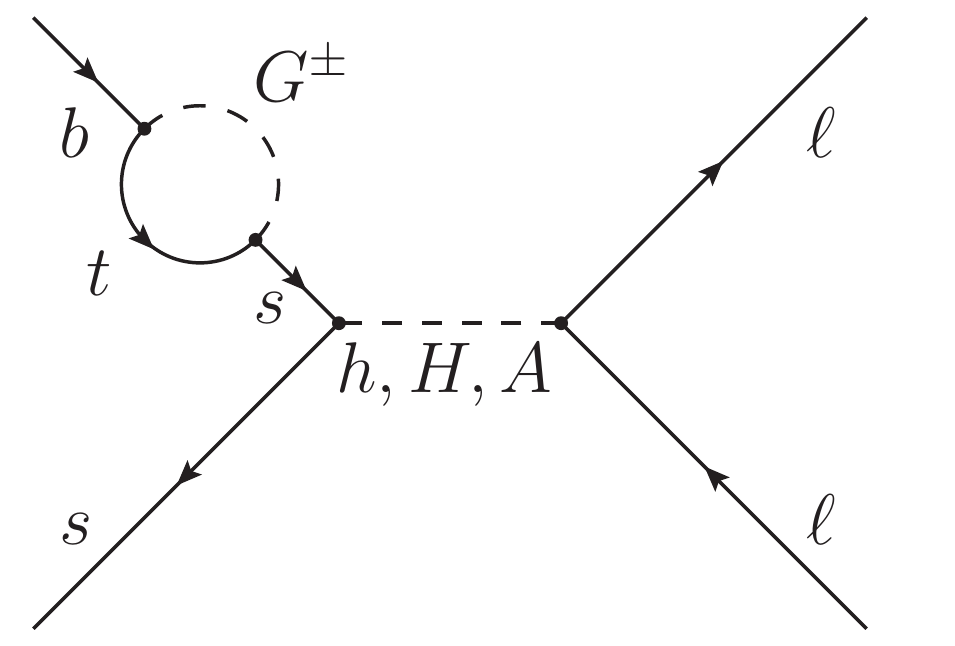}}

\subfloat[5.15]{\includegraphics[scale=0.42]{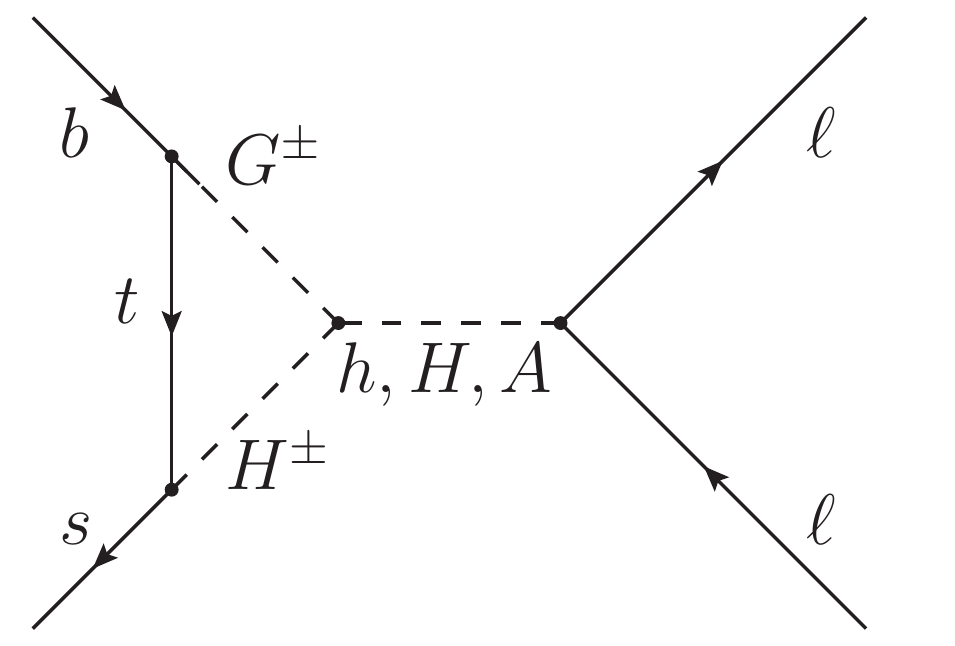}}
\subfloat[5.16]{\includegraphics[scale=0.42]{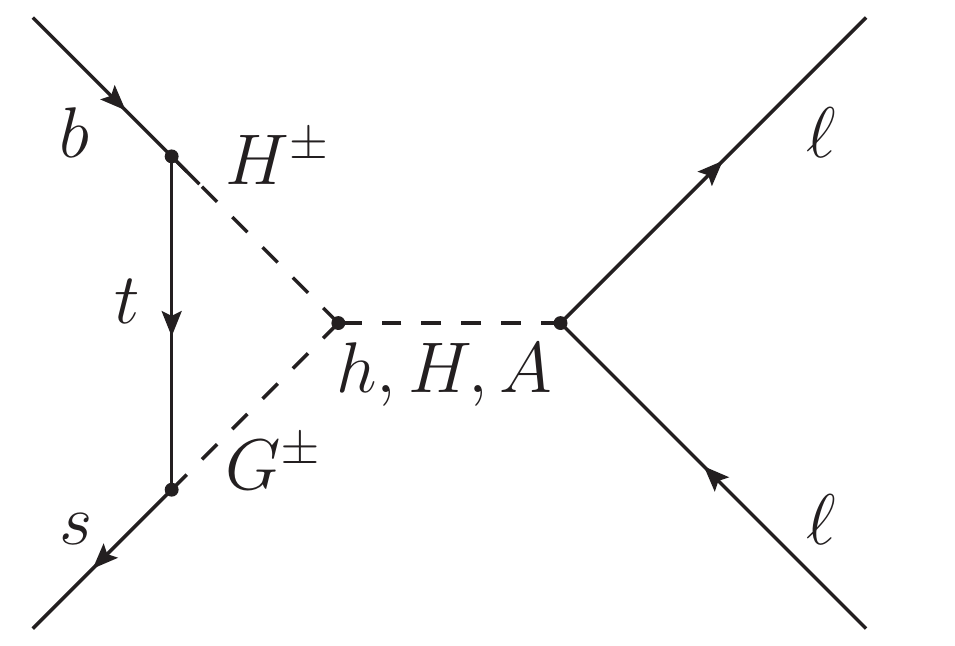}}
\subfloat[5.17]{\includegraphics[scale=0.42]{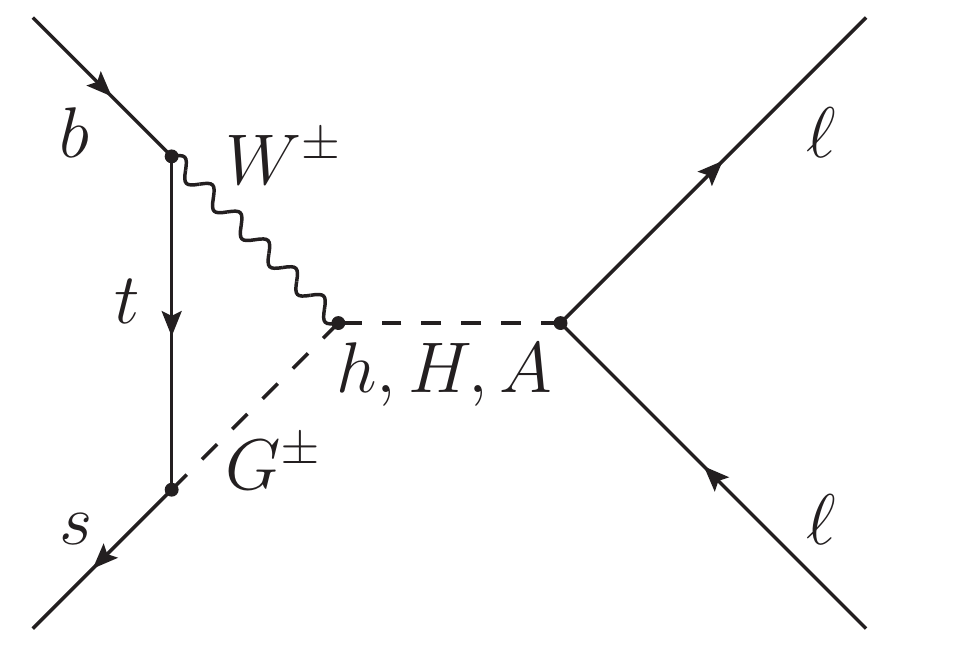}}
\subfloat[5.18]{\includegraphics[scale=0.42]{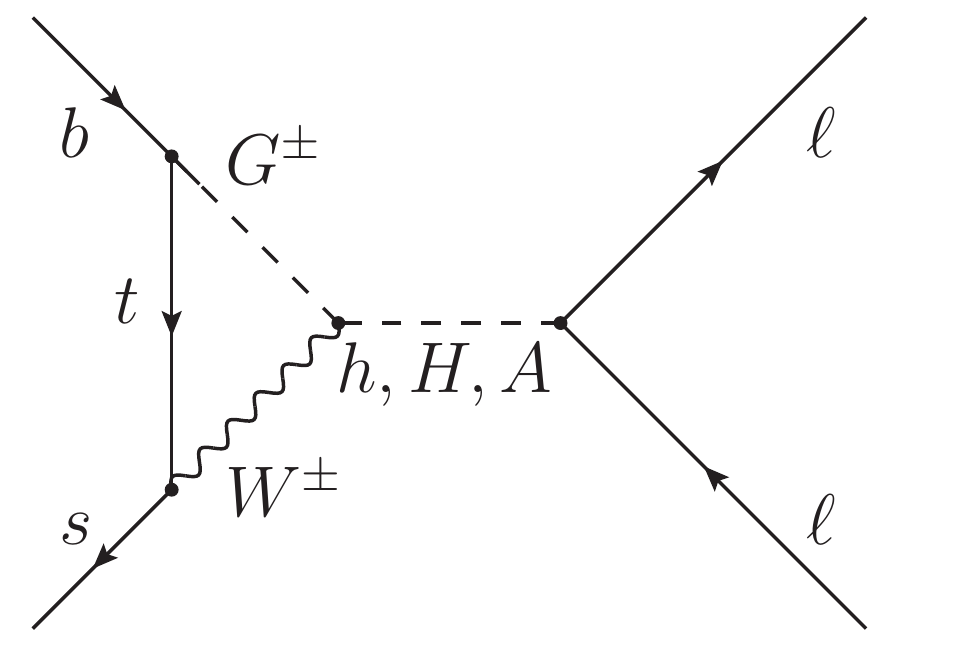}}
 \caption{\sl Higgs penguin diagrams generated by the additional scalars.}
  \label{fig:scalar-penguins}
\end{figure}

We now turn to the effective coefficients $C_P^{\mathrm{NP},\, A}$, $C_S^{\mathrm{NP},\, h}$ and $C_S^{\mathrm{NP},\, H}$, generated by the scalar penguin diagrams shown in Fig.~\ref{fig:scalar-penguins}. We recall that the total ultraviolet divergence coming from these diagrams is proportional to the factor $(1+\zeta_u\zeta_d)(\zeta_u-\zeta_d)$, which vanishes due to the $\mathbb{Z}_2$ symmetry (cf.~Table \ref{tab:y2hdm}).~\footnote{Notice that this is not true in general. For instance, in the A2HDM the divergences are canceled by contributions coming from the radiatively induced misalignment of the Yukawa matrices. The alignment is only preserved at all scales in the context of $\mathbb{Z}_2$-symmetric models~\cite{Li:2014fea}.}

The penguins with the CP-odd Higgs give rise to, 

\begin{align}
\begin{split}
C_P^{\mathrm{NP},\, A}&= -\dfrac{\sqrt{x_\ell x_b}}{\sin^2\theta_W}\dfrac{\zeta_\ell x_t}{2 x_A }\Bigg{\lbrace} \dfrac{\zeta_u^3 x_t}{2} \Bigg{[} \dfrac{1}{x_{H^\pm}-x_t}-\dfrac{x_{H^\pm}}{(x_{H^\pm}-x_t)^2}\log \left(\dfrac{x_{H^\pm}}{x_t}\right)\Bigg{]}\\
&+\frac{\zeta_u}{4}\Bigg{[} -\dfrac{3 x_{H^\pm}x_t-6 x_{H^\pm}-2 x_t^2+5x_t}{(x_t-1)(x_{H^\pm}-x_t)}+\dfrac{x_{H^\pm}(x_{H^\pm}^2-7 x_{H^\pm}+6 x_t)}{(x_{H^\pm}-x_t)^2(x_{H^\pm}-1)}\log x_{H^\pm}\\ 
&-\dfrac{x_{H^\pm}^2(x_t^2-2 x_t+4)+3 x_t^2(2x_t-2 x_{H^\pm}-1)}{(x_{H^\pm}-x_t)^2(x_t-1)^2}\log x_t\Bigg{]}\Bigg{\rbrace},
\end{split}
\end{align}

\noindent where we used that $\zeta_f \in \mathbb{R}$, and $(1+\zeta_u\zeta_d)(\zeta_u-\zeta_d)=0$. Similarly, the penguins with the CP-even Higgs lead to: 

\begin{align}
\label{eq:CS-hH}
\begin{split}C_S^{\mathrm{NP},\, h} &= \dfrac{\sqrt{x_\ell x_b}}{\sin^2 \theta_W}\dfrac{x_t}{2 x_h }\left[\sin(\beta-\alpha)+\cos(\beta-\alpha)\zeta_\ell\right]\\
&\qquad\qquad\quad\times\Bigg{[}g_1\sin (\beta-\alpha)+g_2\cos(\beta-\alpha)-g_0 \frac{2v^2}{m_W^2}\lambda_{H^+ H^-}^h \Bigg{]},\\[0.7em]
C_S^{\mathrm{NP},\, H} &= \dfrac{\sqrt{x_\ell x_b}}{\sin^2 \theta_W}\dfrac{x_t}{2 x_H }\left[\cos(\beta-\alpha)-\sin(\beta-\alpha)\zeta_\ell\right]\\
&\qquad\qquad\quad\times\Bigg{[}g_1\cos (\beta-\alpha)-g_2\sin(\beta-\alpha)-g_0\frac{2v^2}{m_W^2}\lambda_{H^+ H^-}^H\Bigg{]},
\end{split}
\end{align}

\noindent where $\lambda_{H^+H^-}^{\varphi_i^0}$ are the trilinear couplings defined in Appendix \ref{app:feynman-rules}. The functions $g_{0,1,2}$ are given in Appendix \ref{app:scalar-penguins} along with the amplitudes generated by each of the diagrams shown in Fig.~\ref{fig:scalar-penguins}.

\section{Comparison with Other Computations}
\label{sec:compare}

In this Section we compare our Wilson coefficients with the results obtained in previous studies. Before doing so we should emphasize the novelties of the present work: 

\begin{itemize}
	\item[(i)] The result for $C_9$ in a general 2HDM with a $\mathbb{Z}_2$ symmetry is new;
	\item[(ii)] The subleading terms $\mathcal{O}(m_{b})$ to $C_{9,10}$ have been neglected in the previous computations, and they are included here;
	\item[(iii)] We provided an independent computation of the coefficients $C_S$ and $C_P$, and elucidate inconsistencies present in  
	Ref.~\cite{Li:2014fea}, cf. Sec.~\ref{sec:MATCH} where we propose a general prescription for matching procedure when the external momenta are not neglected. 
\end{itemize}

The effective coefficients $C_S$ and $C_P$, in the context of Type~II 2HDM, were first computed in Refs.~\cite{Huang:2000sm,Logan:2000iv,Bobeth:2001sq,Isidori:2001fv,Chankowski:2000ng,Dedes:2008iw}. In these papers $\tan\beta$ was assumed to be very large, which considerably simplifies the computation because in that case only the box diagrams give significant contributions. We agree with these results if we keep only the leading terms in $\tan\beta$ in our expressions, namely,

\begin{align}\label{eq:largeB_Csp}
C_P=-C_S &\simeq \tan^2\beta\dfrac{\sqrt{x_\ell x_b}}{4\sin^2\theta_W}\dfrac{x_t}{x_{H^\pm}-x_t}\log\left(\dfrac{x_{H^\pm}}{x_t}\right).
\end{align}

\noindent Along the same lines, the leading order QCD corrections to the same coefficients were included in Ref.~\cite{Bobeth:2004jz}. 
Recently, the computation of $C_S$ and $C_P$ was extended to the context of a general A2HDM, which comprises all four types of 2HDM with $\mathbb{Z}_2$ symmetry discussed here but without the usual assumption of large $\tan\beta$~\cite{Li:2014fea}. We agree with their general results, except for the expression for $C_P^{\mathrm{NP},\, Z}$ which differs from the one reported in the present paper. The disagreement comes from the fact that the authors of Ref.~\cite{Li:2014fea} worked with the assumption $p_s=0$, which appears not to be fully appropriate.~\footnote{We should emphasize that we were able to reproduce the expression for $C_P^{\mathrm{NP},\, Z}$ reported in Ref.~\cite{Li:2014fea} by taking $p_s=0$, which however is not an appropriate assumption as we argue in the text.} By keeping $p_s\neq 0$ one realizes that the computation of $Z$-penguin leads to two independent terms, one proportional to $p_H=p_b+p_s$ and the other to $q=p_b-p_s$. 
By using equations of motion, $C_{P,S}$ correctly receive contributions from the terms proportional to $q$, but not from those proportional to $p_H$.
With $p_s=0$ only one invariant appears, because $p_H\equiv q$, and thus the resulting $C_{P,S}$ also receive spurious contributions from $p_H$. 

Regarding the other Wilson coefficients, the first computations of $C_7$ for a general 2HDM have been performed in Ref.~\cite{Bertolini:1990if}, then in Refs.~\cite{Ciuchini:1997xe,Degrassi:2006eh}  and~\cite{Hermann:2012fc} where the leading and subleading QCD corrections were included too. Our results are consistent with those, as well as with the expression for $C_{10}$ presented in Ref.~\cite{Chankowski:2000ng} and more recently in Ref.~\cite{Li:2014fea}. The only difference with respect to those results is that we include the subleading terms in $m_b$.

\section{Matching Procedure}
\label{sec:MATCH}

In this section we discuss in more detail the matching of the one-loop amplitudes when the nonzero external momenta are considered. We stress once again that keeping external momenta non-zero is necessary to obtain the correct values for the Wilson coefficients $C_{S,P}$. As we mentioned in Sec.~\ref{sec:eff} the insertion of external momenta result in dimension-seven operators which can be simplified by using equations of motion, except in the cases when the lepton momenta are to be contracted with the quark current and/or the quark momenta to be contracted with the lepton current. 
The amplitudes which need a special treatment, to leading order in external momenta, are:

\begin{align}\label{eq:AMPS}
\begin{split}
\mathcal{A}^\ell_{ij} &= \dfrac{\alpha}{4 \pi} \dfrac{1}{m_W} (\bar{s}(\slashed{p}_- -\slashed{p}_+ )P_i b)(\bar{\ell} P_j \ell), \hspace*{2.1cm} \mathcal{A}_{ij}^q=\dfrac{\alpha}{4\pi}\dfrac{1}{m_W}(\bar{s}P_i b)(\bar{\ell}(\slashed{p}_b +\slashed{p}_s )P_j\ell),\\
\mathcal{A}^{V\ell}_{ij} &= \dfrac{\alpha}{4 \pi}\dfrac{1}{m_W} (\bar{s}(\slashed{p}_- -\slashed{p}_+ )\gamma_\mu P_i b)(\bar{\ell} \gamma^\mu P_j \ell), \hspace*{1.15cm} \mathcal{A}_{ij}^{Vq}=\dfrac{\alpha}{4\pi}\dfrac{1}{m_W}(\bar{s}\gamma_\mu P_i b)(\bar{\ell}(\slashed{p}_b +\slashed{p}_s)\gamma^\mu P_j\ell),
\end{split}
\end{align}

\noindent where $i,j=L,R$ and $s,b,\ell$ are the fermion spinors. Note again that our convention is $b(p_b)\to s(p_s) \ell^-(p_-)\ell^+(p_+)$, and $q=p_b-p_s=p_++p_-$. In order to keep our discussion general, we first extend the Hamiltonian~\eqref{eq:heff} and include the following operators
\begin{align}
\label{eq:heff-deriv}
\mathcal{H}_\mathrm{eff}^{\prime}=  - \frac{4 G_F}{\sqrt{2}}V_{tb}V_{ts}^\ast \sum_{i,j=L,R}\Bigg{(}C_{ij}^{\mathcal{T}\ell}(\mu)\mathcal{O}_{ij}^{\mathcal{T}\ell}(\mu)+C_{ij}^{\mathcal{T}q}(\mu)\mathcal{O}_{ij}^{\mathcal{T}q}(\mu)\Bigg{)}+ \mathrm{h.c.},
\end{align}
where
\begin{align}
\label{eq:ope-deriv-2}
\begin{split}
\mathcal{O}_{ij}^{\mathcal{T}\ell}&=\frac{e^2}{(4\pi)^2}\frac{1}{m_W}(\bar{s}\gamma^\mu P_i b)\partial^\nu(\bar{\ell}\sigma_{\mu\nu} P_j\ell),\\
\mathcal{O}_{ij}^{\mathcal{T}q}&=-\frac{e^2}{(4\pi)^2}\frac{1}{m_W}{\partial^\nu}(\bar{s}\sigma_{\mu\nu} P_i b)(\bar{\ell}\gamma^\mu P_j \ell),
\end{split}
\end{align}
with $i,j=L,R$.~\footnote{Notice that we are not computing the QCD corrections to the Wilson coefficients and therefore, at this order, we do not make distinction between the ordinary and the covariant $SU(3)_c$ derivative.} We reiterate that even though these operators are suppressed by $1/m_W$, they are necessary to unambiguously match the loop induced amplitudes with the effective field theory. The above choice of the basis of dimension-seven operators is convenient since they do not contribute to $\mathcal{B}(B_s\to\mu^+\mu^-)$, while for the other decays their hadronic matrix elements are easy to calculate.

By using the Fierz rearrangement and by applying the field equations, the amplitudes~\eqref{eq:AMPS} are reduced to

\begin{align}
\mathcal{A}^\ell_{LL} &\leftrightarrow - \mathcal{O}_{LL}^{\mathcal{T}\ell}+\mathcal{O}_9\frac{m_\ell}{m_W},\\
\mathcal{A}^\ell_{LR} &\leftrightarrow - \mathcal{O}_{LR}^{\mathcal{T}\ell}+\mathcal{O}_9\frac{m_\ell}{m_W}, \\
\mathcal{A}^{V\ell}_{LL} &\leftrightarrow -\mathcal{O}_{LL}^{\mathcal{T}q}+ \left(\mathcal{O}_S^\prime-\dfrac{\mathcal{O}_T-\mathcal{O}_{T5}}{4}\right)\dfrac{m_\ell}{m_W},\label{eq:ex}\\
\mathcal{A}^{V\ell}_{LR} &\leftrightarrow \mathcal{O}_{LR}^{\mathcal{T}q}+ \left(\mathcal{O}_S^\prime+\dfrac{\mathcal{O}_T-\mathcal{O}_{T5}}{4}\right)\dfrac{m_\ell}{m_W},\\
\mathcal{A}^q_{LL} &\leftrightarrow \mathcal{O}_{LL}^{\mathcal{T}q}+\dfrac{\mathcal{O}_9^\prime-\mathcal{O}_{10}^\prime}{2}\dfrac{m_b}{m_W}+\dfrac{\mathcal{O}_9-\mathcal{O}_{10}}{2}\dfrac{m_s}{m_W}, \\
\mathcal{A}^q_{LR} &\leftrightarrow  \mathcal{O}_{LR}^{\mathcal{T}q}+\dfrac{\mathcal{O}_9^\prime+\mathcal{O}_{10}^\prime}{2}\dfrac{m_b}{m_W}+\dfrac{\mathcal{O}_9+\mathcal{O}_{10}}{2}\dfrac{m_s}{m_W},\\
\mathcal{A}^{Vq}_{LL} &\leftrightarrow \mathcal{O}_{LL}^{\mathcal{T}\ell}+\dfrac{\mathcal{O}_S-\mathcal{O}_P}{2}\dfrac{m_b}{m_W}+\left(\mathcal{O}_S^\prime-\mathcal{O}_P^\prime - \dfrac{\mathcal{O}_T-\mathcal{O}_{T5}}{2}\right)\dfrac{m_s}{2 m_W},\\
\mathcal{A}^{Vq}_{LR} &\leftrightarrow -\mathcal{O}_{LR}^{\mathcal{T}\ell}+\dfrac{\mathcal{O}_S^\prime+\mathcal{O}_P^\prime}{2}\dfrac{m_s}{m_W}+\left(\mathcal{O}_S+\mathcal{O}_P + \dfrac{\mathcal{O}_T+\mathcal{O}_{T5}}{2}\right)\dfrac{m_b}{2 m_W}.
\end{align}

\noindent  To remain completely general, in the above equations we also kept the lepton mass and the mass of $s$-quark different from zero. 
As an example we show the validity of Eq.~\eqref{eq:ex}. 
Using $p_--p_+=2p_--q$, and by the multiple use of field equations, we can write:
\begin{align}
\mathcal{A}^{V\ell}_{LL} &=  \dfrac{\alpha}{4 \pi} \dfrac{2}{m_W}  (\bar{s}\slashed{p}_-  \gamma_\mu P_L b)(\bar{\ell} \gamma^\mu P_L \ell)- \dfrac{\alpha}{4 \pi} \dfrac{1}{m_W}  (\bar{s}\slashed{q}  \gamma_\mu P_L b)(\bar{\ell} \gamma^\mu P_L \ell)\nn\\
& = \dfrac{\alpha}{4 \pi} \dfrac{1}{m_W} \left[ 4  
 (\bar{s} P_L b)(\bar{\ell} \slashed{p}_- P_L \ell) - 2  (\bar{s} \gamma_\mu P_R \underbracket{ \slashed{p}_-  b})(\bar{\ell} \gamma^\mu  P_L \ell)
 \right.\nn\\
 &\left. \qquad +m_s (\bar{s} \gamma_\mu P_L  b)(\bar{\ell} \gamma^\mu  P_L \ell)
  +m_b (\bar{s} \gamma_\mu P_R  b)(\bar{\ell} \gamma^\mu  P_L \ell)
 - 2  (\bar{s}  P_L  b)(\bar{\ell} \slashed{p}_b  P_L \ell)
  \right]\nn\\
  & \stackrel{\text{Fierz}}{=} \dfrac{\alpha}{4 \pi} \dfrac{1}{m_W} \left[ 4  m_\ell 
 (\bar{s} P_L b)(\bar{\ell}  P_L \ell) - 4  (\bar{s}  P_L  \ell )(\bar{\ell}  P_R\slashed{p}_- b) 
 +m_s (\bar{s} \gamma_\mu P_L  b)(\bar{\ell} \gamma^\mu  P_L \ell)
 \right.\nn \\
 &\left. \qquad 
  +m_b (\bar{s} \gamma_\mu P_R  b)(\bar{\ell} \gamma^\mu  P_L \ell)
 -  (\bar{s}  P_L  b)(\bar{\ell} (\slashed{p}_b + \slashed{p}_s)  P_L \ell) + m_\ell   (\bar{s}  P_L  b)(\bar{\ell} \gamma_5 \ell) 
  \right].
\end{align}
By applying the Fierz identity once again, we arrive at,
\begin{align}
\mathcal{A}^{V\ell}_{LL}   & \stackrel{\text{Fierz}}{\to}  \frac{m_\ell}{m_W} \left(  \mathcal{O}_S^\prime - { \mathcal{O}_T - \mathcal{O}_{T5}\over 4}\right) - \mathcal{O}_{LL}^{\mathcal{T}q}.
\end{align}
Clearly, for the appropriate matching of these amplitudes to the effective theory, the operators appearing in Eq.~\eqref{eq:heff} are not enough and the extended basis given in Eq.~\eqref{eq:heff-deriv} is necessary. Once the matching is performed, the operators from Eq.~\eqref{eq:heff} could be neglected since they are $1/m_W$ suppressed with respect to the dominant (dimension six) ones.

This delicate point can then be verified explicitly by computing the Wilson coefficients $C_{RL}^{\mathcal{T}q}$ and $C_{RR}^{\mathcal{T}q}$ which come from the $Z$-penguin diagrams and the coefficients $C^{\mathcal{T}\ell}_{LL}=(C^{\mathcal{T}\ell}_{LR})^\ast$ generated by the box diagrams. 
Their explicit expression is given in Appendix~\ref{app:wc-dim7}.

We can now easily understand the source of our disagreement with Ref.~\cite{Li:2014fea}. 
If one sets $p_s=0$ in $\mathcal{A}_{RR}^q$ of Eq.~\eqref{eq:AMPS}, then just like in Ref.~\cite{Li:2014fea} one could write $\slashed{p}_b+ \slashed{p}_s =\slashed{p}_b=\slashed{q}$ which, by means of equations of motion, yields 
\bea
\mathcal{A}_{RR}^q = \frac{m_\ell}{m_W} 
\dfrac{\alpha}{4\pi}  (\bar{s}P_R b) \left(\bar{\ell} (P_R-P_L)\ell\right) =
\sqrt{x_\ell}\,  \mathcal{O}_P \ ,
\eea
which then in the actual computation gives a contribution to $C_P$. With our procedure, we understand that this contribution does not come from $C_P$ but actually from $\sqrt{x_\ell} C_{RL}^{\mathcal{T}q}$. In other words, and by using our definition of operators and of the effective Hamiltonian, we find~\footnote{ Notice also that the notation of Ref.~\cite{Li:2014fea} is such that their Wilson coefficient $C_P$, which we can call $\widetilde C_P$, is related to our's via $C_P  = \sqrt{x_\ell  x_b} \widetilde C_P/\sin^2\theta_W$.  }
\bea
C_P^\mathrm{Ref.[17]} 
= \left[ C_P +\frac{\sqrt{x_\ell}}{2 \sin^2\theta_W} C_{RR}^{\mathcal{T}q}\right]^\mathrm{(this\,work)}\!\!\!.
\eea
Therefore the Wilson coefficient $C_P$ of Ref.~\cite{Li:2014fea} contains the Wilson coefficient of the operator $\mathcal{O}_{RR}^{\mathcal{T}q}$, the matrix element of which is not equal to the matrix element of the operator $\mathcal{O}_P$ but is, instead, suppressed by $m_W$ as we explicitly check in the next section.  For that reason the Wilson coefficient of Ref.~\cite{Li:2014fea}  is not correct.

\section{$B_s \to \mu^+ \mu^-$ and $B \to K \mu^+\mu^-$ in 2HDM}
\label{sec:pheno0}

In this Section we give the expressions for $\mathcal{B}(B_s\to \mu^+\mu^-)$ and $\mathcal{B}(B\to K \mu^+\mu^-)$ to which we also include the contributions 
of the operators given in Eq.~\eqref{eq:ope-deriv-2}. Those additional operators were necessary for the appropriate matching procedure between the full and the effective theories. However, since they are suppressed by $1/m_W$ they are expected to be negligible with respect to the dominant operators entering the effective Hamiltonian~\eqref{eq:heff}. The purpose of this exercise is to check whether or not the size of the matrix elements of the operators~\eqref{eq:ope-deriv-2} is indeed numerically insignificant for phenomenology. 

\subsection{$B_s \to \mu^+ \mu^-$}

On the basis of Lorentz invariance and invariance of the strong interaction with respect to parity, one can easily verify that $B_s\to \mu^+\mu^-$ is not affected by the operators $\mathcal{O}_{i,j}^{\mathcal{T}q}$ and $\mathcal{O}_{i,j}^{\mathcal{T}\ell}$, with $i,j=L,R$. The expression for the decay rate of this process remains the standard one

\begin{align}
\label{eq:BS}
\mathcal{B}(B_s\to \ell^+\ell^-)^\mathrm{th} &= \tau_{B_s}\dfrac{\alpha^2 G_F^2 m_{B_s}\beta_\ell }{16 \pi^3}  \left| V_{tb}V_{ts}^\ast \right|^2 f_{B_s}^2 m_\ell^2 \Bigg{[}\left|C_{10}-C_{10}^\prime +\dfrac{m_{B_s}^2 (C_P-C_{P}^\prime)}{2 m_\ell(m_b+m_s)} \right|^2 \nonumber \\
&+ \left|C_S-C_S^\prime \right|^2\dfrac{m_{B_s}^2(m_{B_s}^2-4 m_\ell^2)}{4 m_\ell^2(m_b+m_s)^2}\Bigg{]},
\end{align}
where $\beta_\ell=\sqrt{1-4 m_\ell^2/m_{B_s}^2}$. To compare Eq.~\eqref{eq:BS} with the available experimental value, one needs to take into account the effects of $B_s- \overline{B}_s$ oscillations which, to a good approximation, amounts to~\cite{DeBruyn:2012wj}

\begin{equation}
\mathcal{B}(B_s\to\ell^+\ell^-)^\mathrm{exp} \approx \dfrac{1}{1-y_s}\mathcal{B}(B_s\to \ell^+\ell^-)^\mathrm{th},
\end{equation}

\noindent where $y_s=\Delta \Gamma_{B_s}/(2 \Gamma_{B_s})=0.061(9)$, experimentally established by the LHCb Collaboration~\cite{Aaij:2014zsa}. 
As we mentioned before, the dimension-seven operators~\eqref{eq:ope-deriv-2} were chosen in such a way that they do not contribute the $B_s\to \ell^+\ell^-$ decay amplitude.

\subsection{$B \to K \mu^+\mu^-$}

In contrast to $B_s\to \ell^+\ell^-$, the decay $B\to K \ell^+\ell^-$ receives contributions from the operators of the extended basis~\eqref{eq:ope-deriv-2}. 
To write the decay amplitude in a compact form, it is convenient to use the formalism of helicity amplitudes (HA's). In the absence of the (pseudo-)scalar operators, the total amplitude can be schematically written as

\begin{align}
\mathcal{M}=\mathcal{M}_\mu^L \,\bar{\ell} \gamma^\mu P_L \ell+\mathcal{M}_{\mu\nu}^L \,\bar{\ell} \sigma^{\mu\nu} P_L \ell + (L \leftrightarrow R).
\end{align}

\noindent By describing the decay mode as $B\to K V^\ast \to K \ell^+\ell^-$, where $V^\ast$ is a virtual vector boson, one can decompose the total decay amplitude in terms of HA's, 

\begin{align}
A_{m}^{L(R)} = \mathcal{M}_{\mu}^{L(R)}\varepsilon_V^{\mu\ast}(m), \qquad \text{and} \qquad A_{mn}^{L(R)}=\mathcal{M}_{\mu\nu}^{L(R)}\varepsilon_V^{\mu\ast}(m)\varepsilon_V^{\nu\ast}(n),
\end{align}

\noindent where $\varepsilon_V^{\mu}(m)$ (with $m,n=0,t,\pm$) are the $V^\ast$-boson polarization vectors, explicitly defined in Appendix~\ref{app:angular}. We repeat that the above decomposition is valid as long as the scalar and the pseudoscalar operators are not present. To incorporate those contributions unambiguously one can assume the lepton masses to be unequal ($m_{\ell_1}\neq m_{\ell_2}$) and then apply the Ward identities, 

\begin{align}
\bar{\ell}_1 \gamma_5 \ell_2 =\dfrac{q^\mu }{m_{\ell_1}+m_{\ell_2}}\bar{\ell}_1 \gamma_\mu \gamma_5 \ell_2,\qquad \bar{\ell}_1 \ell_2 =\dfrac{q^\mu }{m_{\ell_1}-m_{\ell_2}}\bar{\ell}_1 \gamma_\mu \ell_2 ,
\end{align}
to absorb the (pseudo-)scalar terms in the time-like coefficients ${A}_{t}^{L(R)}$. By taking the limit $m_{\ell_1}=m_{\ell_2}$ in the final expression one ends up with the desired HA's and the total decay amplitude, from which is then easy to compute the decay rate~\cite{Becirevic:2016zri}. Notice that the contributions from $C_{S,P}^{(\prime)}$ enter the amplitudes $A_S$ and $A_t$ defined as,

\begin{align}
A_t &= \lim_{m_{\ell_1}\to m_{\ell_2}}\left(A_t^L - A_t^R \right),\\
A_S &= \lim_{m_{\ell_1}\to m_{\ell_2}} \Bigg{[} \dfrac{m_{\ell_1}-m_{\ell_2}}{\sqrt{q^2}} \left(A_t^L + A_t^R \right) \Bigg{]}.
\end{align}
\noindent More details regarding this point can be found in Ref.~\cite{Becirevic:2016zri}. 
We also need to stress that all the helicity amplitudes are the $q^2$-dependent functions, $A_i\equiv A_i(q^2)$. 
By applying the method briefly sketched above we obtain,
\begin{align}
\label{eq:BK}
\begin{split}
\dfrac{\mathrm{d}}{\mathrm{d}q^2}\mathcal{B}(B \to K &\ell^+ \ell^-)^\mathrm{th} = \dfrac{2(q^2-m_\ell^2)}{3}\left[|A_0^L|^2+|A_0^R|^2\right]+2 m_\ell^2 \left|A_t\right|^2+\dfrac{q^2-4m_\ell^2}{2}|A_S|^2 \\[0.4em]
&+\dfrac{q^2+2 m_\ell^2}{3}\left[|A_{t0}^L-A_{0t}^L|^2+|A_{t0}^R-A_{0t}^R|^2\right]+4m_\ell^2 \mathrm{Re}\left[A_0^{L\ast} A_0^R\right]\\[0.4em]
&+\dfrac{8 (q^2-4m_\ell^2)}{3}\left|A_{T5}\right|^2+\dfrac{4\left(q^2-4 m_\ell^2\right)}{3}\mathrm{Re}\left[A_{T5}^\ast (A_{t0}^L-A_{0t}^L)-(L\leftrightarrow R)\right] \\[0.4em]
&+4 m_\ell^2 \mathrm{Re}\left[A_{0t}^{L\ast}\left(A_{0t}^R-A_{t0}^R\right)-A_{t0}^{L\ast}\left(A_{0t}^R-A_{t0}^R\right)  \right]\\[0.4em] 
&-2 m_\ell \sqrt{q^2}\,\mathrm{Im} \left[\left(A_0^L+A_0^R\right)^\ast \left(A_{t0}^L-A_{0t}^L+(L \leftrightarrow R)\right) \right] ,
\end{split}
\end{align}

\noindent where the explicit expressions for the helicity amplitudes are:

\begin{align}
A_0^{L(R)}(q^2) &= \mathcal{N}_K\dfrac{\lambda_B^{1/2}}{2\sqrt{q^2}} \Bigg{[}f_+(q^2)\left[(C_9+C_9^\prime)\mp(C_{10}+C_{10}^\prime)\right] +f_T(q^2) \dfrac{2 m_b }{m_B+m_K}(C_7+C_7^\prime)\nonumber\\
&-f_T(q^2) \dfrac{q^2}{m_W (m_B+m_K)} \left[C_{L,L(R)}^{\mathcal{T}q}+C_{R,L(R)}^{\mathcal{T}q}\right]  \Bigg{]},\\[0.7em]
A_t (q^2) &= - \mathcal{N}_K f_0(q^2) \dfrac{m_B^2-m_K^2}{\sqrt{q^2}} \Bigg{[}C_{10}+C_{10}^\prime+\dfrac{q^2 \left(C_P+C_P^\prime\right)}{2 m_\ell (m_b-m_s)} \Bigg{]}, \\[0.7em]
A_S(q^2) &= \mathcal{N}_K f_0(q^2) \dfrac{m_B^2-m_K^2}{m_b-m_s}\left( C_S+C_S^\prime\right),\\[0.7em]
A_{0t}^{L(R)} (q^2)&= i \mathcal{N}_K\lambda^{1/2}_B\Bigg{[}f_T(q^2)\dfrac{C_T}{m_B+m_K}+f_+(q^2)\dfrac{C_{L,L(R)}^{\mathcal{T}\ell}+
C_{R,L(R)}^{\mathcal{T}\ell}}{2 m_W}\Bigg{]},\\[0.7em]
A_{t0}^{L(R)} (q^2)&= -i \mathcal{N}_Kf_T(q^2) \dfrac{C_T \lambda_B^{1/2}}{m_B+m_K},\\[0.7em]
A_{T5} (q^2) &\equiv A^{L(R)}_{+-} = i \mathcal{N}_K f_T(q^2) \dfrac{C_{T5} \lambda_B^{1/2}}{m_B+m_K},
\end{align}

\noindent where the normalization factor also accounts for the remaining phase space, namely, 
\begin{equation}
\left|\mathcal{N}_K(q^2)\right|^2 = \tau_{B_d}\dfrac{\alpha_\mathrm{em}^2 G_F^2 \left| V_{tb}V_{ts}^\ast \right|^2}{512 \pi^5 m_B^3}\dfrac{\lambda_q^{1/2}}{q^2}\lambda_B^{1/2}.
\end{equation}
For shortness, in the above formulas, we used $\lambda_q=\lambda(\sqrt{q^2},m_\ell,m_\ell)$ and $\lambda_B=\lambda(m_B,m_K,\sqrt{q^2})$, where $\lambda(a,b,c)\equiv[a^2-(b-c)^2][a^2-(b+c)^2]$. 
The kinematic conventions and the form factor definitions are collected in Appendix~\ref{app:angular}. In the limit in which the derivative operators vanish we retrieve the usual expression for differential branching fraction~\cite{Becirevic:2016zri}. The choice of dimension-seven operators~\eqref{eq:ope-deriv-2}  is convenient also because their matrix elements are proportional to the original hadronic matrix elements multiplied by $i q^\mu$. As it can be seen from the above expressions the coefficients $C_{i,j}^{\mathcal{T}\ell}$ and $C_{i,j}^{\mathcal{T}q}$ enters the above formulas with the explicit $1/m_W$-suppression factor. In other words, with the above formulas and by using the Wilson coefficients presented in the previous Sections, we see that the derivative operators~\eqref{eq:ope-deriv-2} are indeed irrelevant for phenomenology. Their presence is therefore essential for the unambiguous matching procedure in the computation of Wilson coefficients but they do not alter the phenomenological analysis even at the sub-percent level.

\section{Phenomenology and discussion}
\label{sec:pheno}

In this Section we use our results for Wilson coefficients and compare the experimental data for the exclusive $b\to s\ell^+\ell^-$ modes with various types of 2HDM. 
We decided to focus on $\mathcal{B}(B_s\to \mu^+ \mu^-)^\mathrm{exp}=(2.8^{+0.7}_{-0.6})\times 10^{-9}$~\cite{CMS:2014xfa}, 
and $\mathcal{B}(B\to K \mu^+ \mu^-)_{\mathrm{high}\,q^2}^\mathrm{exp}=(8.5\pm 0.3 \pm 0.4)\times 10^{-8}$~\cite{Aaij:2014pli}, where ``high~$q^2$" means that the decay rate has been 
integrated over the interval $q^2\in [15,22]~\mathrm{GeV}^2$. The reason for opting for these decay modes is that the relevant hadronic uncertainties 
are under good theoretical control. 
The hadronic quantity entering the $B_s\to \mu^+ \mu^-$ decay amplitude is the decay constant, $f_{B_s}$. It has been abundantly computed 
by means of numerical simulations of QCD on the lattice (LQCD) and its value is nowadays one of the most accurately computed hadronic quantities as far as $B_{(s)}$-mesons are concerned~\cite{Aoki:2016frl}. The hadronic form factors entering the  $B\to K \mu^+ \mu^-$ decay amplitude 
have been directly computed in LQCD only in the region of large $q^2$'s~\cite{Bouchard:2013pna,Bailey:2015dka}, which explains why we use $\mathcal{B}(B\to K \mu^+ \mu^-)_{\mathrm{high}\,q^2}^\mathrm{exp}$ to do phenomenology.
Furthermore, since the bin corresponding to $q^2\in [15,22]~\mathrm{GeV}^2$ is rather wide and away from the very narrow charmonium resonances, 
the assumption of quark-hadron duality is likely to be valid~\cite{Beylich:2011aq}. 
By using the recent LQCD results for the form factors provided by HPQCD~\cite{Bouchard:2013pna} and MILC Collaborations~\cite{Bailey:2015dka}, 
the SM results are
\begin{align}
	\mathcal{B}(B\to K \mu^+\mu^-)_{\mathrm{high}\,q^2} &=\left\{\biggl. (10.0\pm 0.5)\times 10^{-8}\biggr|_\mathrm{HPQCD} ,\biggl.(10.7\pm 0.5)\times 10^{-8}\biggr|_\mathrm{MILC} \right\} ,
\end{align}
\noindent both being about $2 \sigma$ larger than the experimental value measured at LHCb.~\footnote{In the following we will average the results obtained by using the two sets of form factors obtained in LQCD.} Since the current disagreement between theory and experiment needs to be corroborated by more data, we decided to impose all the constraints to $3\sigma$ accuracy. 
We will then discuss the impact of $\mathcal{B}(B\to K \mu^+ \mu^-)_{\mathrm{high}\,q^2}^\mathrm{exp}$ on 2HDM if the current discrepancy remains,
 i.e. by requiring the 2HDM to compensate the disagreement between theory (SM) and experiment at the level of $2\sigma$ and more. Notice also that the measured $\mathcal{B}(B_s\to \mu^+ \mu^-)^\mathrm{exp}$ is slightly smaller than predicted, $\mathcal{B}(B_s\to \mu^+ \mu^-)^\mathrm{SM} =(3.65\pm 0.23)\times 10^{-9}$\cite{Bobeth:2013uxa}.

\begin{figure}[!htp]
  \centering
  \includegraphics[width=0.5\textwidth]{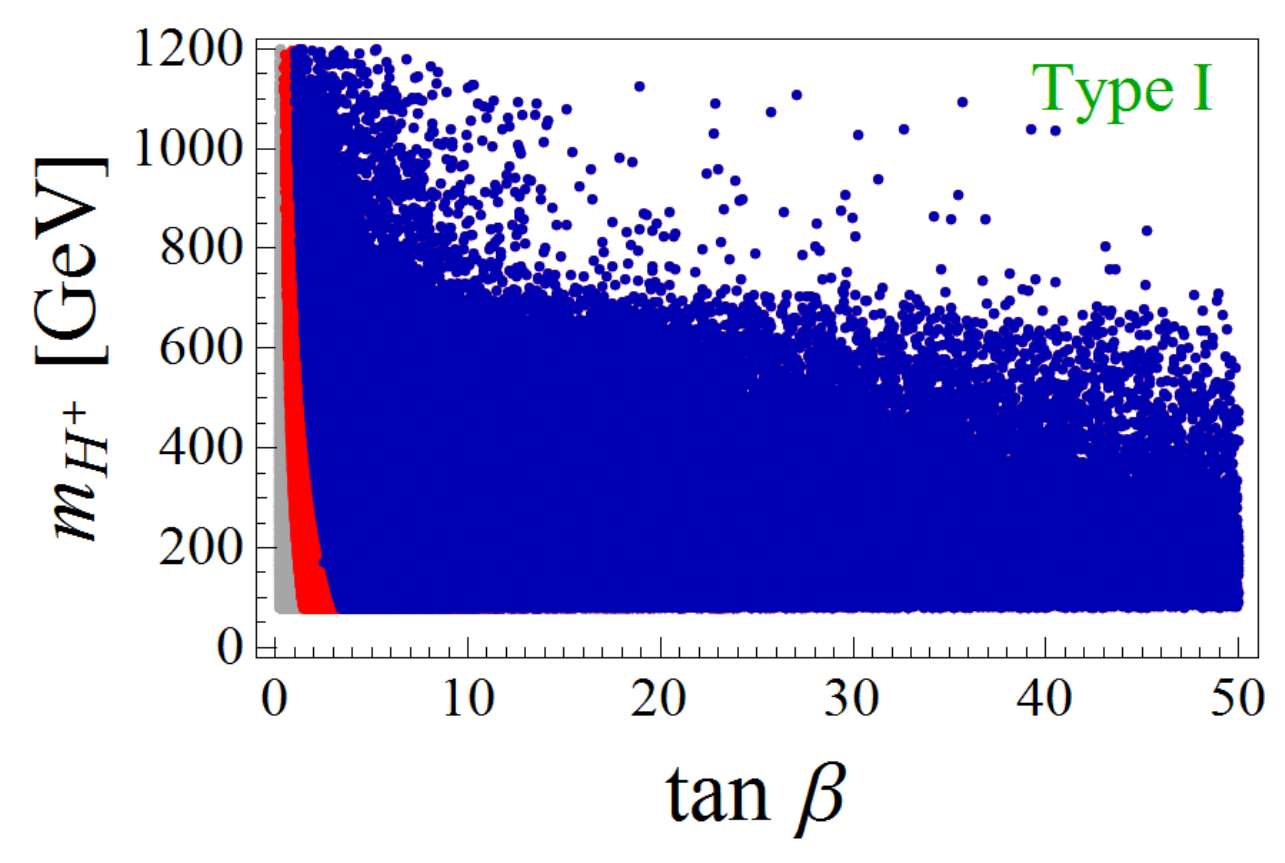}~\includegraphics[width=0.5\textwidth]{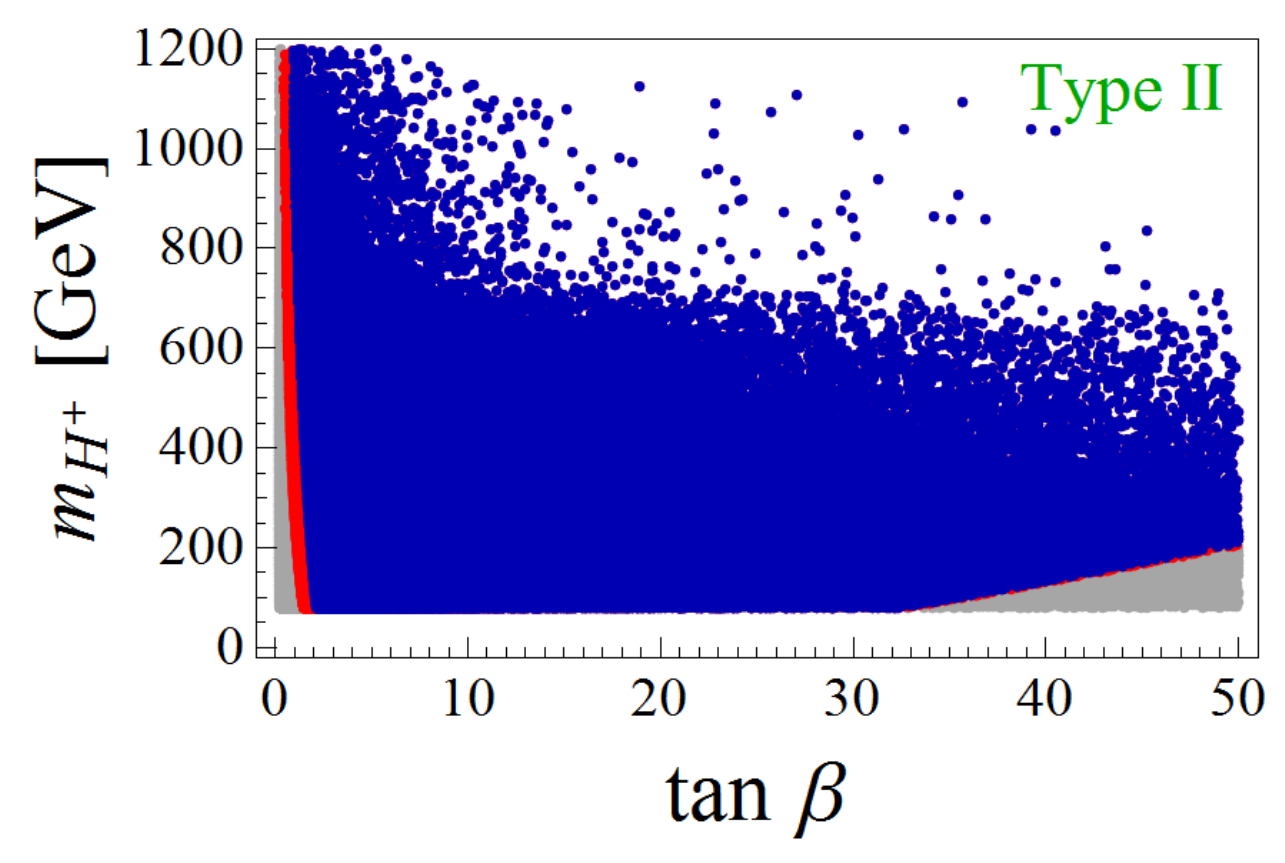}\\
    \includegraphics[width=0.5\textwidth]{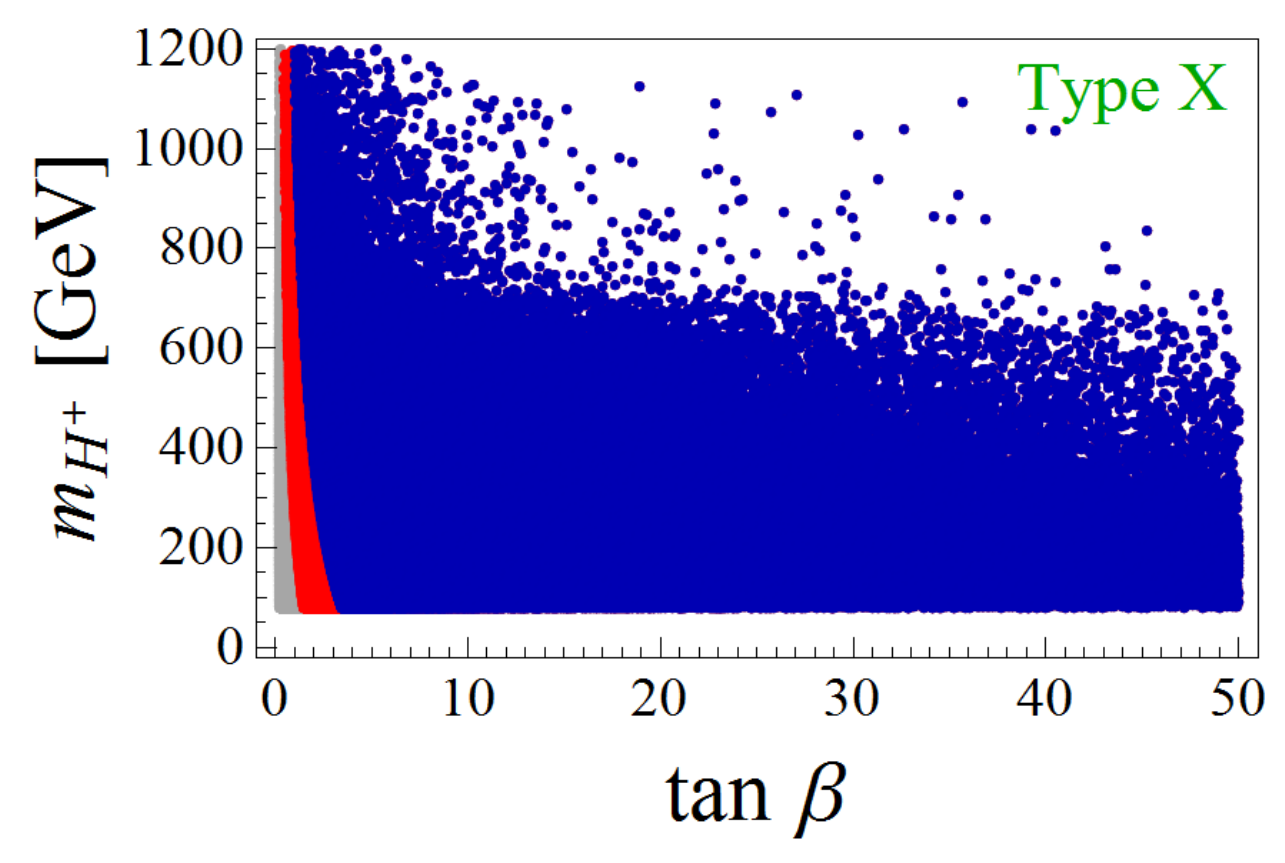}~\includegraphics[width=0.5\textwidth]{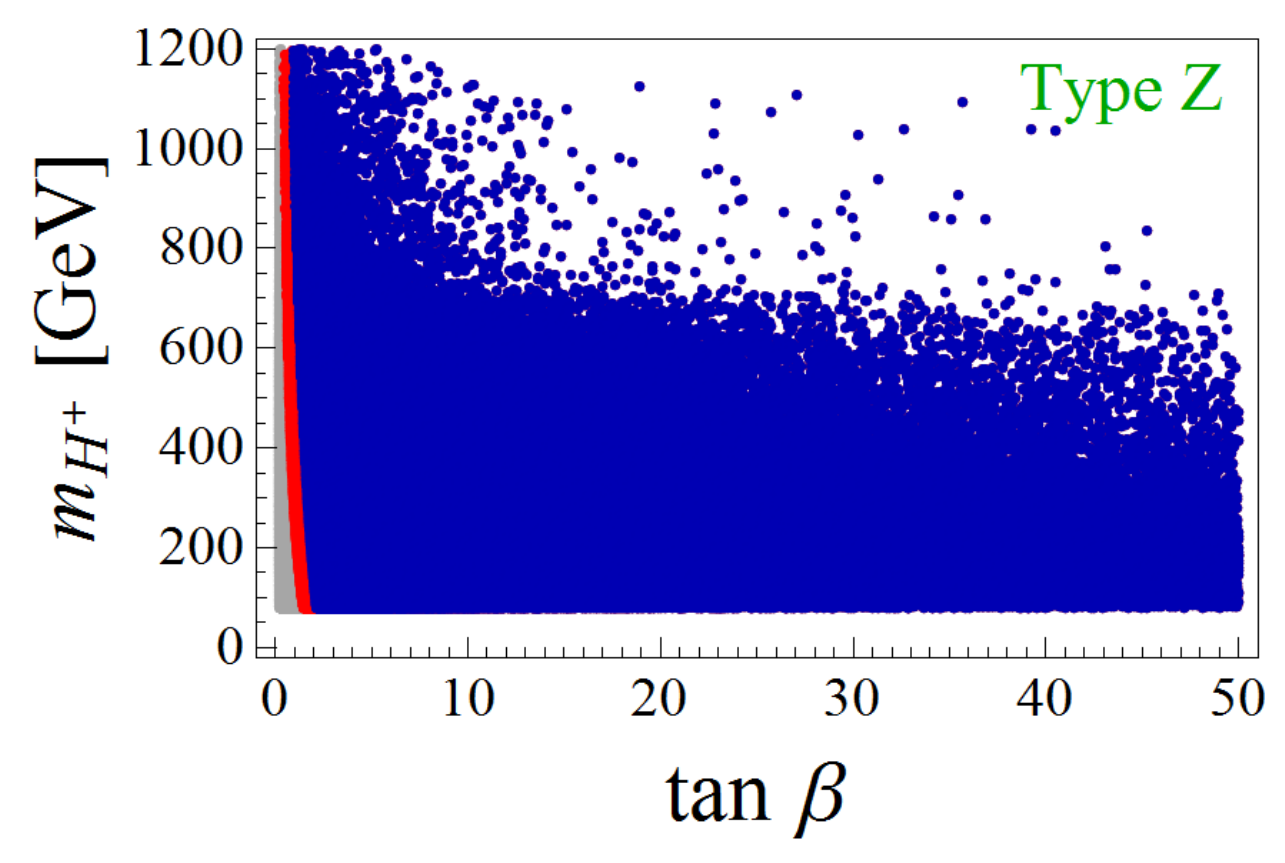}
  \caption{\sl Results of the scan given in Fig.~\ref{fig:scan} after imposing the constraints coming from $\mathcal{B}(B_s\to \mu^+\mu^-)^\mathrm{exp}$ and $\mathcal{B}(B\to K \mu^+ \mu^-)_{\mathrm{high}\,q^2}^\mathrm{exp}$ to $3\sigma$ accuracy. Blue points are allowed by all observables, while gray points are excluded by $\mathcal{B}(B_s\to \mu^+\mu^-)$, and the red ones are excluded by $\mathcal{B}(B\to K \mu^+ \mu^-)_{\mathrm{high}\,q^2}$.}
  \label{fig:scan-3sigma}
\end{figure}

We now use the results of our scan from Sec.~\ref{sec:scan}, require the $3\sigma$ agreement between experiment and theory, which means that we add the generic 2HDM Wilson coefficients derived in the previous Section to the SM values. The result, in the plane $(\tan\beta,m_{H^\pm})$, is shown in Fig.~\ref{fig:scan-3sigma}  for each type of 2HDM discussed in Sec.~\ref{sec:2hdm}. We learn that both $\mathcal{B}(B_s\to \mu^+ \mu^-)$ and $\mathcal{B}(B\to K \mu^+ \mu^-)_{\mathrm{high}\,q^2}$ exclude the low $\tan\beta \lesssim 1$ region regardless of the type of 2HDM considered. The limit of exclusion of low $\tan\beta$ coming from $\mathcal{B}(B\to K \mu^+ \mu^-)_{\mathrm{high}\,q^2}$ is slightly larger than the one arising from $\mathcal{B}(B_s\to \mu^+ \mu^-)$. The limit on low $\tan\beta$ obtained in this way for each of our four models is given in Tab.~\ref{tab:limitstanb}.   

Besides excluding $\tan\beta \lesssim 1$, it may appear as a surprise that the large $\tan \beta$ are not excluded by these data. The reason for that is the fact that the (pseudo-)scalar Wilson coefficient, with respect to the dominant (axial-)vector one, comes with a term proportional to $(m_{B_s}/m_W)^2$ which suppresses the large $\tan
\beta$ values. This feature can be easily verified in the Type~II model for which the coefficients $C_{S,P}$, in the large $\tan\beta$ limit, are given in Eq.~\eqref{eq:largeB_Csp}. This is why only a small number of points have been eliminated from our scan of Type~II model at large $\tan\beta$ but relatively light $m_{H^\pm}$.

\begin{table}[!htbp]
\renewcommand{\arraystretch}{1.5}
\centering
\begin{tabular}{|c|cccc|}
\hline 
Model & Type I & Type II & Type X & Type Z \\ \hline\hline
$\tan\beta$  & $ >1.0$ & $ >0.9$ & $ >1.0$ & $>0.9 $\\ \hline
\end{tabular}
\caption{\label{tab:limitstanb} \sl Allowed values of low $\tan\beta$ (at $99\%$ CL) for the different 2HDMs. See text for details.}
\end{table}

\begin{figure}[ht!]
  \centering
  \includegraphics[width=0.5\textwidth]{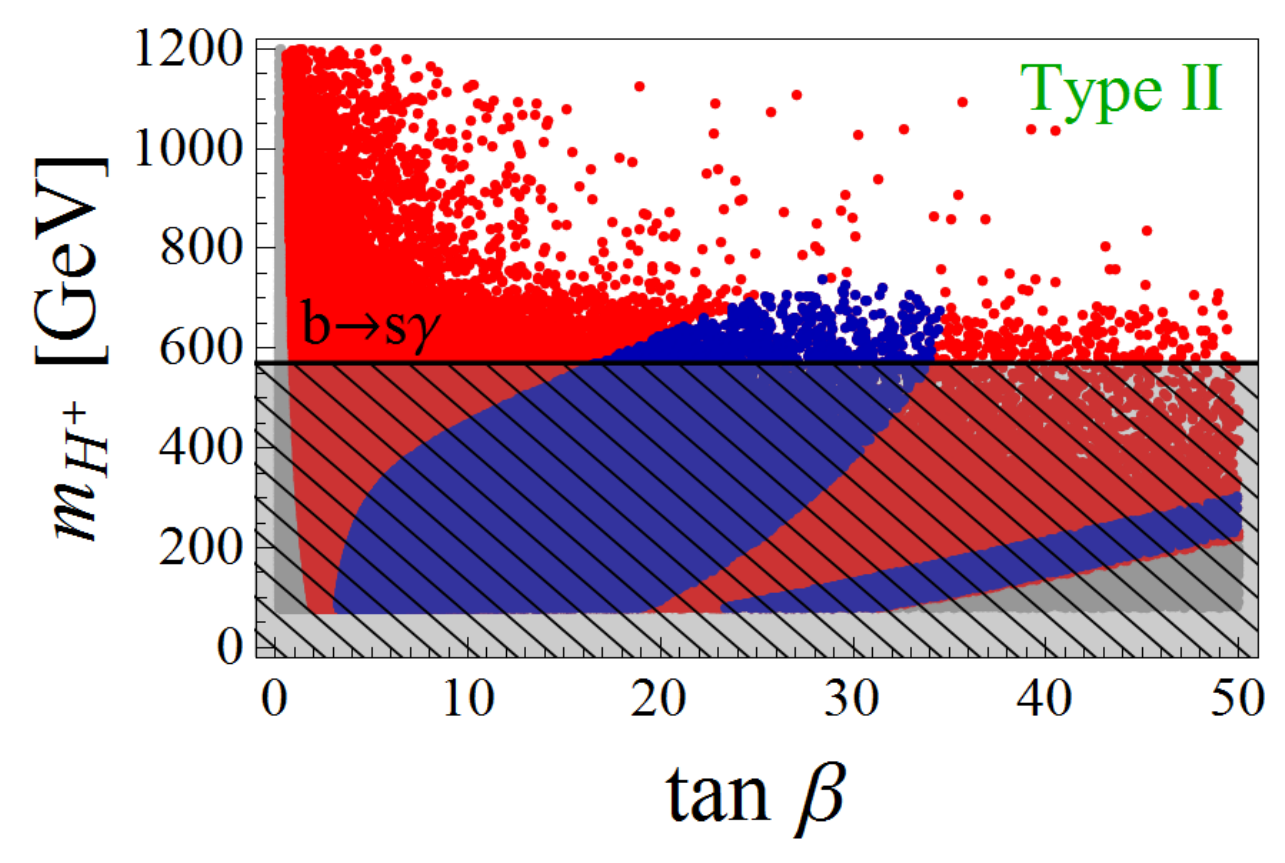}~\includegraphics[width=0.5\textwidth]{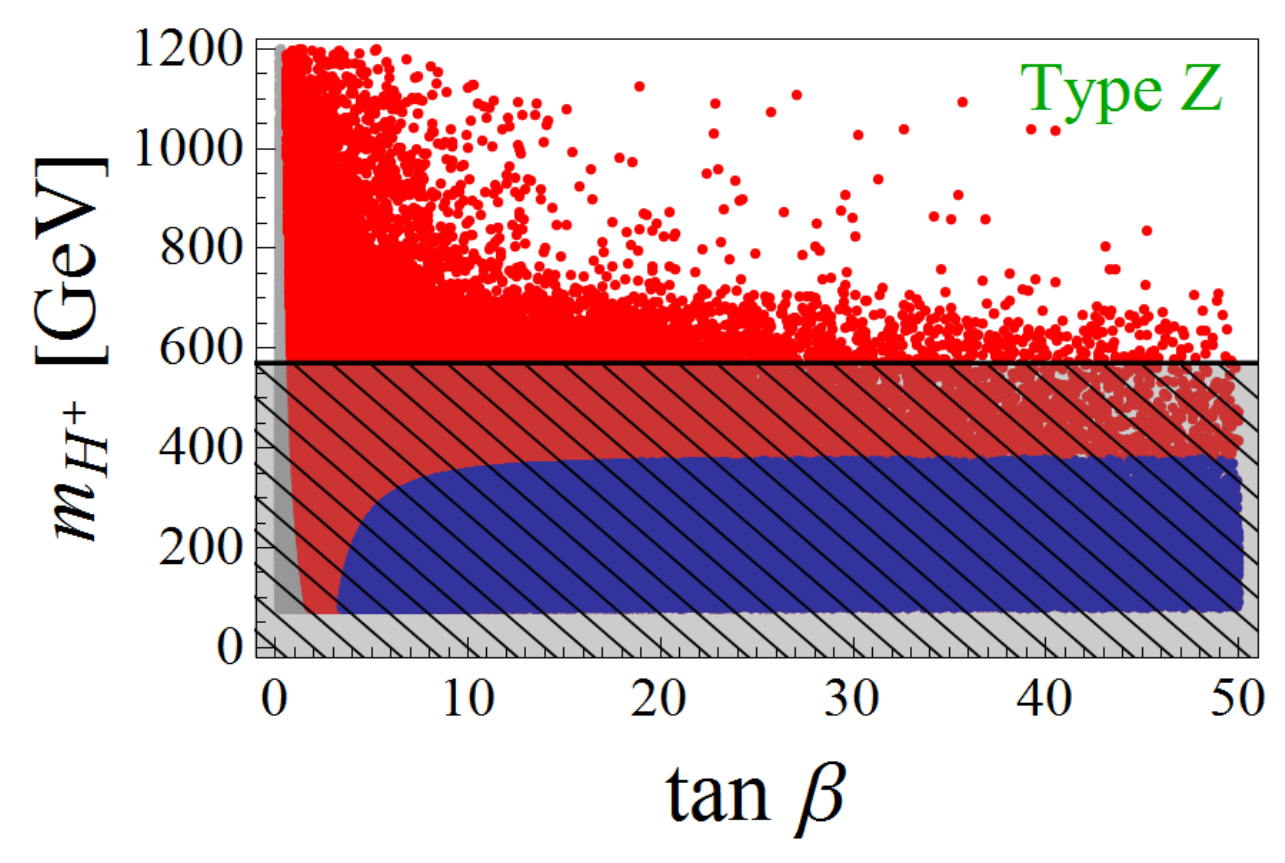}
  \caption{\sl Results of the scan in Fig.~\ref{fig:scan} after imposing the $b\to s$ constraints to $2\sigma$ accuracy. The hatched area is excluded by $\mathcal{B}(B\to X_s \gamma)$ at 95\%~\cite{Misiak:2017bgg}. See Fig.~\ref{fig:scan-3sigma} for the color code.}
  \label{fig:scan-2sigma}
\end{figure}

\begin{figure}[ht]
  \centering
  \includegraphics[width=0.5\textwidth]{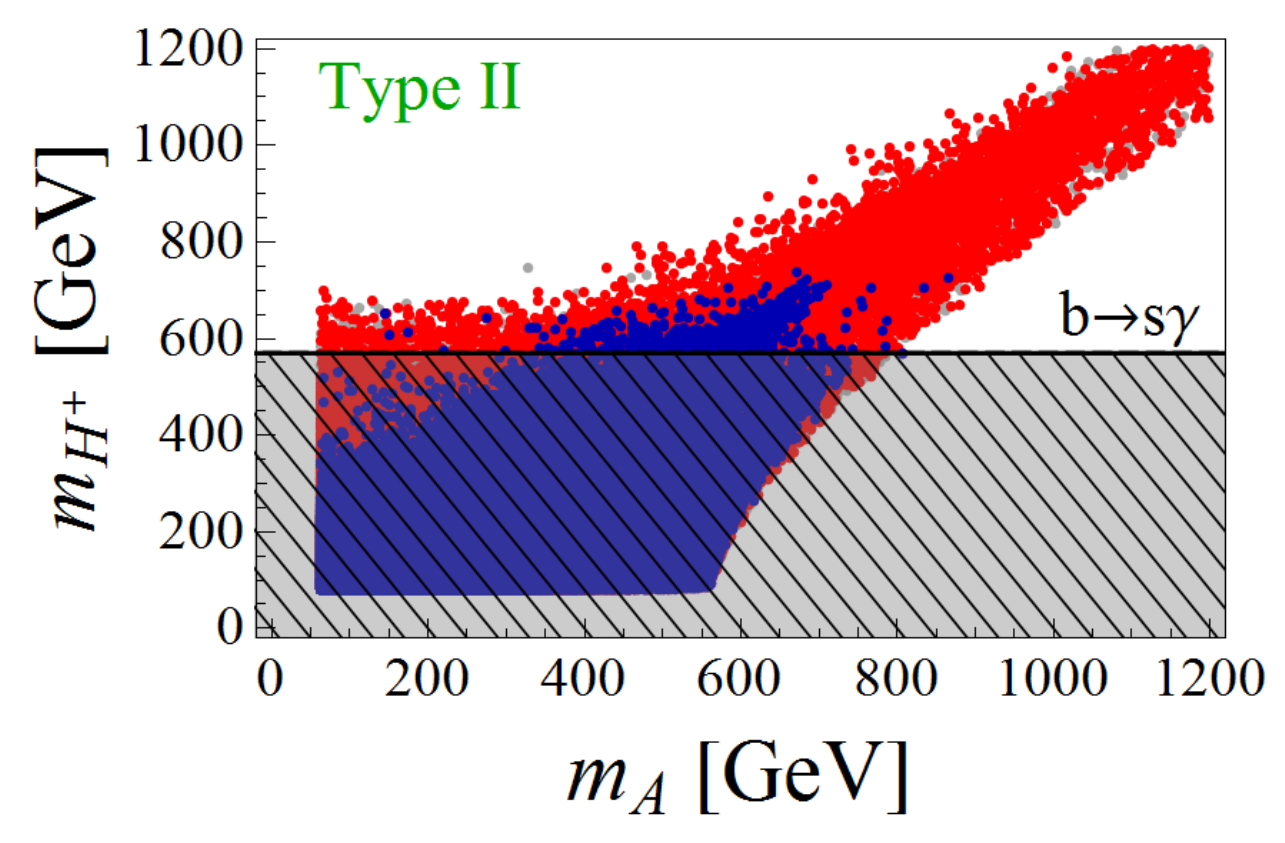}~\includegraphics[width=0.5\textwidth]{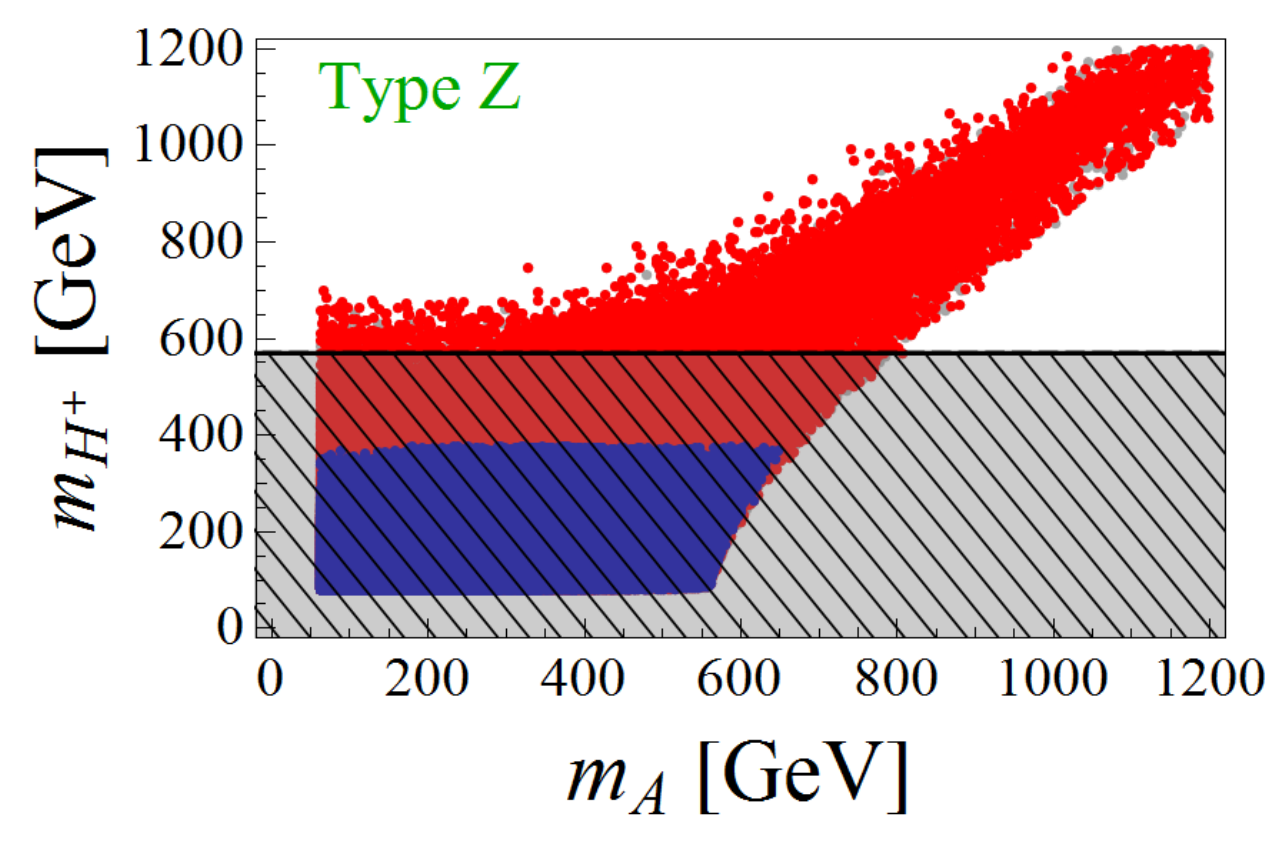}
  \caption{\sl Same as in Fig.~\ref{fig:scan-2sigma} but in the $(m_A,m_{H^\pm})$ plane.}
  \label{fig:scan-2sigma-bis}
\end{figure}

Since the SM value is in slight tension with $\mathcal{B}(B\to K \mu^+ \mu^-)_{\mathrm{high}\,q^2}^\mathrm{exp}$ at the $2.1\sigma$ level, we can now check which of the models discussed in this paper can be made consistent with the experimental data if any disagreement beyond $2\sigma$ between theory (SM) and experiment is to be attributed to 2HDM. It turns out that two such models are Type~II and Type~Z 2HDM, which we illustrate in Fig.~\ref{fig:scan-2sigma}. For the other two scenarios (Type~I and Type~X) the NP contributions are either too small or already in conflict with $\mathcal{B}(B_s\to \mu^+\mu^-)^\mathrm{exp}$. From Figs.~\ref{fig:scan-2sigma} and ~\ref{fig:scan-2sigma-bis} we see that 
in order to explain the discrepancy one needs a relatively light charged scalar: (i) $m_{H^\pm} \lesssim 735~\mathrm{GeV}$ and $\tan\beta > 2.3$ in the Type~II scenario, and (ii) $m_{H^\pm} \lesssim 380~\mathrm{GeV}$ and $\tan\beta > 3.5$ for the Type~Z scenario. Since the masses of the additional scalars are correlated, we see that $m_H$ and $m_A$ become bounded as well, cf.~Fig.~\ref{fig:scan-2sigma-bis}. In the case of Type~II and Type~Z 2HDM an additional bound on the charged Higgs has been recently derived from the inclusive mode $\mathcal{B}(B\to X_s \gamma)$. After comparing the experimental spectra with  theoretical expressions in which the higher order QCD corrections have been included, the lower bound $m_{H^\pm} >570$~GeV (95\% CL) was obtained in Ref.~\cite{Misiak:2017bgg} (c.f. also Ref.\cite{Misiak:2015xwa}). This bound is superposed on our results 
in  Figs.~\ref{fig:scan-2sigma} and~\ref{fig:scan-2sigma-bis}, which then also eliminates Type~Z 2HDM. Furthermore, we can say that the requirement of agreement between theory and experiment to $2\sigma$, for the quantities discussed in this Section, reduces the available space of parameters for Type~II 2HDM to $m_{H^\pm} \in (570,  735)~\mathrm{GeV}$, and $\tan \beta \in (16, 35)$, while the available range of values for the mass of the CP-odd Higgs becomes $m_A\in  (145,  865)~\mathrm{GeV}$.


\begin{figure}[ht]
  \centering
  \includegraphics[width=0.5\textwidth]{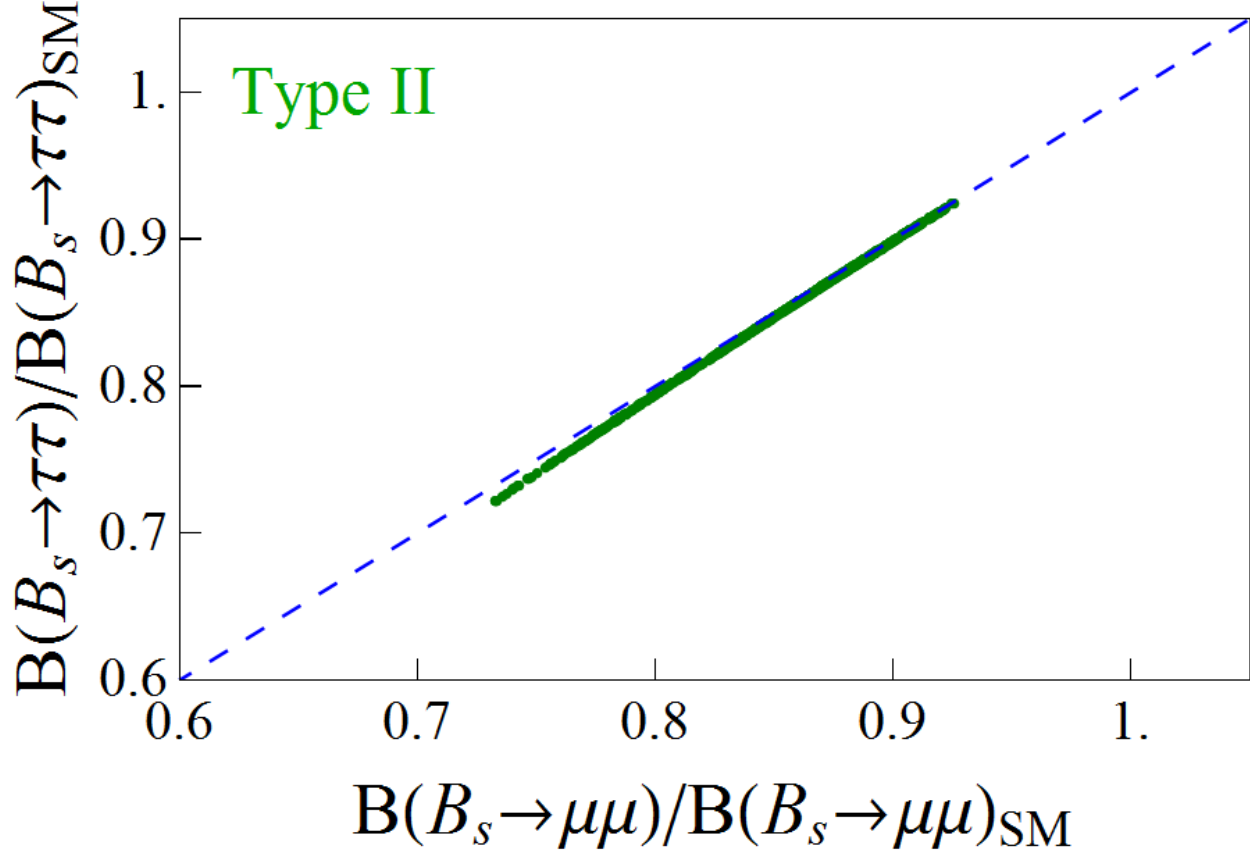}~\includegraphics[width=0.5\textwidth]{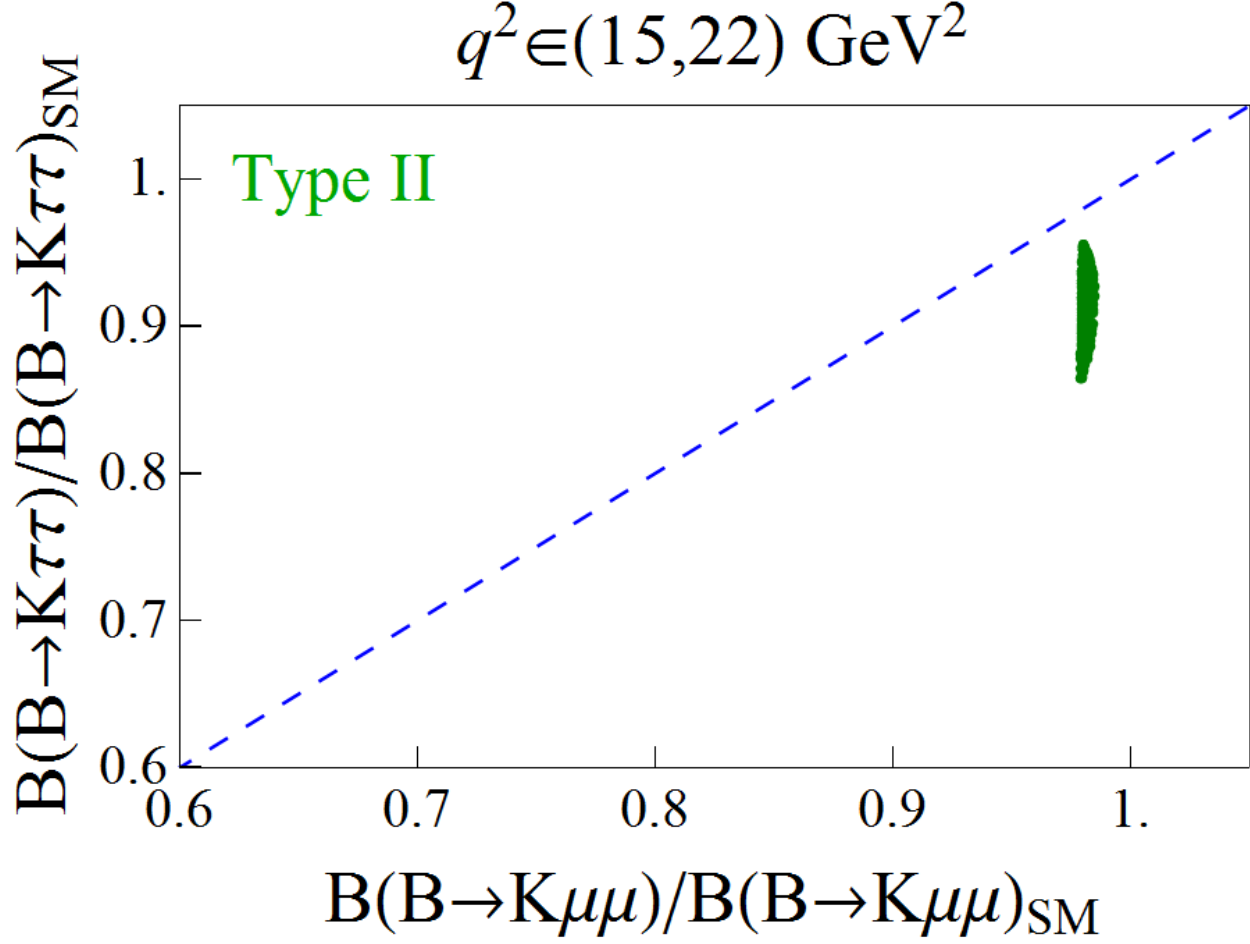}
  \caption{\sl We show the branching fractions of the decay to $\tau$-leptons with respect to their SM predictions, as obtained in the Type~II 2HDM, consistent with experimental results for the decays to muons in the final state. 
  }
  \label{fig:XXX}
\end{figure}

In what follows we will assume that the $2\sigma$ disagreement of the measure $\cb(B\to K\mu^+\mu^-)_{\mathrm{high}\,q^2}^\mathrm{exp}$ with respect to the SM prediction indeed remains as such in the future and discuss 
the consequences on the decays $\cb(B_s\to \tau^+\tau^-)$ and $\cb(B\to K\tau^+\tau^-)_{\mathrm{high}\,q^2}$ if the Type~II 2HDM is used to explain the disagreement.  From Eq.~\eqref{eq:BS} we can see that
\bea
{\cb(B_s\to \tau^+\tau^-)\over \cb(B_s\to \tau^+\tau^-)^{\mathrm{SM}} }= {\cb(B_s\to \mu^+\mu^-)\over \cb(B_s\to \mu^+\mu^-)^{\mathrm{SM}} } - { |C_S^{\tau\tau}|^2\over 
|C_{10}^\mathrm{SM}|^2 } {m_{B_s}^2\over (m_b+m_s)^2}\,,
\eea
where the only remaining $m_\ell$ dependence comes from the last numerator in the factor multiplying  $|C_S-C_S^\prime|^2$ in Eq.~\eqref{eq:BS}. 
In Fig.~\ref{fig:XXX} we illustrate the validity of the above equality. Notice that a tiny departure from equality comes from the large $\tan\beta$ values which enhance the $C_S$ contribution.
In other words, the current experimental result $\cb(B_s\to \mu^+\mu^-)^\mathrm{exp}$, which is slightly lower than the one predicted in the SM, is expected to 
lead to $\cb(B_s\to \tau^+\tau^-)^\mathrm{exp}$ compatible or slightly lower than predicted in the SM. 
The cancellation of the lepton mass in $\cb(B_s\to \ell^+\ell^-)^{\mathrm{2HDM}}$, discussed above,  does not occur in $\cb(B\to K \ell^+\ell^-)^{\mathrm{2HDM}}_{\mathrm{high}-q^2}$. 
As a result we obtain,
\bea
{\cb(B\to K \tau^+\tau^-)^{\mathrm{Type\,II}}\over \cb(B\to K \tau^+\tau^-)^{\mathrm{SM}} } \lesssim {\cb(B\to K \mu^+\mu^-)^{\mathrm{Type\,II}}\over \cb(B\to K\mu^+\mu^-)^{\mathrm{SM}} }\,,
\eea
where we omitted the subscript ``high-$q^2$" to avoid too heavy a notation. Illustration is provided in Fig.~\ref{fig:XXX}. We can rephrase this observation with an 
equivalent statement:
\bea
 {\cb(B\to K \tau^+\tau^-)^{\mathrm{Type~II}}\over \cb(B\to K \mu^+\mu^-)^{\mathrm{Type~II}}} <  
  {\cb(B\to K \tau^+\tau^-)^{\mathrm{SM}}\over \cb(B\to K \mu^+\mu^-)^{\mathrm{SM}}} \,.
\eea
To be fully explicit, we obtain
\bea
\left. {\cb(B\to K \tau^+\tau^-)\over \cb(B\to K \mu^+\mu^-)}\right|_{\mathrm{high}-q^2}\!\! \in (1.12,1.14)_\mathrm{SM}, (1.0,1.1)_{\mathrm{Type~II}}.
\eea
\section{Conclusion}
\label{sec:concl}
In this paper we computed the leading order Wilson coefficients relevant to the exclusive $b\to s\ell^+\ell^-$ decays in the framework of 2HDM with a softly broken $\mathbb{Z}_2$ symmetry. Most of these Wilson coefficients have been computed previously but in the limit of large $\tan\beta$, which we extend here to a generic setup. We also included $\mathcal{O}(m_b)$ corrections, which were neglected in the previous computations. Regarding the (pseudo-)scalar Wilson coefficients, we elucidated the issue of unambiguous matching of the one-loop amplitudes between the full and effective theories which requires extending the basis of operators in the effective theory by including two types of operators suppressed by $1/m_W$ (altogether, eight new operators). 
We pointed out that for the appropriate identification of the $Z$-penguin contribution to the Wilson coefficient $C_P$ it is necessary to keep all external momenta different from zero. 

After having computed the full set of Wilson coefficients we were able to make a phenomenological analysis by focusing on $\cb(B_s\to \mu^+\mu^-)$ and $\cb(B\to K \mu^+\mu^-)_{\mathrm{high}-q^2}$, the quantities which are measured at LHC and for which the hadronic uncertainties are under good theoretical control (computed in LQCD). After carefully scanning the parameter space of 2HDM with a softly broken $\mathbb{Z}_2$ symmetry, we tested various types of 2HDM against the experimental data for $\cb(B_s\to \mu^+\mu^-)^\mathrm{exp}$ and $\cb(B\to K \mu^+\mu^-)_{\mathrm{high}-q^2}^\mathrm{exp}$,
 and found that to $3\sigma$ the values of low $\tan\beta \lesssim 1$ are excluded for all types of 2HDM's considered here.

 If, instead, we require the $2\sigma$ agreement with experiment, then only Type~II and Type~Z models provide a viable description of the data. After combining ours with the bound on the charged Higgs deduced from the inclusive $b\to s\gamma$ decay, we find that the Type~Z model can be discarded and 
\begin{align}
&\mathrm{Type ~II}\ : & m_{H^\pm} \in (570,735)~\gev, && m_A\in (145,865)~\gev, && \tan \beta \in (16,35).  
\end{align}
We also discussed the repercussions of the current results on the decays $\cb(B_s\to \tau^+\tau^-)$ and $\cb(B\to K \tau^+\tau^-)_{\mathrm{high}-q^2}$.

\section*{Acknowledgments}
P.~A. and F.~M. acknowledge the financial support from  FPA2016-76005-C2-1-P, 2014-SGR-104, and project MDM-2014-0369 of ICCUB (Unidad de Excelencia {\em Maria de Maeztu}). F.~M. have further been supported by project FPA2014-61478-EXP.
This project has received funding from the European Union's Horizon 2020 research and innovation program under 
the Marie Sklodowska-Curie grant agreements No. 690575 and No. 674896. 

\clearpage

\appendix

\section{Conventions and Kinematics}
\label{app:angular}

\subsection*{Angular Conventions}
We adopt the same angular conventions for $B(p)\to K (k) \ell^+(p_+)\ell^-(p_-)$ as those used in Ref.~\cite{Becirevic:2016zri,Gratrex:2015hna}. In the $B$-meson rest frame, the leptonic and hadronic four-momenta are chosen as $q^\mu=p_+^\mu+p_-^\mu=(q_0,0,0,q_z)$ and $k^\mu=(k_0,0,0,-q_z)$, where
\begin{align*}
q_0 = \dfrac{m_B^2+q^2-m_K^2}{2 m_B},\qquad k_0=\dfrac{m_B^2+m_{K}^2-q^2}{2m_B},\quad\mathrm{and}\quad q_z=\dfrac{\lambda^{1/2}(m_B,m_K,\sqrt{q^2})}{2 m_B}.
\end{align*}

\noindent In the dilepton rest frame the components of the leptonic four-vectors are given by
\begin{align*}
p_-^\mu &= (E_\ell,|p_\ell|\sin\theta_\ell,0,|p_\ell|\cos\theta_\ell),\\
p_+^\mu &= (E_\ell,-|p_\ell|\sin\theta_\ell,0,-|p_\ell|\cos\theta_\ell),
\end{align*}
where $E_\ell=\sqrt{q^2}/2$, and $\theta_\ell$ is the angle between $\ell^-$ (in the dilepton rest frame) and the line of flight of the two leptons (in the $B$-meson rest frame). The momentum $p_\ell$ is given by
\begin{equation}
|p_\ell|=\dfrac{\lambda^{1/2}(\sqrt{q^2},m_\ell,m_\ell)}{2 m_B}.
\end{equation}

\subsection*{Polarization Vectors}

In the $B$-meson rest frame we take the polarization vectors of the virtual vector boson $V^\ast$ to be:
\begin{align}
\varepsilon^\mu_V(\pm) &= \dfrac{1}{\sqrt{2}}(0,\pm 1,i,0),\quad 
\varepsilon^\mu_V(0)   = \dfrac{1}{\sqrt{q^2}}(q_z,0,0,q_0),\quad 
\varepsilon^\mu_V(t)	&= \dfrac{1}{\sqrt{q^2}}(q_0,0,0,q_z).
\end{align}

\noindent These vectors are orthonormal and satisfy the completeness relation

\begin{equation}
\sum_{n,n^\prime} \varepsilon_V^{\mu\ast}(n)\varepsilon_V^\nu(n^\prime)g_{nn^\prime}=g^{\mu\nu},
\end{equation}
\noindent where $n,n^\prime\in \lbrace t,0,\pm \rbrace$, and $g_{n n^\prime}=\mathrm{diag}(1,-1,-1,-1)$. 

\subsection*{Hadronic matrix elements}
For completeness we also give the definitions of the decay constant ($ f_{B_s}$) and of the form factors [$f_{+,0,T}(q^2)$], quantities which parametrize the hadronic matrix elements relevant to the processes discussed in this paper: 
\begin{align}
\langle 0 | \bar{b} \gamma_\mu \gamma_5 s | B_s(p) \rangle &= i p_\mu f_{B_s},\nn\\
\langle \bar K(k)|\bar{s}\gamma_\mu b|\bar B(p)\rangle &= \Big{[}(p+k)_\mu- \frac{m_B^2-m_K^2}{q^2}q_\mu \Big{]}f_+(q^2)+\frac{m_B^2-m_K^2}{q^2} q_\mu f_0(q^2),\nn\\
\langle  \bar K(k)|\bar{s}  b|\bar B(p)\rangle  &=
{1 \over m_b-m_s}q^\mu  \langle \bar K(k)|\bar{s} \gamma_\mu b|\bar B(p)\rangle  =  {m_B^2-m_K^2 \over m_b-m_s}f_0(q^2) ,\nn \\
\langle \bar K(k)|\bar{s}\sigma_{\mu\nu} b|\bar B(p)\rangle &= -i (p_\mu k_\nu-p_\nu k_\mu)\frac{2 f_T(q^2,\mu)}{m_B+m_K},
\end{align}
where for $B\to K\ell^+\ell^-$ the kinematically accessible $q^2$ values lie in the interval $4m_\ell^2\leq q^2\leq  (m_B-m_K)^2$. Notice that we do not write explicitly the scale dependence of the quark masses, nor of the scalar and tensor densities and of the form factor $f_T(q^2)$. In the actual computations the $\msbar$ values of these quantities are taken at $\mu =m_b$.
\section{Feynman rules}
\label{app:feynman-rules}

In this Appendix we collect the Feynman rules used in our computation. For the couplings between the gauge bosons and the scalars we have

\vspace*{0.3cm}

\begin{minipage}{.3\textwidth}
  \hspace*{0.2cm} \includegraphics[width=0.8\textwidth]{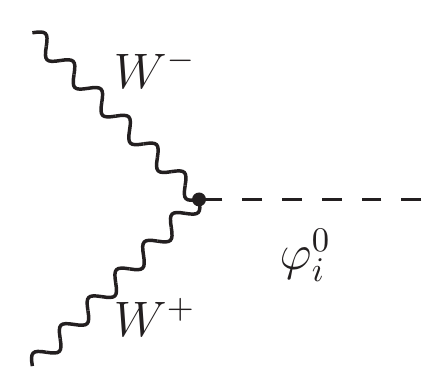}
\end{minipage}%
\begin{minipage}{.7\textwidth}
	\begin{equation}
	\label{eq:WWphi}
		i g m_W \lambda_{W^+W^-}^{\varphi_i^0}\,g^{\mu\nu},
	\end{equation}
\end{minipage}

\vspace*{0.3cm}

\noindent where $\lambda_{W^+W^-}^h=\sin (\beta-\alpha)$, $\lambda_{W^+W^-}^H=\cos (\beta-\alpha)$ and $\lambda_{W^+W^-}^A=0$. Similarly, we also have

\vspace*{0.3cm}

\begin{minipage}{.3\textwidth}
  \centering
  \includegraphics[width=0.8\textwidth]{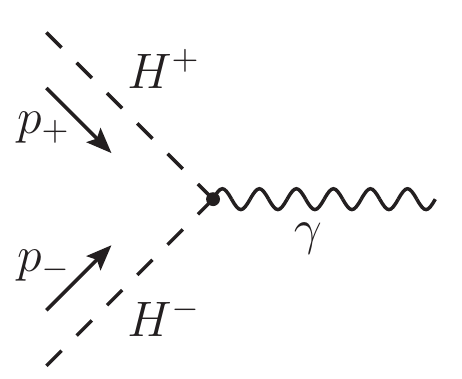}
\end{minipage}%
\begin{minipage}{.7\textwidth}
	\begin{equation}
		i e (p_- - p_+)^\mu,
	\end{equation}
\end{minipage}

\vspace*{0.3cm}

\begin{minipage}{.3\textwidth}
 \hspace*{0.2cm} \includegraphics[width=0.8\textwidth]{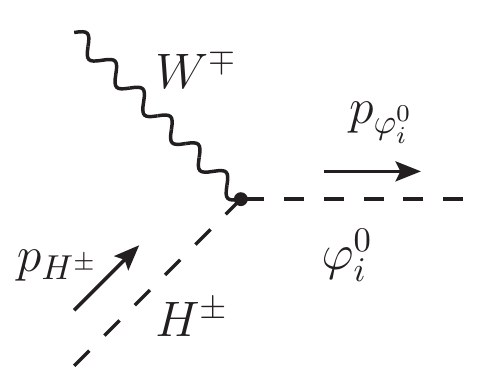}
\end{minipage}%
\begin{minipage}{.7\textwidth}
	\begin{equation}
	\label{eq:WHphi}
		\pm \dfrac{i g}{2}\lambda_{H^\pm W^\mp}^{\varphi_i^0}(p_{H^\pm}+p_{\varphi_i^0})^\mu,
	\end{equation}
\end{minipage}

\vspace*{0.3cm}

\noindent where $\lambda_{H^\pm W^\mp }^h=\cos (\beta-\alpha)$, $\lambda_{H^\pm W^\mp}^H=-\sin (\beta-\alpha)$, and $\lambda_{H^\pm W^\mp }^A=\mp i$, depending on the charges of the initial particles. For the trilinear scalar interactions, we have

\vspace*{0.3cm}

\begin{minipage}{.3\textwidth}
  \hspace*{0.5cm}\includegraphics[width=0.8\textwidth]{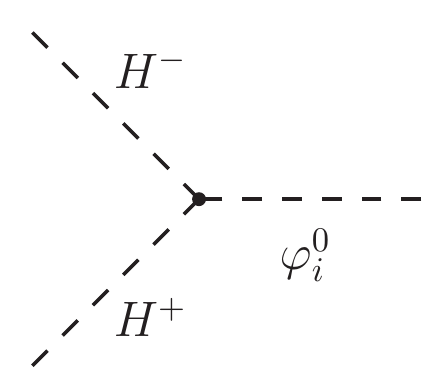}
\end{minipage}%
\begin{minipage}{.7\textwidth}
	\begin{equation}
		i v \lambda_{H^+ H^-}^{\varphi^0_i}
	\end{equation}
\end{minipage}

\vspace*{0.3cm}

\noindent where the trilinear couplings read

\begin{align}
	\lambda_{H^+ H^-}^h &=- \dfrac{m_h^2[3 \cos(\alpha+\beta)+\cos(\alpha-3\beta)]+4\sin(2\beta) \sin(\beta-\alpha)m_{H^\pm}^2-4M^2\cos(\alpha+\beta)}{2 v^2 \sin(2\beta)},\nn \\
	\lambda_{H^+ H^-}^H &= - \dfrac{m_H^2[3 \sin(\alpha+\beta)+\sin(\alpha-3\beta)]+4\sin(2\beta) \cos(\beta-\alpha)m_{H^\pm}^2-4M^2\sin(\alpha+\beta)}{2 v^2 \sin(2\beta)},\nn \\
	\lambda_{H^+ H^-}^A &= 0.
\end{align}

\noindent These results agree with the ones given in Refs.~\cite{Li:2014fea,Kanemura:2004mg} after the appropriate change of basis and/or conventions.~\footnote{Notice that our $\lambda$ is $-\lambda$ of Ref.~\cite{Li:2014fea}.}

\section{Scalar penguins and auxiliary functions}
\label{app:scalar-penguins}

In this Appendix we give the expressions for the Wilson coefficients generated by each diagram shown in Fig.~\ref{fig:scalar-penguins}. We also give the expressions for the auxiliary functions ($f_i$ and $g_i$) used in this paper. 

The penguins arising from coupling to $\varphi_i^0\in \lbrace h,H,A\rbrace$ contribute to the effective coefficient $C_{S,P}$ and can be generically written as

\begin{align}
C_{S}^{\mathrm{NP},\varphi_i^0} &= \dfrac{\sqrt{x_b x_\ell}}{\sin^2\theta_W} \sum_{k=1}^{18} \dfrac{m_t^2}{m_{\varphi_i^0}} \mathrm{Re}\left(\xi_\ell^{\varphi_i^0}\right) \widehat{C}^{k,\varphi_i^0},\\
C_{P}^{\mathrm{NP},\varphi_i^0} &= \dfrac{\sqrt{x_b x_\ell}}{\sin^2\theta_W}\sum_{k=1}^{18} \dfrac{m_t^2}{m_{\varphi_i^0}} i \,\mathrm{Im}\left(\xi_\ell^{\varphi_i^0}\right)\widehat{C}^{k,\varphi_i^0},
\end{align}

\noindent where $\widehat{C}^{k,\varphi_i^0}$ is a common coefficient generated by the diagram $k$, with $k=1,\dots,18$. Since, in our framework,  $\zeta_{\ell}^h,\zeta_{\ell}^H\in \mathbb{R}$ and $\zeta_{\ell}^A\in i\times \mathbb{R}$, it is clear that the CP-even scalars $h$ and $H$ contribute only to $C_S$, whereas the CP-odd Higgs $A$ contributes only to $C_P$, as expected from the assumption of CP conservation. We obtain in the unitary gauge,

\begin{align}
	\widehat{C}^{1,\varphi_i^0} &= \dfrac{\xi_u^{\varphi_i^0}}{4}\Bigg{\lbrace}\zeta_d \zeta_u^\ast \dfrac{x_t}{x_{H^\pm}-x_t}\Bigg{[}1- \dfrac{x_{H^\pm}}{x_{H^\pm}-x_t}\log \left(\dfrac{x_{H^\pm}}{x_t}\right)\Bigg{]}\\[0.25em]
	&\hspace*{1.cm}+|\zeta_u|^2 \dfrac{x_t}{2(x_{H^\pm}-x_t)^2}\Bigg{[}\dfrac{3 x_t-x_{H^\pm}}{2}+\dfrac{x_{H^\pm}(x_{H^\pm}-2x_t)}{x_{H^\pm}-x_t}\log\left(\dfrac{x_{H^\pm}}{x_t}\right)\Bigg{]}\Bigg{\rbrace}\nonumber\\[0.25em]
	&+ \dfrac{\xi_u^{\varphi_i^0\ast}}{4}\Bigg{\lbrace}\zeta_d\zeta_u^\ast \Bigg{[}\Lambda-\dfrac{x_t}{x_{H^\pm}-x_t}-\dfrac{x_{H^\pm}^2}{(x_{H^\pm}-x_t)^2}\log x_{H^\pm}+\dfrac{x_t(2 x_{H^\pm}-x_t)}{(x_{H^\pm}-x_t)^2}\log x_t\Bigg{]}\nonumber\\[0.25em]
	&\hspace*{1cm}+|\zeta_u|^2\dfrac{x_t}{2(x_{H^\pm}-x_t)^2}\Bigg{[}\dfrac{3 x_{H^\pm}-x_t}{2}-\dfrac{x_{H^\pm}^2}{x_{H^\pm}-x_t}\log\left(\dfrac{x_{H^\pm}}{x_t}\right)\Bigg{]}\Bigg{\rbrace},\nonumber\\[0.5em]
	\widehat{C}^{2,\varphi_i^0} &= -\varepsilon_{\varphi_i^0} \dfrac{\sin^2\theta_W\lambda_{H^+H^-}^{\varphi_i^0}}{4\pi\alpha (x_{H^\pm}-x_t)}\Bigg{\lbrace}\zeta_d\zeta_u^\ast\Bigg{[}\dfrac{x_t}{x_{H^\pm}-x_t}\log \left(\dfrac{x_{H^\pm}}{x_t}\right)-1\Bigg{]}\\[0.25em]
	&\hspace{3.5cm}+|\zeta_u|^2\Bigg{[}\dfrac{x_t^2}{2(x_{H^\pm}-x_t)^2}\log \left(\dfrac{x_{H^\pm}}{x_t}\right)+\dfrac{x_{H^\pm}-3 x_t}{4(x_{H^\pm}-x_t)}\Bigg{]}\Bigg{\rbrace},\nonumber\\[0.5em]
	\widehat{C}^{3,\varphi_i^0} &= \dfrac{\xi_d^{\varphi_i^0}}{4}\zeta_d \zeta_u^\ast \Bigg{[}-\Lambda +\dfrac{x_{H^\pm}}{x_{H^\pm}-x_t}\log x_{H^\pm}-\dfrac{x_t}{x_{H^\pm}-x_t}\log x_t \Bigg{]},\\[0.5em]
	\widehat{C}^{4,\varphi_i^0} &= 0,\\[0.5em]
	\widehat{C}^{5,\varphi_i^0} &= \dfrac{1}{4}\Bigg{\lbrace}\xi_u^{\varphi^0_{i}\ast}\Bigg{[}\Lambda-\dfrac{5x_t^2-13 x_t-2}{4(x_t-1)^2}-\dfrac{2x_t^3-6x_t^2+9x_t-2}{2(x_t-1)^3}\log x_t\Bigg{]}\\[0.25em]
	&\hspace*{1cm}+\xi_u^{\varphi^0_{i}}\Bigg{[}\dfrac{\Lambda}{2}-\dfrac{2x_t^2-x_t-7}{4(x_t-1)^2}-\dfrac{x_t^3-3x_t^2+3x_t+2}{2(x_t-1)^2}\log x_t\Bigg{]}\Bigg{\rbrace},\nonumber\\[0.5em]
	\widehat{C}^{6,\varphi_i^0} &= \varepsilon_{\varphi_i^0}\dfrac{\lambda_{W^+W^-}^{\varphi_i^0}}{8}\Bigg{[}-3\Lambda+\dfrac{x_t^2-2x_t-11}{2(x_t-1)^2}+\dfrac{3 x_t(x_t^2-3x_t+4)}{(x_t-1)^3}\log x_t\Bigg{]},\\[0.5em]
	\widehat{C}^{7,\varphi_i^0} &= \widehat{C}^{8,\varphi_i^0}=0,\\[0.5em]
	\widehat{C}^{9,\varphi_i^0} &= \dfrac{\lambda_{H^+W^-}^{\varphi_i^0}}{8}\zeta_u^\ast \Bigg{[}\dfrac{1}{2}-\Lambda + \dfrac{x_{H^\pm}(x_{H^\pm}+2)\log x_{H^\pm}}{(x_{H^\pm}-1)(x_{H^\pm}-x_t)}-\dfrac{x_t(x_t+2)\log x_t}{(x_t-1)(x_{H^\pm}-x_t)}\Bigg{]},\\[0.5em]
	\widehat{C}^{10,\varphi_i^0} &= \dfrac{\lambda_{H^+W^-}^{\varphi_i^0\,\ast}}{4}\Bigg{\lbrace} -\dfrac{\zeta_u}{2}\Bigg{[}\dfrac{x_t(x_{H^\pm}x_t-4x_{H^\pm}+3x_t)}{(x_t-1)(x_{H^\pm}-x_t)^2}\log x_t-\dfrac{x_{H^\pm}(x_{H^\pm}x_t-3 x_{H^\pm}+2 x_t)}{(x_t-1)(x_{H^\pm}-x_t)^2}\log x_{H^\pm}\nonumber\\[0.25em]
	&\hspace*{3.cm}+\dfrac{x_{H^\pm}}{x_{H^\pm}-x_t}\Bigg{]}+\zeta_d\Bigg{[}-\Lambda+\dfrac{x_{H^\pm}\log x_{H^\pm}}{x_{H^\pm}-x_t}-\dfrac{x_t\log x_t}{x_{H^\pm}-x_t}\Bigg{]}\Bigg{\rbrace},
\end{align}

\noindent where the couplings $\lambda_{W^+W^-}^{\varphi_i^0}$ and $\lambda_{H^\pm W^\mp}^{\varphi_i^0}$ are defined below Eq.~\eqref{eq:WWphi} and Eq.~\eqref{eq:WHphi}, respectively.  The coefficient $\varepsilon_{\varphi_i^0}=-1$ for $\varphi_i^0 =A$, and $+1$ otherwise. Moreover, $\Lambda=- \frac{2 \mu^{D-4}}{D-4}-\gamma_E+\log 4\pi -\log\left(\frac{m_W^2}{\mu^2}\right)+1$ contains an ultraviolet divergence which cancels out after summing up all the diagrams. The diagrams $(9.11)$--$(9.18)$ do not contribute in our computation, owing to the fact that we work in the unitary gauge. To make sure that our resulting (total) expressions are gauge independent we performed the computation in the Feynman gauge too. In comparison with Ref.~\cite{Li:2014fea}, we only disagree with one of the signs in the expression for $\widehat{C}^{5,\varphi_i^0}$, which we believe is a typo.

\

The auxiliary functions $g_{0,1,2}$ used in Eq.~\eqref{eq:CS-hH} are defined by

\begin{align}
g_0 &= \dfrac{1}{4(x_{H^\pm}-x_t)}\Bigg{\lbrace} \zeta_d \zeta_u^\ast \Bigg{[} \dfrac{x_t}{x_{H^\pm}-x_t}\log \left(\dfrac{x_{H^\pm}}{x_t}\right)-1\Bigg{]} \\
&\qquad\qquad\qquad+ |\zeta_u|^2 \Bigg{[} \dfrac{x_t^2}{2(x_{H^\pm}-x_t)^2}\log \left(\dfrac{x_{H^\pm}}{x_t}\right)+\dfrac{x_{H^\pm}-3 x_t}{4(x_{H^\pm}-x_t)} \Bigg{]} \Bigg{\rbrace} \nonumber \\[0.5em]
g_1 &= -\dfrac{3}{4}+\zeta_d \zeta_u^\ast \dfrac{x_t}{x_{H^\pm}-x_t}\Bigg{[} 1-\dfrac{x_{H^\pm}}{x_{H^\pm}-x_t}\log \left(\dfrac{x_{H^\pm}}{x_t}\right)\Bigg{]}\\
&\qquad\qquad\qquad+|\zeta_u|^2 \dfrac{x_t}{2(x_{H^\pm}-x_t)^2}\Bigg{[} \dfrac{x_{H^\pm}+x_t}{2}-\dfrac{x_{H^\pm}x_t}{x_{H^\pm}-x_t} \log\left(\dfrac{x_{H^\pm}}{x_t}\right)\Bigg{]}, \nonumber \\[0.5em]
g_2 &= \zeta_d(\zeta_d\zeta_u^\ast+1) f_1 (x_t,x_{H^\pm})+\zeta_d\left(\zeta_u^\ast\right)^2 f_2 (x_t,x_{H^\pm})+\zeta_d \left| \zeta_u\right|^2 f_3 (x_t,x_{H^\pm})\nonumber\\
&+\zeta_u \left| \zeta_u\right|^2 f_4 (x_t,x_{H^\pm})- \zeta_u^\ast \left| \zeta_u\right|^2 f_5 (x_t,x_{H^\pm})+\zeta_u f_6 (x_t,x_{H^\pm})-\zeta_u^\ast f_7 (x_t,x_{H^\pm}),
\end{align}

\noindent with

\begin{align}
f_1(x_t,x_{H^\pm}) &=\frac{1}{2(x_{H^\pm}-x_t)}[-x_{H^\pm}+x_t+x_{H^\pm}\log x_{H^\pm}-x_t \log x_t],\\
f_2(x_t,x_{H^\pm}) &=\frac{1}{2(x_{H^\pm}-x_t)}\Bigg{[}x_t-\frac{x_{H^\pm}x_t}{x_{H^\pm}-x_t}\log \left( \dfrac{x_{H^\pm}}{x_t}\right)\Bigg{]},\\
f_3(x_t,x_{H^\pm}) &=\frac{1}{2(x_{H^\pm}-x_t)}\Bigg{[}x_{H^\pm}-\frac{x_{H^\pm}^2 \log x_{H^\pm}}{x_{H^\pm}-x_t} + \frac{x_t(2 x_{H^\pm}-x_t)\log x_t}{x_{H^\pm}-x_t}\Bigg{]},\\
f_4(x_t,x_{H^\pm}) &= \frac{1}{4(x_{H^\pm}-x_t)^2} \Bigg{[} \frac{x_t (3 x_{H^\pm}-x_t)}{2}-\frac{x_{H^\pm}^2 x_t}{x_{H^\pm}-x_t}\log \left( \dfrac{x_{H^\pm}}{x_t}\right)\Bigg{]} ,\\
f_5(x_t,x_{H^\pm}) &=  \frac{1}{4(x_{H^\pm}-x_t)^2} \Bigg{[} \frac{x_t (x_{H^\pm}-3 x_t)}{2} - \frac{x_{H^\pm} x_t(x_{H^\pm}-2x_t)}{x_{H+}-x_t}\log \left( \dfrac{x_{H^\pm}}{x_t}\right)\Bigg{]},\\
f_6(x_t,x_{H^\pm}) &= \frac{1}{2(x_{H^\pm}-x_t)}\Bigg{[}\frac{x_t(x_t^2-3 x_{H^\pm} x_t+9 x_{H^\pm}-5 x_t-2)}{4(x_t-1)^2}\\
&+\frac{x_{H^\pm}(x_{H^\pm}x_t-3 x_{H^\pm}+2x_t)\log x_{H^\pm}}{2 (x_{H^\pm}-1)(x_{H^\pm}-x_t)}\nonumber\\
&+\frac{x_{H^\pm}^2(-2 x_t^3+6 x_t^2-9 x_t+2)+3 x_{H^\pm}x_t^2(x_t^2-2 x_t+3)-x_t^2(2 x_t^3-3 x_t^2+3x_t+1)}{2 (x_t-1)^3(x_{H^\pm}-x_t)}\log x_t\Bigg{]},\nonumber\\
f_7(x_t,x_{H^\pm}) &= \frac{1}{2(x_{H^\pm}-x_t)}\Bigg{[}\frac{(x_t^2+x_t-8)(x_{H^\pm}-x_t)}{4(x_t-1)^2}-\frac{x_{H^\pm}(x_{H^\pm}+2)}{2(x_{H^\pm}-1)}\log x_{H^\pm} \\&+\frac{x_{H^\pm}(x_t^3-3x_t^2+3x_t+2)+3(x_t-2)x_t^2}{2(x_t-1)^3}\log x_t\Bigg{]}\nonumber.
\end{align}

\noindent Notice that in the above expressions we assumed the couplings $\zeta_f \in \mathbb{C}$ in order to keep our formulas as general as possible, 
although in the body of the paper we consistently used $\zeta_f\in \mathbb{R}$.

\subsection{Wilson Coefficients for the Derivative Operators}
\label{app:wc-dim7}

In this subsection we present the explicit expressions for the Wilson coefficients relevant to the derivative operators given in Eq.~\eqref{eq:ope-deriv-2}. 
From the $Z$-penguins we obtain,
\begin{align}
\begin{split}
C_{RR}^{\mathcal{T}q} &= |\zeta_u|^2\dfrac{\sqrt{x_b}x_t}{72}\Bigg{\lbrace}\frac{3(x_{H^\pm}^2-5 x_{H^\pm}x_t-2x_t^2)}{(x_{H^\pm}-x_t)^3}+\frac{18 x_{H^\pm}x_t^2}{(x_{H^\pm}-x_t)^4}\log \left( \frac{x_{H^\pm}}{x_t}\right)\\
&-2 \sin^2\theta_W \Bigg{[}\dfrac{7 x_{H^\pm}^2-5 x_{H^\pm}x_t-8 x_t^2}{(x_{H^\pm}-x_t)^3}-\dfrac{6 x_{H^\pm} x_t(2 x_{H^\pm}-3x_t)}{(x_{H^\pm}-x_t)^4}\log \left( \frac{x_{H^\pm}}{x_t}\right)\Bigg{]} \Bigg{\rbrace}\\
&+\zeta_u^\ast \zeta_d\dfrac{\sqrt{x_b}x_t}{24}\Bigg{\lbrace}\dfrac{3(x_{H^\pm}-3 x_t)}{(x_{H^\pm}-x_t)^2}-\dfrac{6 x_{H^\pm}(x_{H^\pm}- 2 x_t)}{ x_{H^\pm}-x_t}\log \left( \frac{x_{H^\pm}}{x_t}\right)\\
&+4\sin^2\theta_W\left[\dfrac{5 x_t-3 x_{H^\pm}}{(x_{H^\pm}-x_t)^2}+\dfrac{2 x_{H^\pm}(2x_{H^\pm}-3x_t)}{(x_{H^\pm}-x_t)^3}\log \left( \frac{x_{H^\pm}}{x_t}\right)\right]\Bigg{\rbrace},
\end{split}
\end{align}
and $C_{RL}^{\mathcal{T}q} = C_{RR}^{\mathcal{T}} \left(1-\dfrac{1}{2 \sin^2\theta_W}\right)$. 
Similarly, from the box diagrams we get
\begin{align}
\begin{split}
C^{\mathcal{T}\ell}_{LL}= &-\zeta_u\zeta_\ell^\ast \frac{\sqrt{x_\ell}x_t}{4(x_{H^\pm}-x_t)\sin^2\theta_W}\Bigg{[}-\frac{1}{(x_{H^\pm}-1)}+\dfrac{x_{H^\pm}(1-x_{H^\pm})\log x_t}{(x_{H^\pm}-x_t)(x_t-1)(x_{H^\pm}-1)}\\
&-\dfrac{x_{H^\pm}(x_t+1-2 x_{H^\pm})\log x_{H^\pm}}{(x_{H^\pm}-x_t)(x_{H^\pm}-1)^2}\Bigg{]},
\end{split}
\end{align}
and $C^{\mathcal{T}\ell}_{LL}=(C^{\mathcal{T}\ell}_{LR})^\ast$.

\subsection{Wilson Coefficients Suppressed by $m_\ell$}
\label{app:wc-xl-suppr}
In addition to the Wilson coefficients given in Section~\ref{sec:2hdm}, in the computation of the box diagrams one gets contributions suppressed by the lepton mass. 
For completeness, we give these contributions here. We obtain:
\begin{align}
\begin{split}
C_{T(5)}^\mathrm{NP,\,box} =\zeta_u^\ast\zeta_\ell \; &\dfrac{\sqrt{x_b x_\ell}x_t}{32(x_{H^\pm}-x_t)\sin^2\theta_W} \times \\ 
&\Bigg{[} \frac{x_t \log x_t}{(x_t-1)(x_{H^\pm}-x_t)}-\frac{x_{H^\pm} \log x_{H^\pm}}{(x_{H^\pm}-1)(x_{H^\pm}-x_t)}+\frac{x_t-\log x_t-1}{(x_t-1)^2}\Bigg{]},
\end{split}
\end{align}
and
\begin{align}
C_9^\mathrm{NP,\,box} &= \frac{x_\ell x_t}{16\sin^2\theta_W}\Bigg{\lbrace}|\zeta_u|^2 |\zeta_\ell|^2\Bigg{[}-\frac{1}{x_{H^\pm}-x_t}+\frac{x_t}{(x_{H^\pm}-x_t)^2}\log\left(\frac{x_{H^\pm}}{x_t}\right)\Bigg{]}\\
&+2 \mathrm{Re}[\zeta_u\zeta_\ell^\ast] \Bigg{[}\dfrac{(x_{t}+2)\log x_t}{(x_{H^\pm}-x_t)(x_t-1)} -\dfrac{(x_{H^\pm}+2)\log x_{H^\pm}}{(x_{H^\pm}-x_t)(x_{H^\pm}-1)} \Bigg{]}\Bigg{\rbrace} +2\sqrt{x_\ell} \,\mathrm{Re}\left(C_{LL}^{\mathcal{T}\ell}\right),\nonumber\\
C_{10}^\mathrm{NP,\,box} &=  \frac{x_\ell x_t}{16\sin^2\theta_W}\Bigg{\lbrace}|\zeta_u|^2 |\zeta_\ell|^2\Bigg{[}-\frac{1}{x_{H^\pm}-x_t}+\frac{x_t}{(x_{H^\pm}-x_t)^2}\log\left(\frac{x_{H^\pm}}{x_t}\right)\Bigg{]}\\
&+2 \mathrm{Re}[\zeta_u\zeta_\ell^\ast] \Bigg{[}\dfrac{(x_{t}-2)\log x_t}{(x_{H^\pm}-x_t)(x_t-1)} -\dfrac{(x_{H^\pm}-2)\log x_{H^\pm}}{(x_{H^\pm}-x_t)(x_{H^\pm}-1)} \Bigg{]}\Bigg{\rbrace}.\nonumber
\end{align}

\end{document}